\documentclass[aps,pra,nobalancelastpage,superscriptaddress,twocolumn,reprint]{revtex4-2}
\usepackage[table]{xcolor}
\usepackage{standalone, gensymb, tensor, amsmath,amsfonts,amssymb,amsthm,bbm,bm, braket}
\usepackage[T1]{fontenc}
\usepackage[inline]{enumitem}
\usepackage{scalerel, physics}
\usepackage{wasysym}
\usepackage{comment}
\usepackage{enumerate}
\usepackage{graphicx}
\usepackage{mathtools}
\usepackage{soul}
\usepackage{xparse}
\usepackage{dsfont}
\usepackage{ifthen}
\usepackage{tensor}
\usepackage{tikz}
\usepackage{pgfplots}
\usepackage{pgfplotstable}

\usepackage{tikz-network}
\usepackage{simplewick} 
\usepackage[tight,FIGTOPCAP]{subfigure}
\usetikzlibrary{patterns,decorations.pathreplacing}
\theoremstyle{plain}        

\theoremstyle{plain}

\usepackage{color}
\definecolor{OliveGreen}{RGB}{85,107,47}
\definecolor{NavyBlue}{RGB}{0,0,128}
\usepackage{tensor}
\newcommand{\ttot}{\tau}
\newcommand{\tA}{\tau_A}
\newcommand{\alessandro}[1]{{\color{teal}#1}}
\newcommand{\tAb}{{\tau_{\bar A}}}
\newcommand{\be}{\begin{equation}}
	\newcommand{\ee}{\end{equation}}
\newcommand{\ba}{\begin{aligned}}
	\newcommand{\ea}{\end{aligned}}
\newcommand{\bw}{\begin{widetext}}
	\newcommand{\ew}{\end{widetext}}
\newcommand{\1}{\mathbbm{1}}

\newtheorem{theorem}{Theorem}
\theoremstyle{plain}

\theoremstyle{plain}

\usepackage[colorlinks,bookmarks=false,citecolor=NavyBlue,linkcolor=OliveGreen,urlcolor=blue]{hyperref}
\usepackage{nicematrix}
\newcommand{\mcirc}{\mathbin{\scalerel*{\fullmoon}{G}}}
\newcommand{\msquare}{\mathord{\scalerel*{\square}{G}}}

\newcommand{\dualave}[1]{\mathbb{E}_{{\boldsymbol u}}\left[#1\right]}

\newcommand{\I}{\mathds{1}}
\newcommand{\udots}{\mathbin{\text{\rotatebox[origin=c]{45}{${\cdot}{\cdot}{\cdot}$}}}}

\newtheorem{asp}{Assumption}

\graphicspath{{./}{./images/}}

\usepackage{xifthen}
\usepackage{xargs}

\newcommandx{\fineq}[5][1=-.8ex,2=1,3=1,5=0]{
	\begin{tikzpicture}[baseline={([yshift=#1]current  bounding  box.center)}, scale = #2, every node/.style={scale = #3},rotate around={#5:(0,0)},every node/.style={transform shape}]
		#4
	\end{tikzpicture}
}

\definecolor{bertinired}{RGB}{232,102,102}
\definecolor{bertiniblue}{RGB}{101,147,245}
\definecolor{bertinigreyblue}{RGB}{166,218,149}
\definecolor{bertinigreyred}{RGB}{232,102,102}
\definecolor{bertinivioletc}{RGB}{45,130,60}
\definecolor{bertinigreen}{RGB}{166,218,149}
\definecolor{bertiniorange}{RGB}{255, 116, 23}
\definecolor{OliveGreen}{RGB}{85,107,47}
\definecolor{NavyBlue}{RGB}{0,0,128}
\definecolor{bertiniviolet}{RGB}{210,145,178}
\definecolor{bertinigrey1}{RGB}{98,98,98}
\definecolor{bertinigrey2}{RGB}{211,211,211}
\definecolor{bertinigrey3}{RGB}{192,192,192}
\definecolor{bertinigrey4}{RGB}{169,169,169}

\newcommandx{\tikzdiagup}{
	\tikz {\draw[thick] (0,0)--(0.15,0.15); \draw (0,0) rectangle (0.15,0.15);}
}

\newcommandx{\gatecross}[1][1=0.5]{
	\pgfmathparse{#1/2.0}
	\let\x\pgfmathresult
	\draw[thick] (-\x,-\x) -- (\x,\x);
	\draw[thick] (\x,-\x) -- (-\x,\x);
}
\newcommandx{\gatesqu}[2][1=0.25,2=]{
	\pgfmathparse{#1/2.0}
	\let\x\pgfmathresult
	\ifthenelse{\equal{#2}{}}{
		\draw[thick, fill=white, rounded corners=2pt] (-\x,\x) rectangle (\x,-\x);
	}{
		\draw[thick, fill=#2, rounded corners=2pt] (-\x,\x) rectangle (\x,-\x);
	}
}

\newcommandx{\gatemark}[2][1=0.075,2=tr]{
	\pgfmathparse{#1}
	\let\l\pgfmathresult
	\ifthenelse{\equal{#2}{topleft}}{
		\draw[thick] (0,\l) -- ++(-\l,0) --++ (0,-\l);
	}{}
	\ifthenelse{\equal{#2}{topright}}{
		\draw[thick] (0,\l) -- ++(\l,0) --++ (0,-\l);
	}{}
	
	\ifthenelse{\equal{#2}{bottomleft}}{
		\draw[thick] (0,-\l) -- ++(-\l,0) --++ (0,\l);
	}{}
	\ifthenelse{\equal{#2}{bottomright}}{
		\draw[thick] (0,-\l) -- ++(\l,0) --++ (0,\l);
	}{}
	
}

\newcommandx{\roundgate}[5][1=0,2=0,3=1,4=topright,5=white]{
	\pgfmathparse{#3}
	\let\l\pgfmathresult
	\begin{scope}[shift={(#1,#2)}]
		\gatecross[\l]
		
		\pgfmathparse{\l/2.0}
		\let\s\pgfmathresult
		\gatesqu[\s][#5]
		
		\pgfmathparse{\l*0.15}
		\let\m\pgfmathresult
		\gatemark[\m][#4]
	\end{scope}
}
\newcommandx{\wcirc}[2]{\begin{scope}
		\draw[fill=white] (#1,#2) circle (0.15);	\end{scope}} 
\newcommandx{\wcircc}[2]{\begin{scope}
		\draw[fill=white] (#1,#2) circle (0.13);	\end{scope}} 

\newcommandx{\wsqr}[2]{\begin{scope}
		\draw[fill=white,shift={(#1,#2)}] (-.13,.13) rectangle (.13,-.13);	\end{scope}} 
\newcommandx{\wsqrr}[2]{\begin{scope}
		\draw[fill=white,shift={(#1,#2)}] (-.11,.11) rectangle (.11,-.11);	\end{scope}} 
\newcommandx{\bcirc}[2]{\begin{scope}
		\draw[fill=black] (#1,#2) circle (0.15);	\end{scope}} 
\newcommandx{\thetastate}[4][1=0,2=0,3=1,4=]{
	\pgfmathparse{#3/2}
	\let\l\pgfmathresult
	\pgfmathparse{\l*0.15}
	\let\m\pgfmathresult
	\begin{scope}[shift={(#1,#2)}]
		\draw[thick] (0,0)--(\l,\l);
		\draw[thick] (0,0)--(-\l,\l);
		\ifthenelse{\equal{#4}{}}{
			\draw[fill=white] (0,0) circle (0.15);
		}{
			\draw[thick, fill=#4] (0,0) circle (0.15);
		}
	\end{scope}
}

\newcommandx{\thetastateflipped}[4][1=0,2=0,3=1,4=]{
	\pgfmathparse{#3/2}
	\let\l\pgfmathresult
	\pgfmathparse{\l*0.15}
	\let\m\pgfmathresult
	\begin{scope}[shift={(#1,#2)}]
		\draw[thick] (0,0)--(\l,-\l);
		\draw[thick] (0,0)--(-\l,-\l);
		\ifthenelse{\equal{#4}{}}{
			\draw[fill=white] (0,0) circle (0.15);
		}{
			\draw[thick, fill=#4] (0,0) circle (0.15);
		}
	\end{scope}
}

\newcommandx{\vertgate}[5][1=0,2=0,3=4,4=bertiniorange,5=topright]
{
	\begin{scope}[shift={(#1,#2)}]
		\ifthenelse{\equal{#3}{1}}{
			\roundgate[0][0][1][#5][#4]
		}{
			\foreach \n[evaluate=\n as \y using {2*\n-2}] in {1,...,#3}{
				\roundgate[0][\y][1][#5][#4]
			}
		}
	\end{scope}
}

\newcommandx{\tsfmatV}[8][1=0,2=0,3=l,4=4,5=tr,6=init,7=bertiniorange,8=topright]{
	\begin{scope}[shift={(#1,#2)}]
		\ifthenelse{\equal{#3}{l}}{
			\pgfmathsetmacro{\flag}{0}
		}{
			\pgfmathsetmacro{\flag}{1}
		}
		
		\foreach \y[evaluate=\y as \x using {mod(\y+\flag,2)}] in {1,...,#4}{
			\roundgate[\x][\y][1][#8][#7]
		}
		\ifthenelse{\equal{#5}{tr}}{
			\foreach \y[evaluate=\y as \x using {mod(\y+\flag,2)}] in {#4}{
				\draw [fill=white] (\x-0.5,\y+0.5) circle (0.15);
				\draw [fill=white] (\x+0.5,\y+0.5) circle (0.15);
			}
		}{}
		\ifthenelse{\equal{#6}{init}}{
			\thetastate[\flag][0][1][#7]
		}{}
	\end{scope}
}

\newcommandx{\leftriangle}[5][1=0,2=0,3=4,4=bertiniorange,5=topright]{
	\begin{scope}[shift={(#1,#2)}]
		\pgfmathsetmacro{\t}{#3}
		\pgfmathsetmacro{\steps}{ceil(\t/2)}
		\foreach \i[evaluate=\i as \x using -\t+2*\i-1, evaluate=\i as \ylim using \t-2*\i+2] in {1,...,\steps}{
			\foreach \y[evaluate=\y as \thisx using {\x+\y-1}] in {1,...,\ylim}{
				\roundgate[\thisx][\y][1][#5][#4]
			}
		}
	\end{scope}
}

\newcommandx{\rightriangle}[5][1=0,2=0,3=4,4=bertiniorange,5=topright]{
	\begin{scope}[shift={(#1,#2)}]
		\pgfmathsetmacro{\t}{#3}
		\pgfmathsetmacro{\steps}{ceil(\t/2)}
		\foreach \i[evaluate=\i as \x using -\t+2*\i-1, evaluate=\i as \ylim using \t-2*\i+2] in {1,...,\steps}{
			\foreach \y[evaluate=\y as \thisx using {-\x-\y+1}] in {1,...,\ylim}{
				\roundgate[\thisx][\y][1][#5][#4]
			}
		}
	\end{scope}
}

\newcommandx{\eigenVL}[8][1=0,2=0,3=l,4=5,5=tr,6=init,7=bertiniorange,8=topright]{
	\begin{scope}[shift={(#1,#2)}]
		\pgfmathsetmacro{\t}{#4}
		\leftriangle[0][0][\t][#7][#8]
		
		\ifthenelse{\equal{#6}{init}}{
			\drawinitstate[0][0][l][\t][#7]
		}{}
		
		\ifthenelse{\equal{#5}{tr}}{
			\draw[fill=white] \foreach \x in {0,...,\t} {(\x-0.5-\t,0.5+\x) circle (0.15)};
			\ifthenelse{\equal{#3}{r}}{
				\draw[fill=white] (0.5,\t+0.5) circle (0.15);
			}{}
		}{}
		\ifthenelse{\equal{#5}{parttr}}{
			\draw[fill=white] \foreach \x in {0,...,\t} {(\x-0.5-\t,0.5+\x) circle (0.15)};
		}{}
	\end{scope}
}

\newcommandx{\eigenVR}[8][1=0,2=0,3=l,4=5,5=tr,6=init,7=bertiniorange,8=topright]{
	\begin{scope}[shift={(#1,#2)}]
		\pgfmathsetmacro{\t}{#4}
		\rightriangle[0][0][\t][#7][#8]
		
		\ifthenelse{\equal{#6}{init}}{
			\drawinitstate[0][0][r][\t][#7]
		}{}
		
		\ifthenelse{\equal{#5}{tr}}{
			\draw[fill=white] \foreach \x in {0,...,\t}{(-\x+0.5+\t,0.5+\x) circle (0.15)};
			\ifthenelse{\equal{#3}{l}}{
				\draw[fill=white] (-0.5,\t+0.5) circle (0.15);
			}{}
		}{}
	\end{scope}
}

\newcommandx{\tra}[2][1]{\underset{#1}{\text{tr}}\left[#2\right]}

\newcommandx{\tsfmatDgate}[7][1=0,2=0,3=l,4=4,5=tr,6=bertiniorange,7=topright]
{
	\begin{scope}[shift={(#1,#2)}]
		\ifthenelse{\equal{#3}{l}}{
			\pgfmathsetmacro{\flag}{-1}
		}{
			\pgfmathsetmacro{\flag}{1}
		}
		\pgfmathsetmacro{\t}{#4}
		\foreach \i[evaluate=\i as \x using {\flag*\i}, evaluate=\i as \y using \i] in {1,...,\t}{
			\roundgate[\x][\y][1][#7][#6]
		}
		
		\ifthenelse{\equal{#5}{tr}}{
			\foreach \i[evaluate=\i as \x using {\flag*\i}, evaluate=\i as \y using \i] in {\t}{
				\draw [fill=white] (\x-0.5,\y+0.5) circle (0.15);
				\draw [fill=white] (\x+0.5,\y+0.5) circle (0.15);
			}  
		}{}
	\end{scope}
	
}

\newcommandx{\tsfmatD}[8][1=0,2=0,3=l,4=4,5=tr,6=init,7=bertiniorange,8=topright]{
	\begin{scope}[shift={(#1,#2)}]
		\ifthenelse{\equal{#6}{init}}{
			\thetastate[0][0][1][#7]
		}{}
		
		\ifthenelse{\equal{#3}{l}}{
			\pgfmathsetmacro{\flag}{-1}
		}{
			\pgfmathsetmacro{\flag}{1}
		}
		
		\pgfmathsetmacro{\t}{#4}
		\foreach \i[evaluate=\i as \x using {\flag*\i}, evaluate=\i as \y using \i] in {1,...,\t}{
			\roundgate[\x][\y][1][#8][#7]
		}
		
		\ifthenelse{\equal{#5}{tr}}{
			\foreach \i[evaluate=\i as \x using {\flag*\i}, evaluate=\i as \y using \i] in {\t}{
				\draw [fill=white] (\x-0.5,\y+0.5) circle (0.15);
				\draw [fill=white] (\x+0.5,\y+0.5) circle (0.15);
			}  
		}
		\ifthenelse{\equal{#5}{parttr}}{
			\foreach \i[evaluate=\i as \x using {\flag*\i}, evaluate=\i as \y using \i] in {\t}{
				\draw [fill=white] (\x+0.5*\flag,\y+0.5) circle (0.15);
			}  
		}
		{}
	\end{scope}
}

\newcommandx{\drawinitstate}[5][1=0,2=0,3=l,4=4,5=bertiniorange]{
	\pgfmathsetmacro{\t}{#4}
	\begin{scope}[shift={(#1,#2)}]
		\pgfmathsetmacro{\steps}{ceil((\t-1)/2)}
		\ifthenelse{\equal{#3}{l}}{
			\foreach \i[evaluate=\i as \x using -\t+2*\i] in {0,...,\steps}{
				\thetastate[\x][0][1][#5]
			}
		}{
			\foreach \i[evaluate=\i as \x using -\t+2*\i] in {0,...,\steps}{      
				\thetastate[-\x][0][1][#5]
			}
		}
	\end{scope}
}

\newcommandx{\drawinitstateflipped}[5][1=0,2=0,3=l,4=4,5=bertiniorange]{
	\pgfmathsetmacro{\t}{#4}
	\begin{scope}[shift={(#1,#2)}]
		\pgfmathsetmacro{\steps}{ceil((\t-1)/2)}
		\ifthenelse{\equal{#3}{l}}{
			\foreach \i[evaluate=\i as \x using -\t+2*\i] in {0,...,\steps}{
				\thetastateflipped[\x][0][1][#5]
			}
		}{
			\foreach \i[evaluate=\i as \x using -\t+2*\i] in {0,...,\steps}{      
				\thetastateflipped[-\x][0][1][#5]
			}
		}
	\end{scope}
}

\newcommandx{\eigenDL}[6][1=0,2=0,3=l,4=4,5=bertiniorange,6=topright]{
	\begin{scope}[shift={(#1,#2)}]
		\pgfmathsetmacro{\t}{#4}
		\ifthenelse{\equal{#3}{l}}{
			\eigenVL[0][0][l][\t][tr][init][#5][#6]
			\pgfmathsetmacro{\t}{#4-1}
			\rightriangle[1][0][\t][#5][#6]
			\drawinitstate[1][0][r][\t][#5]
		}{
			\begin{scope}[shift={(-0.5,0.5)}]
				\foreach \i[evaluate=\i as \x using \i, evaluate=\i as \y using \i] in {0,...,\t}{      
					\draw (\x,\y)--++(0.5,0);
					\draw[fill=white] (\x,\y) circle (0.15);
				}
			\end{scope}
		}
	\end{scope}
}

\newcommandx{\eigenDR}[6][1=0,2=0,3=l,4=4,5=bertiniorange,6=topright]{
	\begin{scope}[shift={(#1,#2)}]
		\pgfmathsetmacro{\t}{#4}
		\ifthenelse{\equal{#3}{r}}{
			\eigenVR[0][0][r][\t][tr][init][#5][#6]
			\pgfmathsetmacro{\t}{#4-1}
			\leftriangle[-1][0][\t][#5][#6]
			\drawinitstate[-1][0][l][\t][#5]
		}{
			\begin{scope}[shift={(0.5,0.5)}]
				\foreach \i[evaluate=\i as \x using \i, evaluate=\i as \y using \t-\i] in {0,...,\t}{      
					\draw (\x,\y)--++(0.5,0);
					\draw[fill=white] (\x+0.5,\y) circle (0.15);
				}
			\end{scope}
		}
	\end{scope}
}

\newcommandx{\idonpurity}[2][1=0,2=0]
{
	\begin{scope}[shift={(#1,#2)}]
		\draw[thick] (-0.5,0)--++(-0.1,0.1)--++(0,0.2)--++(0.1,-0.1);
		\draw[thick] (-0.5,0.4)--++(-0.1,0.1)--++(0,0.2)--++(0.1,-0.1);
		\draw[thick] (0.5,0)--++(0.1,0.1)--++(0,0.2)--++(-0.1,-0.1);
		\draw[thick] (0.5,0.4)--++(0.1,0.1)--++(0,0.2)--++(-0.1,-0.1);
	\end{scope}
}

\newcommandx{\swaponpurity}[2][1=0,2=0]
{
	\begin{scope}[shift={(#1,#2)}]
		\draw[thick] (-0.5,0)--++(-0.2,0.2)--++(0,0.6)--++(0.2,-0.2);
		\draw[thick] (-0.5,0.2)--++(-0.075,0.075)--++(0,0.2)--++(0.075,-0.075);
		\draw[thick] (+0.5,0)--++(+0.2,0.2)--++(0,0.6)--++(-0.2,-0.2);
		\draw[thick] (+0.5,0.2)--++(+0.075,0.075)--++(0,0.2)--++(-0.075,-0.075);
	\end{scope}
}

\newcommandx{\hook}[4][1=0,2=0,3=t,4=l]{
	\begin{scope}[shift={(#1,#2)}]
		\ifthenelse{\equal{#3}{t}}{
			\ifthenelse{\equal{#4}{l}}{\draw[thick] (0.5,-0.5) arc (45:-90:0.15);}{\draw[thick] (0.5,-0.5) arc (45:270:0.15);}
		}{\ifthenelse{\equal{#4}{l}}{\draw[ thick] (0.5,-0.5) arc (-45:90:0.15);}{\draw[ thick] (0.5,-0.5) arc (315:90:0.15);}
		}
	\end{scope}
}

\newcommandx{\hhook}[4][1=0,2=0,3=t,4=l]{
	\begin{scope}[shift={(#1,#2)}]
		\ifthenelse{\equal{#3}{t}}{
			\ifthenelse{\equal{#4}{l}}{\draw[thick] (0.5,-0.5) arc (-45:175:0.15);}{\draw[thick] (0.5,-0.5) arc (225:0:0.15);}
		}{\ifthenelse{\equal{#4}{l}}{\draw[ thick] (0.5,-0.5) arc (-45:180:-0.15);}{\draw[ thick] (0.5,-0.5) arc (45:-180:0.15);}
		}
	\end{scope}
}

\definecolor{FcolU}{rgb}{0.71,0.78,0.91}
\definecolor{colLines}{rgb}{0.31,0.31,0.31}
\definecolor{colVMPSLines}{rgb}{0.11,0.11,0.11}
\definecolor{IcolUc}{rgb}{0.71,0.41,0.42}
\definecolor{IcolU}{rgb}{0.71,0.8,0.76}
\definecolor{IcolVMPSc}{rgb}{0.73,0.69,0.7}
\definecolor{IcolVMPS}{rgb}{0.81,0.77,0.78}
\definecolor{colObs}{rgb}{1.,1.,1.}

\def\dx{0.3}
\def\r{0.08}

\newcommand\gridLine[4]{
	\draw [colLines] ({(#2)*\dx},{-(#1)*\dx}) -- ({(#4)*\dx},{-(#3)*\dx});
}

\newcommand\mpsWire[4]{
	\draw [double,colVMPSLines]({(#2)*\dx},{-(#1)*\dx}) -- ({(#4)*\dx},{-(#3)*\dx});
}

\newcommand\mps[3]{
	\draw [thick,rounded corners=0.5,colVMPSLines,fill=#3] ({\dx*(#2-0.2)},{-\dx*(#1-0.2)}) rectangle  ({\dx*(#2+0.2)},{-\dx*(#1+0.2)});
}

\newcommand\bCircle[3]{
	\draw [thick,colLines,fill=#3] ({\dx*#2},{-\dx*#1}) circle ({1*\r});
}

\newcommand\TEsheet[2]{
	\draw [thick,colLines,fill=bertiniblue,rounded corners=0.5] ({(#2)*\dx},{-(#1)*\dx}) rectangle ({((#2)+7)*\dx},{(-(#1)+4.25)*\dx});
	\draw [thick,colVMPSLines,fill=IcolVMPS,rounded corners=0.5] ({((#2)-0.01)*\dx},{(-(#1)-0.01)*\dx}) rectangle ({((#2)+7+0.01)*\dx},{(-(#1)+0.31)*\dx});
}

\newcommand\TECsheet[2]{
	\draw [thick,colLines,fill=bertinired,rounded corners=0.5] ({(#2)*\dx},{-(#1)*\dx}) rectangle ({((#2)+7)*\dx},{(-(#1)+4.25)*\dx});
	\draw [thick,colVMPSLines,fill=IcolVMPSc,rounded corners=0.5] ({((#2)-0.01)*\dx},{(-(#1)-0.01)*\dx}) rectangle ({((#2)+7+0.01)*\dx},{(-(#1)+0.31)*\dx});
}

\newcommand\TEsheetrotated[2]{
	\draw [thick,colLines,fill=bertiniblue,rounded corners=0.5] ({(#2)*\dx},{-(#1)*\dx}) rectangle ({((#2)+7)*\dx},{(-(#1)+4.25)*\dx});
	\draw [thick,colVMPSLines,fill=IcolVMPSc,rounded corners=0.5] ({((#2)-0.01)*\dx},{(-(#1)+4.25-0.31)*\dx}) rectangle ({((#2)+7+0.01)*\dx},{(-(#1)+4.25+0.01)*\dx});
}

\newcommand\nME[2]{
	\draw [thick,fill=white] ({#2*\dx},{-#1*\dx}) circle ({0.5*\r});
}

\newcommand\sSquare[3]{
	\draw [thick,rounded corners=0.5,colLines,fill=#3] ({\dx*#2-0.5*\r},{-\dx*#1-0.5*\r}) rectangle ({\dx*#2+0.5*\r},{-\dx*#1+0.5*\r});
}

\newcommand\connection[4]{
	\draw [thick,colLines,fill=white,rounded corners=0.5] ({((#4)-0.01+7)*\dx},{-((#3)+0.01-4)*\dx}) rectangle ({((#4)+0.01)*\dx},{-((#3)-0.25-4)*\dx});
	\draw [thick,colLines,fill=white,rounded corners=0.5] ({((#4)-0.01+7)*\dx},{-((#3)-0.25-4)*\dx}) -- ({((#2)-0.01+7)*\dx},{-((#1)-0.25-4)*\dx}) -- ({((#2)+0.01)*\dx},{-((#1)-0.25-4)*\dx}) -- ({((#4)+0.01)*\dx},{-((#3)-0.25-4)*\dx}) -- cycle;
	\draw [thick,colLines,fill=white,rounded corners=0.5] 
	({((#2)-0.01+7)*\dx},{-((#1)-0.25-4)*\dx}) --
	({((#2)+0.01)*\dx},{-((#1)-0.25-4)*\dx}) -- 
	({((#2)+0.01)*\dx},{-((#1)-0.01-4)*\dx}) -- 
	({((#2)-0.01+7)*\dx},{-((#1)-0.01-4)*\dx}) -- cycle;
}

\newcommand\rotatedsquare[4]{
	\draw [thick,colLines,fill=bertinigrey1,rounded corners=0.5] ({((#4)-0.01+.5)*\dx},{-((#3)-0.25-4)*\dx}) -- ({((#2)-0.01+.5)*\dx},{-((#1)-0.25-4)*\dx}) -- ({((#2)+0.01)*\dx},{-((#1)-0.25-4)*\dx}) -- ({((#4)+0.01)*\dx},{-((#3)-0.25-4)*\dx}) -- cycle;
	\draw [thick,colLines,fill=bertinigrey1,rounded corners=0.5] 
	({((#2)-0.01+.5)*\dx},{-((#1)-0.25-4)*\dx}) --
	({((#2)+0.01)*\dx},{-((#1)-0.25-4)*\dx}) -- 
	({((#2)+0.01)*\dx},{-((#1)-0.01-4)*\dx}) -- 
	({((#2)-0.01+.5)*\dx},{-((#1)-0.01-4)*\dx}) -- cycle;
}

\newcommandx{\eightlegs}[2][1=0,2=0]{
	\begin{scope}[shift={(#1,#2)}]
		\foreach \x in {1,...,8}{
			\draw (\x, 0)--++(0,0.25);
			\draw[fill] (\x,0) circle (0.05);
		}
		\foreach \x in {1,3}{
			\pgfmathsetmacro\result{2*\x-1} 
			\node () at (\result,-0.5) {$i_{\x}$};
			\pgfmathsetmacro\result{2*\x}
			\node () at (\result,-0.5) {$j_{\x}$};	
		}
		\foreach \x in {2,4}{
			\pgfmathsetmacro\result{2*\x} 
			\node () at (\result,-0.5) {$i_{\x}$};
			\pgfmathsetmacro\result{2*\x-1}
			\node () at (\result,-0.5) {$j_{\x}$};	
		}
	\end{scope}
}

\begin{document}
	\title{Temporal Entanglement in Chaotic Quantum Circuits}
	\author{Alessandro Foligno}
	\affiliation{School of Physics and Astronomy, University of Nottingham, Nottingham, NG7 2RD, UK}
	\affiliation{Centre for the Mathematics and Theoretical Physics of Quantum Non-Equilibrium Systems,
		University of Nottingham, Nottingham, NG7 2RD, UK}
	\author{Tianci Zhou}
	\affiliation{Center for Theoretical Physics, Massachusetts Institute of Technology, Cambridge, Massachusetts 02139, USA}
	\author{Bruno Bertini}
	\affiliation{School of Physics and Astronomy, University of Nottingham, Nottingham, NG7 2RD, UK}
	\affiliation{Centre for the Mathematics and Theoretical Physics of Quantum Non-Equilibrium Systems,	University of Nottingham, Nottingham, NG7 2RD, UK}

	\begin{abstract}
		{The concept of space-evolution (or space-time duality) has emerged as a promising approach for studying quantum dynamics. The basic idea involves exchanging the roles of space and time, evolving the system using a space transfer matrix rather than the time evolution operator. The infinite-volume limit is then described by the fixed points of the latter transfer matrix, also known as influence matrices. To establish the potential of this method as a bona fide computational scheme, it is important to understand whether the influence matrices can be efficiently encoded in a classical computer. Here we begin this quest by presenting a systematic characterisation of their entanglement --- dubbed temporal entanglement --- in chaotic quantum systems. We consider the most general form of space-evolution, i.e., evolution in a generic space-like direction, 
			and present two fundamental results. First, we show that temporal entanglement always follows a volume law in time. Second, we identify two marginal cases --- (i) pure space evolution in generic chaotic systems (ii) any space-like evolution in dual-unitary circuits --- where R\'enyi entropies with index larger than one are sub-linear in time while the von Neumann entanglement entropy grows linearly. 
			We attribute this behaviour to the existence of a product state with large overlap with the influence matrices
			. This unexpected structure in the temporal entanglement spectrum might be the key to an efficient computational implementation of the space evolution.} 
	\end{abstract}	
	\preprint{MIT-CTP/5368}

	\maketitle
	
	\section{Introduction}	
	The first two decades of the new millennium witnessed extraordinary experimental progress in measuring dynamical properties of quantum many-body systems. Experiments are now able to probe, for instance, local relaxation of isolated systems~\cite{bloch2008many, kinoshita2006quantum, langen2015experimental} and out-of-equilibrium transport~\cite{schemmer2019generalized, jepsen2020spin, malvania2021generalized, wei2022quantum, joshi2022observing} over surprisingly long time scales. Theoreticians, however, can very rarely provide independent predictions to compare with these experiments, especially concerning dynamics beyond intermediate time scale. Indeed, characterising a quantum many-body system out-of-equilibrium, or even simulating its state on a classical computer, remains to date a formidable task. 
	
	{ The situation is slightly more favourable in one dimension, where one can use an extension of the celebrated DMRG algorithm~\cite{white1992density, white1993density} to provide a faithful representation of the time-evolving quantum state~\cite{schollwoeck2011the}}. The initial state is represented as a matrix product state (MPS) and a suitable evolution algorithm (e.g.\ tDMRG~\cite{daley2004time, white2004real} or TEBD~\cite{vidal2003efficient, vidal2004efficient}) { finds an MPS approximation of the state at time $t$ for a given level of accuracy}. {  The problem, however, is that the amount of resources required for such an approximation grows exponentially with the entanglement of the state and, in the absence of localisation or other ergodicity-breaking mechanisms, the latter builds up very quickly as time elapses. In practice this means that one needs an exponentially growing amount  of resources for an accurate representation of the state. This ``entanglement barrier'' is physical and cannot be avoided whenever one tries to characterise the whole quantum state. The key question}, however, is whether or not it is necessary to simulate the evolution of the whole quantum state to compute its experimentally accessible properties, e.g., its correlation functions.
	
	\begin{figure}
		\ifdef{0}{}{ \scalebox{0.875}{
				\begin{tikzpicture}
					[baseline={([yshift=-0.6ex]current bounding box.center)},scale=1.75]
					\node at ({-9.2*\dx},{16.5*\dx}){$(a)$}; 
					\node at ({-9.2*\dx},{4.5*\dx}){$(b)$}; 
					\node at ({3.05*\dx},{5.45*\dx}){$\mathcal T_t$}; 
					\draw [thick,colVMPSLines,fill=white,rounded corners=0.5] (-8.5*\dx,11.5*\dx) rectangle (-1.5*\dx,12.5*\dx);
					\draw [thick,colVMPSLines,fill=bertinigrey1,rounded corners=0.5] (-5*\dx,11.25*\dx) rectangle (-4.5*\dx,12.75*\dx);
					\mpsWire{-7.5}{0.25}{-7.5}{8.75}
					\mpsWire{-16.5}{0.25}{-16.5}{8.75}
					\foreach \t in {-8.25,...,-11.25}{
						\gridLine{\t}{0.25}{\t}{8.75}
					}
					\foreach \t in {-12.75,...,-15.75}{
						\gridLine{\t}{0.25}{\t}{8.75}
					}
					\foreach \x in {1,...,8}{
						\gridLine{-7.5}{\x}{-16.5}{\x}
					}
					\foreach \x in {1,...,8}{
						\mps{-7.5}{\x}{IcolVMPS}
						\bCircle{-8.25}{\x}{bertinired}
						\bCircle{-9.25}{\x}{bertinired}
						\bCircle{-10.25}{\x}{bertinired}
						\bCircle{-11.25}{\x}{bertinired}
					} 
					\foreach \x in {1,...,8}{
						\mps{-16.5}{\x}{IcolVMPS}
						\bCircle{-15.75}{\x}{bertiniblue}
						\bCircle{-14.75}{\x}{bertiniblue}
						\bCircle{-13.75}{\x}{bertiniblue}
						\bCircle{-12.75}{\x}{bertiniblue}
					}    
					\TEsheetrotated{-12.35}{1-9.5}
					\TECsheet{-7.35}{1-9.5} 
					\node at ({-.25*\dx},{12*\dx}){$=$};
					\node at ({-4*\dx},{12*\dx}){$a$};
					\node at ({5.5*\dx},{12*\dx}){$a$};
					\sSquare{-12}{5}{bertinigrey1}
					\draw[->] ({(0.25-9.5)*\dx},{(7)*\dx}) -- ({(0.25-9.5)*\dx},{(10.5)*\dx}) node [midway,xshift=-5pt,rotate=90] {time};
					\draw[->] ({(0.25-9.5)*\dx},{(7)*\dx}) -- ({(8.5-9.5)*\dx},{(7)*\dx}) node [midway,yshift=-5pt] {space};
					\draw[|-latex, thick] ({-.25*\dx},{6.5*\dx}) -- ({-.25*\dx},{5.5*\dx}) node [midway,xshift=20pt] {folding};
					\node at ({-.25*\dx},{2.5*\dx}){$=$};
					\node at ({-4.75*\dx},{5.25*\dx}){$a$};
					\node at ({5*\dx},{5.25*\dx}){$a$};
					\TEsheet{0.5-1}{1.5-9.5}
					\TECsheet{1-1}{1-9.5}
					\connection{-1+2-1}{1-9.5}{-1.5+2-1}{1.5-9.5}
					\rotatedsquare{-1+2-1}{1+3-9.5}{-1.5+2-1}{1.5+3-9.5}
					\mpsWire{-0.25}{0.25}{-0.25}{8.75}
					\foreach \t in {-1,...,-4}{
						\gridLine{\t}{0.25}{\t}{8.75}
					}
					\foreach \x in {1,...,8}{
						\gridLine{0}{\x}{-4.75}{\x}
					}
					\foreach \x in {1,...,8}{
						\mps{-0.25}{\x}{IcolVMPS}
						\bCircle{-1}{\x}{bertiniorange}
						\bCircle{-2}{\x}{bertiniorange}
						\bCircle{-3}{\x}{bertiniorange}
						\bCircle{-4}{\x}{bertiniorange}
					}
					\foreach \x in {1,...,8}{
						\nME{-4.75}{\x}}
					\sSquare{-4.75}{5}{bertinigrey1}
					\draw [semithick,black,fill=gray,opacity=0.2,rounded corners=0.5] (2.5*\dx,-.15*\dx) rectangle (3.5*\dx,5.05*\dx);
				\end{tikzpicture}
			}
		}
		\includegraphics[scale=0.875]{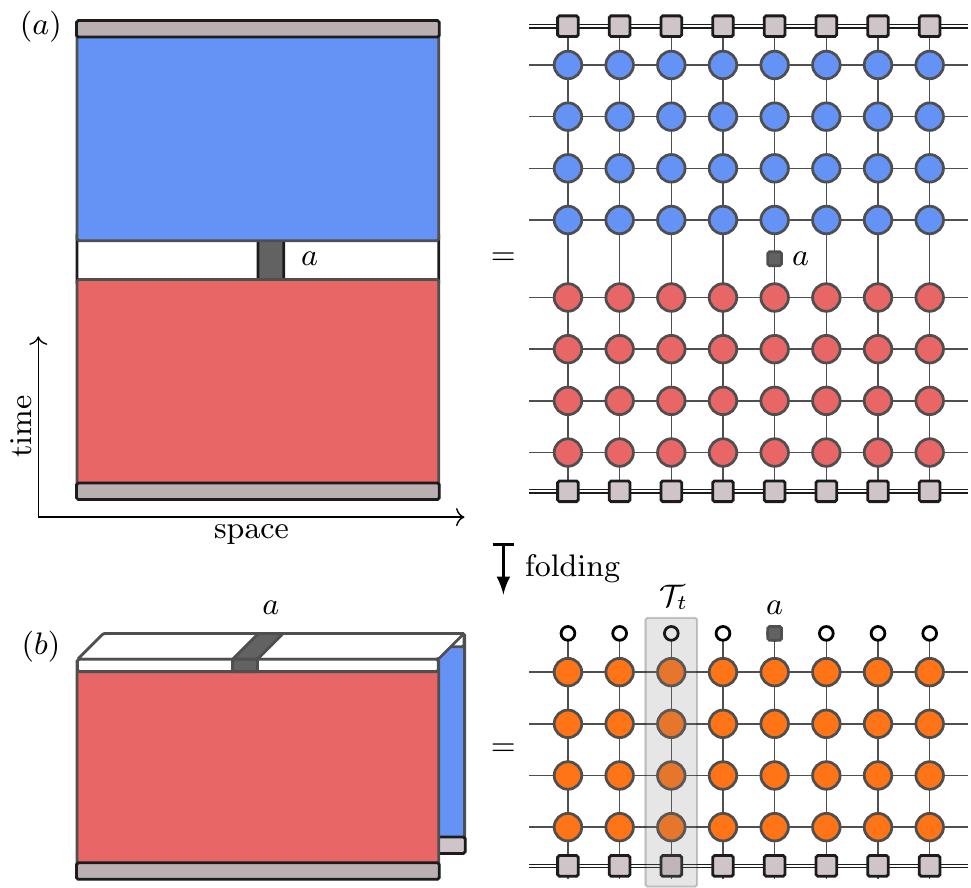}
		\caption{One point function of the local operator $a$ after a quantum quench and its tensor-network representation. Forward and backward time sheets are respectively depicted in red and blue. The lower panel depicts the same one-point function after folding: in the folded tensor network the number of local degrees of freedom is doubled and the white circles at the top of the tensor network denote a loop. The grey shaded box highlights the space transfer matrix $\mathcal T_t$ (acting from left to right on a lattice of $t$ sites).}
		\label{fig:TN}
	\end{figure}
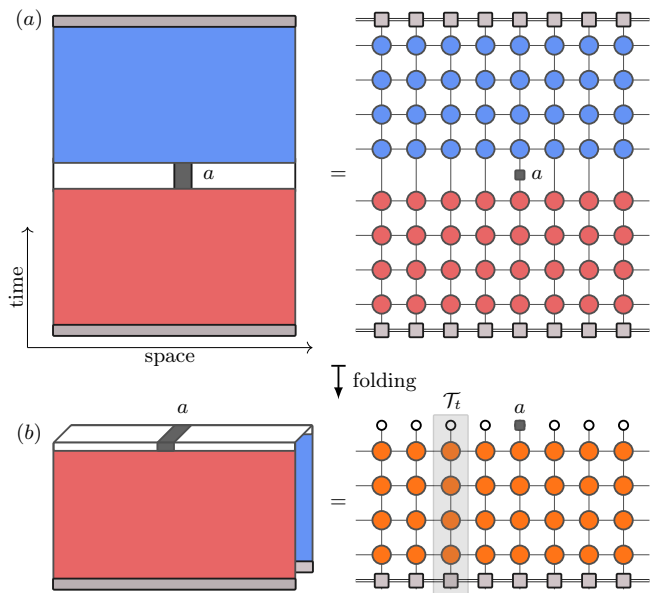
	
	In the course of the past decade several algorithms have been proposed to circumvent the fast entanglement growth of non-equilibrium states~\cite{prosen2007is, haegeman2011time, haegeman2016unifying, leviatan2017quantum, kloss2018time, white2018quantum, znidaric2019nonequilibrium, krumnow2019overcoming, rakovszky2020dissipationassisted, vonkeyserlingk2022operator, banuls2009matrix}. The common theme is to sidestep the problem by exploiting the fact that one is typically only interested in the correlation functions of special observables, for instance, those that are \emph{local} in space. { A promising one, which motivates our work, is the so called ``folding algorithm'' or ``transverse folding algorithm''  proposed in Ref~\cite{banuls2009matrix} (see also Refs.~\cite{muellerhermes2012tensor, hastings2015connecting, sonner2022characterizing, frias2022light, lerose2023overcoming}), whose main idea is to evolve the system in space, rather than in time.} Taking the one-point function in Fig.~\ref{fig:TN} as an example, this means that one has to contract its tensor-network representation horizontally, by means of an appropriate space transfer matrix { rather than vertically using the time evolution operator.} The name of the algorithm derives from the fact that this operation becomes much more efficient when considering the ``folded representation'' of the correlator, i.e., when folding the tensor network around the centre as shown in Fig.~\ref{fig:TN} (b), which doubles the local degrees of freedom { but keeps the correlations short-ranged}. Physically, the vertical column of tensors beneath the observable implements the unitary evolution of the subsystem of interest -- the one where the observable acts -- while the sections on its two sides encode the non-unitary action exerted on the subsystem by the rest of the system, i.e, the environment. For instance, in the example of Fig.~\ref{fig:TN} the system is a single spin (or qudit). Inspired by the Feynman-Vernon influence functional approach~\cite{feynman1963the}, Ref.~\cite{lerose2021influence} proposed to dub ``influence matrices'' the portions of the tensor network describing the action of the environment. Note that when the environment becomes very large, the influence matrices become equal to the left and right fixed points of the space transfer matrix $\mathcal T_t$, see Fig.~\ref{fig:TN}.
	
	The idea of exchanging space and time to describe infinite systems at finite times proved to be very successful and over the last few years has found interesting applications to the study of spectral statistics and quantum chaos~\cite{bertini2018exact, bertini2021random, flack2020statistics, garratt2021local, fritzsch2021eigenstate, garratt2021manybody}, entanglement dynamics~\cite{bertini2019entanglement, ippoliti2021postselectionfree, ippoliti2021fractal, bertini2022growth}, impurity problems~\cite{thoenniss2022an}, and even full-counting statistics of many-body observables~\cite{bertini2022nonequilibrium} and Loschmidt echo~\cite{pozsgay2013dynamical, piroli2017quantum, piroli2018non}. When considered as a computational tool for computing correlation functions, however, the folding algorithm has an important limitation: it can only deal with cases where the operator insertions break the translation symmetry in a single spatial point, i.e., one-point functions and, more generally, auto-correlations. In this way one cannot access, for instance, generic two-point functions --- such as those needed to compute transport coefficients~\cite{mahan1981many, altland2010condensed, bertini2021finitetemperature} --- as they feature two operators separated in both time and space.
	
	
	\footnotetext[20]{Not to be confused with the ``timelike entanglement'' studied in the context of conformal field theory and AdS/CFT correspondence, see Refs.~\cite{doi2022pseudo, doi2023timelike}.}

	Another outstanding question concerns the computational complexity of the folding algorithm. Namely, how hard it is to implement this algorithm on a classical computer for increasingly large times. To answer this question one needs to understand what features of the influence matrices have to be retained to correctly describe expectation values of local operators and what is the amount of resources required to do so. An intuitive estimate can be obtained by studying their entanglement, dubbed ``temporal entanglement''~\cite{hastings2015connecting, Note20}. Indeed, roughly speaking, if the latter does not grow too fast one can efficiently approximate the influence matrices with matrix product states for arbitrarily high fidelity~\cite{verstraete2006matrix, schuch2008entropy}. { Following Refs.~\cite{hastings2015connecting,lerose2021influence}, one can argue that temporal entanglement should be small for generic systems. Indeed, the dephasing caused by the environment tends to align corresponding spins in the forward and backward copies (cf. Fig.~\ref{fig:TN}) producing configurations that are diagonal and hence classical. Although plausible, this picture can be proven only in a few special cases. These include certain special chaotic quantum systems --- dual-unitary circuits~\cite{bertini2019exact} --- prepared in a special family of initial states~\cite{bertini2019entanglement, piroli2020exact} and in certain special classes of integrable models~\cite{klobas2021exact, klobas2021exact2, lerose2021scaling, giudice2022temporal, frias2022light}. In generic cases the temporal entanglement is observed to grow in time, even though its growth appears slower than that of spatial entanglement~\cite{banuls2009matrix, lerose2021influence}.} 
	
	

	\begin{figure}[t]
		\centering
		\includegraphics[width=0.9\columnwidth]{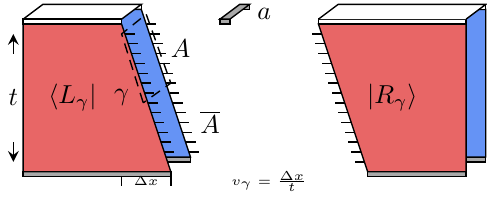}
		\caption{Generalised influence matrices on a temporal slice $\gamma$. The (anti-)slope of the path $\gamma$ is $v_{\gamma} = \Delta x / t$. When $v_\gamma = 0$, they correspond to are the regular influence matrices. A one point or two-point function can be evaluated by contracting the left state $\langle L_\gamma|$, some relevant operators inserted along the path, and the right state $|R_\gamma\rangle$. The temporal entanglement is the larger among the entanglement of $\langle L_\gamma|$ and that of $|R_\gamma\rangle$ for a partition $A, \bar{A}$ on a temporal slice.} 
		\label{fig:TE_intro}
	\end{figure}

	\begin{table*}[t]\centering
		\begin{tabular}{|c|c|c|} 
			\hline
			& $\max(S^{(1)}_{L,A}(\gamma),S^{(1)}_{R,A}(\gamma))$
			& $\max(S^{(2)}_{L,A}(\gamma),S^{(2)}_{R,A}(\gamma))$ 
			\\           
			\hline
			generic circuit ($v_{\gamma} \ne 0 $) &  \textcolor{blue}{$\sim t$} &\textcolor{blue}{$\sim t $} \\ \hline			generic circuit ($v_{\gamma} = 0)$ & \textcolor{red}{$\sim t$} & \textcolor{blue}{$\lesssim \log t$} \\ \hline
			generic dual unitary circuit & \textcolor{blue}{$\sim t$} &  \textcolor{blue}{$\sim 1$} \\ 
			\hline 
		\end{tabular}
		
		\caption{Scaling of the temporal entanglement. We take the second R\'enyi entropy as a representative of higher R\'enyi entropies. The maximum is taken over $L$, $R$ states and the possible contiguous regions $A$ on $\gamma$ for a given slope $v_{\gamma}$. All the circuits have a brickwork architecture (see Sec.~\ref{sec:setting}). Only vertical cuts or dual unitary circuits give sub-linear growth in higher R\'enyi entropy. { Blue denotes analytical results (obtained by membrane picture/exact calculation), red denotes numerical evidence.}} 
		\label{tab:results}
	\end{table*}

	{ In this work we fill both the aforementioned gaps: (1) We extend the folding algorithm to compute generic two-point functions and (2) we characterise the scaling of temporal entanglement in generic quantum many-body systems. 
		
		The main idea for extending the folding algorithm is to embed the two operators in the same system defined on a {\it time-like surface}, or path, $\gamma$, see the illustration in Fig.~\ref{fig:TE_intro}, and evolve it in the orthogonal space-like direction.} In a relativistic field theory one can imagine to implement our construction by boosting to a reference frame where the operators are measured at the same position and then use the usual folding algorithm. This setup allows us to treat generic two-point functions, and also gives the option to optimise the evaluation of one-point functions by varying the path on which the influence matrices are evaluated. Note that the extreme case of a time-like surface corresponding with the light cone edge has been considered in Ref.~\cite{gopalakrishnan2019unitary} and, for this case, Ref.~\cite{giudice2022temporal} characterised the complexity of the corresponding influence matrices for integrable dual-unitary circuits.
	
	To characterise the scaling of temporal entanglement, we compute the entanglement entropies of the generalised influence matrices $\bra{L_\gamma}$ and $\ket{R_\gamma}$ across contiguous bipartitions of $\gamma$. We respectively denote them by  
	\begin{equation}
		S^{(\alpha)}_{L,A}(\gamma),\qquad \text{and}\qquad S^{(\alpha)}_{R,A}(\gamma),
		\label{eq:TE}
	\end{equation}
	for subregion $A$ and R\'enyi index $\alpha$. Our findings are summarised in Tab.~\ref{tab:results}.

	{ Overall we find that the temporal von Neumann entropy (${\alpha=1}$) always grows linearly in time after a quench from a generic initial state. Nevertheless, we find cases in which R\'enyi entropies with index ${\alpha > 1}$ (higher R\'enyi entropies from now on) grow sublinearly. In particular, the higher R\'enyi entropies of vertical influence matrices (the regular ones) in {\it any} chaotic system are logarithmic in time, while those of {\it any} influence matrix in a dual-unitary circuit are bounded by a constant. In these cases the slope of growth of von Neumann entropy is strictly smaller than that of regular state entanglement. These statements are proven analytically for dual-unitary circuits, while in the case of generic circuits they result from the combination of entanglement membrane theory~\cite{nahum2017quantum, zhou2019emergent} and numerical observations.

		The observed linear growth of von-Neumann entanglement entropy rules out an efficient high-fidelity approximation of the influence matrices via matrix product states~\cite{schuch2008entropy}. Our findings, however, suggest that there are physically relevant cases where the temporal entanglement spectrum displays a strong separation of scales: There are a few large (at most linearly decaying) Schmidt values and many exponentially small ones. This remarkable structure might be the key for an efficient implementation of the folding algorithm.
		
		In the following subsection, we sketch the key steps to obtain the scalings in Tab.~\ref{tab:results} and discuss their consequences. A complete description of our setup begins in Sec.~\ref{sec:setting}.}

	\subsection{Summary of Approaches and Results}
	\label{sec:mainresults}
	
	We consider generic quantum many-body systems with local interactions modelled by local brickwork quantum circuits~\cite{fisher_random_2022}. { This is a class of locally interacting systems in discrete time that has recently played a key role in 
		understanding many-body quantum dynamics. The enormous complexity of the latter implies that the theoretical description, or even the mere numerical simulation, of quantum matter out of equilibrium is practically possible only in the short-time regime. Brickwork quantum circuits simplify the picture by imposing strictly local interactions over a finite time step and give rare examples where local observables and information theoretical quantities can be determined at all times. The results obtained in these systems, for instance through random averaging~\cite{nahum2017quantum, vonKeyserlingk2018operator,rakovszky2018diffusive, zhou2020entanglement,khemani2018operator} and/or space-time duality~\cite{piroli2020exact,bertini2019entanglement,bertini2020operator, bertini2020operator2, claeys2020maximum, bertini2020scrambling, jonay2021triunitary,claeys2020maximum, zhou2022maximal,gopalakrishnan2019unitary, foligno2022growth}, have significantly advanced our understanding of universal properties of the dynamics. 
		Applications include, for instance, operator dynamics and information spreading~\cite{nahum2017quantum, vonKeyserlingk2018operator,  chan2018solution, khemani2018operator, rakovszky2018diffusive, zhou2020entanglement,reid2021entanglement, wang2019barrier}, statistical properties of the spectrum~\cite{friedman2019spectral, bertini2018exact,chan2018spectral, flack2020statistics, bertini2021random, fritzsch2021eigenstate, kos2021thermalization, bertini2022exact, garratt2021manybody, garratt2021local}, and more broadly thermalisation~\cite{piroli2020exact, claeys2021ergodic, suzuki2022computational,klobas2021exact, klobas2021exact2, klobas2021entanglement,kos2021thermalization}. We also note that quantum circuits are vital tools for experimental simulation of quantum systems and quantum computation. For instance, they can be used to demonstrate quantum advantage~\cite{arute_quantum_2019,boixo_characterizing_2018,deshpande_dynamical_2018,muraleedharan_quantum_2019}, to perform randomised benchmarking~\cite{liuBenchmarkingNeartermQuantum2021,magesan_characterizing_2012,magesan_scalable_2011-1,proctor_what_2017}, randomised measurements~\cite{brydges_probing_2019,elben_many-body_2019,elben_renyi_2018,pichler_measurement_2016,vermersch_probing_2019,vermersch_unitary_2018}, shadow tomography~\cite{aaronson_shadow_2018,huang_predicting_2020,ohliger_efficient_2013}, and, more generally, to study non-equilibrium dynamics of Floquet systems~\cite{keenan2022evidence, morvan2022formation}.}

	{The structure of these circuits look like a Suzuki-Trotter~\cite{trotter1959product, suzuki1991general} approximation of (local) Hamiltonian evolution}, but the unitary gates are not necessarily infinitesimal in time or close to the identity: They can be arbitrary unitaries (see the detailed illustration of our setup in Sec.~\ref{sec:setting}). 
	
	To understand the behaviour of $S^{(\alpha)}_{L/R,A}(\gamma)$ in generic circuits, we take the gates forming the brickwork structure to be independent Haar random matrices. By averaging over the random gates the calculation of entanglement related quantities is mapped into that of the free energy of a statistical mechanical model of emergent spins~\cite{nahum2017quantum,zhou2019emergent}. In particular, we find that the averaged temporal purity 
	\begin{equation}
		\overline{\exp( - S^{(2)}_{L/R,A}(\gamma))},
	\end{equation}
	is the difference of free energies of the same statistical model subjected to different boundary conditions. By minimising the free energies we find domain-wall configurations that give (cf.\ Sec.~\ref{sec:TEinRU}) 
	\begin{equation}
		\label{eq:purity_bound}
		\overline{\exp( - S^{(2)}_{L/R,A}(\gamma)) }
		\le e^{-v_{\rm TE}^{(2)} t \log d},
	\end{equation}
	where the linear coefficient $v_{\rm TE}^{(2)}\geq0$ is determined by the line tension $\mathcal{E}_H(v)$ of the membrane {  separating the different domains. The line tension is an intrinsic function of the membrane, which, in translational invariant systems (at least after disorder average), only depends on the space time slope $v$.} Although the explicit expression is complicated, we have a useful condition 
	\begin{equation}
		\label{eq:v_TE}
		v_{\rm TE}^{(2)}=0\quad \Leftrightarrow\quad \mathcal{E}_H( v_\gamma) = \mathcal{E}_H(0), 
	\end{equation}
	where $v_{\gamma}$ is the anti-slope \footnote{The dimension of slope in space time is time/distance = 1/velocity. } of the path $\gamma$, see Fig.~\ref{fig:TE_intro}.

	Eq.~\eqref{eq:purity_bound} results in a lower bound of the typical growth rates of the temporal entanglement entropies. In particular we have
	\begin{equation}
		\begin{aligned}
			\overline{S^{(1)}_{L/R,A}(\gamma)} &\ge 
			v_{\rm TE}^{(2)} \log (d) t, \\
			\overline{S^{(\alpha>1)}_{L/R,A}(\gamma)} &\ge 
			\frac{1}{2} v_{\rm TE}^{(2)} \log (d) t . 
		\end{aligned}
		\label{eq:genericboundHaar}
	\end{equation}
	In fact, following Ref.~\cite{zhou2020entanglement}, we argue that this conclusion can be applied to generic Floquet circuits \emph{even in the absence of randomness}. In this case the entanglement dynamics is still described by an emergent statistical mechanical model and the tension ${\mathcal{E}}(v)$ of the associated membrane can be determined perturbatively, dressing the random unitary one~\cite{zhou2020entanglement}. In practice this means that one can apply \eqref{eq:genericboundHaar} without the average by replacing $\mathcal{E}_{H}(v)$ by ${\mathcal{E}}(v)$. Therefore, for generic Floquet circuits and generic paths $\gamma$ the temporal entanglement entropies with R\'enyi index $\alpha \ge 1$ grow linearly in time~\footnote{Here we assumed reflection symmetry of the gate (hence $\tilde{\mathcal{E}}(v) = \tilde{\mathcal{E}}(-v)$). In the absence of this property Eq.~\eqref{eq:v_TE} takes a different form but our conclusion about linear growth of temporal entanglement continues to apply.}. 
	
	Eq.~\eqref{eq:v_TE}, however, also suggests that there are two interesting marginal cases where temporal entanglement entropies can be sub-linear 
	\begin{itemize}
		\item[(I)] {\bf  Constant line tension}, i.e., 
		\begin{equation}
			\label{eq:margin_case_1}
			{\mathcal{E}}(v)=\text{constant};
		\end{equation}
		\item[(II)] {\bf  Vertical path}, i.e., 
		\begin{equation}
			\label{eq:margin_case_2}
			v_{\gamma}=0;    
		\end{equation}
	\end{itemize}
	
	{ Condition (I) provides a very stringent constraint. Indeed, invoking general properties of the line tension function~\cite{jonay2018coarsegrained}, one can conclude that a constant line tension has to be equal to one. This in turn implies a maximal growth rate of the regular {\it spatial} entanglement after a quantum quench in the circuit. As shown in Refs.~\cite{zhou2020entanglement,zhou_maximal_2022} circuits with this property have to be {\it dual unitary}.
		
		On the contrary, Condition (II) does not involve the line tension function, it only requires the temporal surface to be vertical (i.e.\ it holds for regular influence matrices). This means that, intriguingly, the vanishing of the linear coefficient at ${v_{\gamma}=0}$ should occur for generic circuits.}

	The two marginal cases (I) and (II) are studied in detail in Secs.~\ref{sec:TEinDU} and~\ref{sec:TEinRUvertical}. There we show that in both cases higher R\'enyi entropies ($\alpha>1$) display a \emph{sub-linear growth in time}. Nevertheless, their von Neumann entropy $(\alpha = 1)$ grows \emph{linearly} { (second and third rows of Tab.~\ref{tab:results})}. Namely, one cannot evaluate the scaling of von Neumann entropy via a replica trick as the replica limit does not commute with the large time limit. Interestingly, a similar discrepancy in the scaling of R\'enyi entropies was also observed in Refs.~\cite{rakovszky2019sub,huang2020dynamics} for the behaviour of the ``regular'' spatial entanglement in circuits with conservation laws. 
	
	At the level of entanglement spectrum the mechanism driving the observed sub-linear scaling is the same in both cases (I) and (II): the influence matrices have  large overlap with a product state of the form $\ket{\Psi'_A}\otimes\ket{\Psi''_{\bar A}}$ on $A$ and $\bar{A}$. Then, an immediate application of Eckart-Young's Theorem~\cite{eckart1936approximation} implies that the reduced density matrices
	\begin{equation}
		\label{eq:rho_L_rho_R}
		\rho_{L, \gamma, A} = \frac{\tr_A ( \ketbra{L_\gamma})}{\norm{\bra{L_\gamma}}^2},\quad 
		\rho_{R, \gamma, A} = \frac{\tr_A ( \ketbra{R_\gamma})}{\norm{\ket{R_\gamma}}^2},
	\end{equation}
	have at least one slowly decaying eigenvalue. This eigenvalue determines the slow growth of higher R\'enyi entropies. Meanwhile, we find exponentially many other eigenvalues of $\rho_{L/R,\gamma,A}$ that decay exponentially fast with time $t$. This produces a linearly growing entanglement entropy $S^{(1)}_{L/R,A}(\gamma))$. {  Fig.~\ref{plot:eigenvalue_distribution}  shows the separation of scales in the entanglement spectrum of the temporal state in two representative examples. 
		
		\begin{figure}
			\includegraphics[scale=1.4]{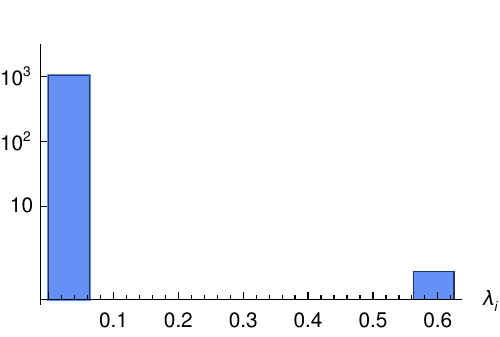}
			\includegraphics[scale=1.4]{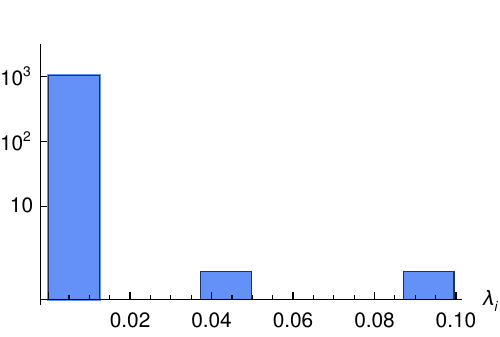}
			\caption{{ Schmidt values $\lambda_i$ for the bipartition of $\bra{L_\gamma}$ in two subsystems of equal size for $t=5$. The y-axis reports the number of Schmidt values within a small bin centred on $\lambda_i$. The top panel corresponds to the marginal case (I) (cf. Eq.~\eqref{eq:margin_case_1}) while the bottom panel to the marginal case (II) (cf.~\eqref{eq:margin_case_2}).} 
			}
			\ifdef{0}{}{
				\begin{tikzpicture}[scale=1,remember picture]
					\begin{semilogyaxis}[grid=major,
						legend columns=2,
						ymin=0.001, ymax=1,
						legend style={at={(0.6,0.8)},anchor=south east,font=\scriptsize,draw= none, fill=none},
						mark size=1.3pt,	
						xlabel=$i$,
						ylabel=$\lambda_i     $,
						y label style={at={(axis description cs:.05,.5)},anchor=south,font=\normalsize	},		
						tick label style={font=\normalsize	},	
						x label style={font=\normalsize	},	
						]
						
						\addplot[
						smooth,
						thick,
						mark=*,
						blue,
						dashed
						] table [x expr=(\thisrowno{0})*1024, y expr=(\thisrowno{1})]
						{plots_data/dataeig1.dat};
						\addlegendentry{$t=5$}
						\addplot[
						smooth,
						thick,
						mark=*,
						green,
						dashed
						] table [x expr=(\thisrowno{0})*246, y expr=(\thisrowno{1})]
						{plots_data/dataeig2.dat};
						\addlegendentry{$t=4$}
						\addplot[
						smooth,
						thick,
						mark=*,
						red,
						dashed
						] table [x expr=(\thisrowno{0})*64, y expr=(\thisrowno{1})]
						{plots_data/dataeig3.dat};
						\addlegendentry{$t=3$}
					\end{semilogyaxis}	
				\end{tikzpicture}
				\caption{\alessandro{Schimdt values associated with a bipartition of $\bra{L_\gamma}$ in two subsystems of equal size, for various values of $t$. In the plot, we ordered the eigenvalues: $\lambda_i \le \lambda_j \implies i\ge j$, and normalized the indices $i$ such that they are always in $[0,1]$.
						The plot shows a qualitative change of behavior from the first eigenvalues an the latter, with a vertical tangent point around $i=0$.}
				}
			}
			\label{plot:eigenvalue_distribution}
		\end{figure}
		In this situation one might be inclined to conclude that the singular states corresponding to the large Schmidt values represent the dominant part of the state. The linear growth of the von Neumann entropy, however, excludes the possibility of constructing a high-fidelity approximation of the state by keeping a polynomial number of Schmidt eigenvectors. Nevertheless, if the objective is to only approximate special observables, for instance the one-point function $\tr\left[\rho_0 a_{x}(t)\right]$, the answer might be different. More generally, it interesting to ask how much physically relevant information can be extracted faithfully from the first few Schmidt eigenstates in cases with such a strong separation of scales. We leave these questions to future research. 
	}

	\begin{figure}[t]
		\ifdef{0}{}{\scalebox{0.875}{
				\begin{tikzpicture}
					[baseline={([yshift=-0.6ex]current bounding box.center)},scale=1.75]
					\node at ({2.5*\dx},{3.5*\dx}){$a$}; 
					\node at ({4.5*\dx},{6.8*\dx}){$b$}; 
					\node at ({4.5*\dx},{7.5*\dx}){$\tilde{\mathcal T}_{\gamma, t}$}; 
					\node at ({6.5*\dx},{7.5*\dx}){$\mathcal T_{\gamma, t}$}; 
					\node at ({15*\dx},{3.5*\dx}){$= \tr[\mathcal T_{\gamma, t}^{L-2} \tilde{\mathcal T}_{\gamma, t}]$}; 
					\mpsWire{-0.25}{0.25}{-0.25}{12.75}
					\foreach \t in {-1,...,-6}{
						\gridLine{\t}{0.25}{\t}{12.75}
					}
					\foreach \x in {1,...,12}{
						\gridLine{0}{\x}{-6.75}{\x}
					}
					\foreach \x in {1,...,12}{
						\mps{-0.25}{\x}{IcolVMPS}
						\bCircle{-1}{\x}{bertiniorange}
						\bCircle{-2}{\x}{bertiniorange}
						\bCircle{-3}{\x}{bertiniorange}
						\bCircle{-4}{\x}{bertiniorange}
						\bCircle{-5}{\x}{bertiniorange}
						\bCircle{-6}{\x}{bertiniorange}
					}
					\foreach \x in {1,...,12}{
						\nME{-6.75}{\x}}
					\sSquare{-6.75}{5}{bertinigrey1}
					\sSquare{-3.5}{2}{bertinigrey1}
					\draw [semithick,black,fill=gray,opacity=0.2,rounded corners=0.5] (1.5*\dx,-.15*\dx) --  (1.5*\dx,4.5*\dx) -- (2.5*\dx,4.5*\dx) -- (2.5*\dx,5.5*\dx) -- (3.5*\dx,5.5*\dx) -- (3.5*\dx,7.05*\dx) -- (5.5*\dx,7.05*\dx) -- (5.5*\dx,5.5*\dx) -- (4.5*\dx,5.5*\dx) -- (4.5*\dx,4.5*\dx) -- (3.5*\dx,4.5*\dx) -- (3.5*\dx,-.15*\dx) -- (1.5*\dx,-.15*\dx);
					\draw [semithick,black,fill=bertiniblue,opacity=0.2,rounded corners=0.5] ({(2+1.5)*\dx},{-.15*\dx}) --  ({(2+1.5)*\dx},{4.5*\dx}) -- ({(2+2.5)*\dx},{4.5*\dx}) -- ({(2+2.5)*\dx},{5.5*\dx}) -- ({(2+3.5)*\dx},{5.5*\dx}) -- ({(2+3.5)*\dx},{7.05*\dx}) -- ({(2+5.5)*\dx},{7.05*\dx}) -- ({(2+5.5)*\dx},{5.5*\dx}) -- ({(2+4.5)*\dx},{5.5*\dx}) -- ({(2+4.5)*\dx},{4.5*\dx}) -- ({(2+3.5)*\dx},{4.5*\dx}) -- ({(2+3.5)*\dx},{-.15*\dx}) -- ({(2+1.5)*\dx},{-.15*\dx});
				\end{tikzpicture}
			}
		}
		\includegraphics[scale=0.875]{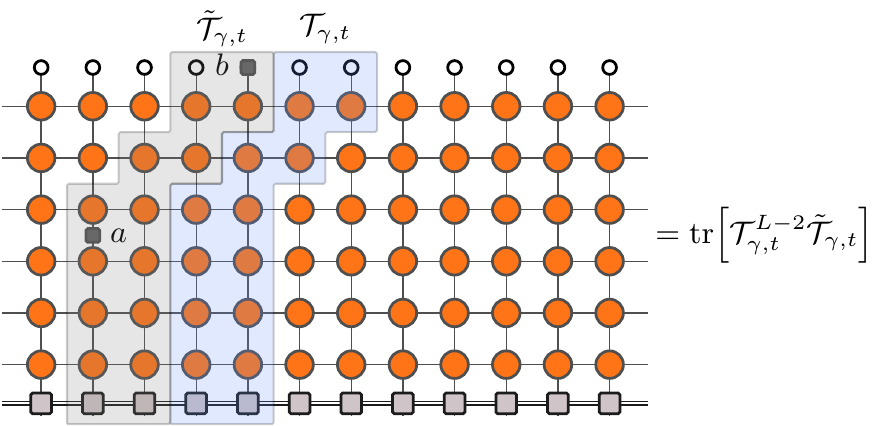}
		\caption{Folded tensor network representation of a dynamical two-point function of two operators $a$ and $b$ after a quantum quench. The tensor network can be contracted using non-vertical transfer matrices that follow the time-like path $\gamma$ and propagate in the space-like direction orthogonal to $\gamma$. The two different transfer matrices used are highlighted in the boxes.}
		\label{fig:TN2}
	\end{figure}

	The rest of this paper is laid out as follows. In Sec.~\ref{sec:setting} we introduce the precise setting considered in this work. In Sec.~\ref{sec:folding} we discuss the folding algorithm and explain its extension to non-vertical cuts. In Sec.~\ref{sec:TEinRU} we determine the scaling of temporal entanglement in generic quantum circuits using the entanglement-membrane approach of Refs.~\cite{nahum2017quantum, zhou2019emergent, zhou2020entanglement}. The two marginal cases with sub-linear growth of higher R\'enyi entropies are analysed in Secs.~\ref{sec:TEinDU} and~\ref{sec:TEinRUvertical}. In particular, in Sec.~\ref{sec:TEinDU} we discuss the scaling of temporal entanglement in dual-unitary circuits evolving from generic initial states, while in Sec.~\ref{sec:TEinRUvertical} we study the scaling of regular temporal entanglement, i.e.\ of the influence matrix on the vertical cut, in generic circuits. In Sec.~\ref{sec:temporalvsspatial} we compare the growth of temporal entanglement and that of regular state entanglement. Our conclusions and final remarks are reported in Sec.~\ref{sec:conclusions}.

	\section{Setting}
	\label{sec:setting}

	We consider the quantum dynamics generated by local quantum circuits acting on a chain of $2L$ qudits ($d$ internal states) placed at half integer positions. These circuits have nearest-neighbour interactions, and are often dubbed ``brickwork'' quantum circuits. The operator performing one step of evolution alternatively evolves the even and odd sublattices 
	\begin{equation}
		{\mathbb U}={\mathbb U}_1\cdot {\mathbb U}_2\,.
		\label{eq:U}
	\end{equation}
	Here we introduced 
	\begin{align}
		\label{eq:U1U2}
		{\mathbb U}_1=\bigotimes_{x\in\mathbb Z_L} U_{x,x+1/2},\qquad {\mathbb U}_2=\bigotimes_{x\in\mathbb Z_L+1/2} U_{x,x+1/2},
	\end{align}
	with $U_{x,x+1/2}$ acting non-trivially, as the $d^2\times d^2$ unitary matrix $U$, only on the qudits at positions $x$ and $x+1/2$. The matrix $U$ is known as ``local gate'' and specifies the local interactions. Local gates can in principle be different at each space-time point, i.e.
	\begin{equation}
		U\mapsto U(t,x),
	\end{equation}
	representing a disordered system undergoing aperiodic quantum dynamics. In contrast, in the special case where all local gates coincide, the quantum circuit constitutes a clean (two-site shift invariant), periodically driven system. 
	
	{  A useful property of the local gate, which we use later to identify quantum circuits with similar dynamical features, is its \emph{entangling power}. Roughly speaking, the latter is a  measure of the ability of the gate to entangle two qubits~\cite{rather2020creating}. Normalising it to be in $[0,1]$, the entangling power can be expressed as~\cite{rather2020creating}
		\begin{equation}
			\!\!\!\!\! p(U)\!=\!\frac{d^4\!+\!d^2\!-\!\tr\!\smash{({U}^R{U}^{R\,\dag})^2}\!-\!\tr\!\smash{(\! {(US)}^R {(US)}^{R\,\dag}\!)^2}}{d^2(d^2-1)}.\label{eq:entanglingpowerdef}
		\end{equation}
		Here $U^R$ indicates the gate obtained by rotating the original one by a right angle
		\begin{align}
			\mel{i j}{U^R}{l k}=\mel{l i}{U}{k j}, \qquad i,j,k,l=1,\ldots d,
		\end{align}
		and $S$ denotes the SWAP gate
		\begin{align}
			\mel{i j}{S}{l k}= \delta_{i,k} \delta_{j,l}\,.
		\end{align}
	}
	
	Here we are interested in the evolution of the system for $t>0$. At $t=0$ the system is prepared in a generic ``pair-product'' state
	\begin{equation}
		\ket{\Psi_0}=\frac{1}{d^{L/2}}\bigotimes_{x=1}^L \left(\sum_{i,j=0}^{d-1} m_{ij} \ket{i}\otimes\ket{j}    \right),
		\label{eq:initialstate}
	\end{equation}
	where $\{\ket{i}\}_{i=0}^{d-1}$ is a basis of the configuration space of a single qudit --- the ``local'' Hilbert space. The matrix $m$, with elements $m_{ij}$, fulfils 
	\begin{equation}
		{\tr}[{m m^\dag}]=d,
		\label{eq:normalization}
	\end{equation}
	which ensures that $\ket{\Psi_0}$ is normalised to one. We consider general pair-product states, rather than simple product states, to keep the staggered structure of the brickwork quantum circuit. Note that a product state is recovered by the choice 
	\begin{equation}
		m_{ij}\propto \delta_{i,i_0}\delta_{j,j_0},\qquad i_0,j_0\in\mathbb \{0,\ldots,d-1\},
		\label{eq:mprod}
	\end{equation}
	while generically one can think of \eqref{eq:initialstate} as a product state which has been subject to half a step of evolution. 
	
	The evolution in a quantum circuit can be conveniently illustrated using a tensor-network-inspired graphical representation~\cite{cirac2021matrix}. In particular, depicting the components of the local gate and the initial state matrix as
	\begin{align}
		U_{(k,l);(i,j)}=\fineq[-0.8ex][1][1]{
			\tsfmatV[0][-0.5][r][1][][][bertinired]
			\node at (-0.5,-0.2) {$i$};
			\node at (0.5,-0.15) {$j$};
			\node at (-0.5,1.2) {$k$};
			\node at (0.5,1.25) {$l$};},
		\qquad
		m_{ij}=
		\fineq[-0.8ex][1][1]{
			\drawinitstate[1][0.5][l][1][bertinired]
			\node at (-0.5,1.15) {$i$};
			\node at (0.5,1.2) {$j$};},
		\label{eq:Uloc}
	\end{align}
	we can represent the state of the system at time $t$ as follows 
	\begin{equation}
		\ket{\Psi_t}= \frac{1}{d^{L/2}}\fineq[-0.8ex][0.5][1]{
			\foreach \i in {0,...,4}{
				\tsfmatV[2*\i][0][r][6][][init][bertinired][topright]}
			\foreach \i in {2,4,6}{
				\hook[9][\i][t][l]
				\hook[9][\i-1][b][l]
				\hook[-1][\i][t][r]
				\hook[-1][\i-1][b][r]}
			\foreach \i in {-2,...,0}
			{
				\draw[thick, dashed] (-.5,2*\i+5.25)--(9.5,2*\i+5.25);
				\draw[thick, dashed] (-.5,2*\i+4.75)--(9.5,2*\i+4.75);
			}			
		}\,,
	\end{equation}
	where we considered ${t=3}$. As illustrated in the above diagram, we depicted the periodic boundary conditions by connecting left and right boundaries, and use the convention that when legs of different tensors are joined together the index of the corresponding local space is summed over. Moreover, we drop the indices to represent the full vector rather than its components. We will use this convention whenever it does not lead to confusion.

	Let us consider the evolution of the reduced density matrix of a finite region $A$. Representing it diagrammatically we have
	\begin{align}
		\!\!\rho_{A}(t)\!=\frac{1}{d^{L}}\fineq[-0.8ex][0.5][1]{\foreach \i in {0,...,2}
			{
				\draw[thick, dashed] (-.5,2*\i+5.25)--(9.5,2*\i+5.25);
				\draw[thick, dashed] (-.5,2*\i+4.75)--(9.5,2*\i+4.75);
			}\foreach \i in {-2,...,-4}
			{
				\draw[thick, dashed] (-.5,2*\i+6.25)--(9.5,2*\i+6.25);
				\draw[thick, dashed] (-.5,2*\i+5.75)--(9.5,2*\i+5.75);
			}
			\foreach \i in {0,1,4}{
				\draw[thick, dashed] (.75+2*\i,-3.5)--(.75+2*\i,10.5);	
				\draw[thick, dashed] (1.25+2*\i,-3.5)--(1.25+2*\i,10.5);	
			}
			\foreach \i in {0,...,4}{
				\tsfmatV[2*\i][4][r][6][][init][bertinired][topright]
				\drawinitstateflipped[2*\i+2][3][l][1][bertiniblue]
				\tsfmatV[2*\i][-4][l][6][][][bertiniblue][topright]
			}
			\foreach \i in {6,8,10}{
				\hook[9][\i][t][l]
				\hook[9][\i-1][b][l]
				\hook[-1][\i][t][r]
				\hook[-1][\i-1][b][r]}
			\foreach \i in {-1,1,3}{
				\hook[9][\i][t][l]
				\hook[9][\i-1][b][l]
				\hook[-1][\i][t][r]
				\hook[-1][\i-1][b][r]}
			\foreach \i in {0,2,8}{
				\hhook[\i][-3][b][l]
				\hhook[\i+1][-3][b][r]
				\hhook[\i+1][11][t][l]
				\hhook[\i][11][t][r]}
		}	
		\label{eq:reducedAdiag}
	\end{align}
	where we took $A=\{1,3/2,2,5/2\}$ and introduced a diagrammatic representation for the Hermitian conjugate of the local gate
	\begin{equation}
		U^\dag= \fineq[-0.8ex][.85][1]{
			\tsfmatV[0][0][r][1][][][bertiniblue][topright]
		},
		\label{eq:Ulocdag}
	\end{equation}
	and the complex conjugate of the initial state matrix
	\begin{equation}
		m^*=\fineq[-0.8ex][1][1]{
			\drawinitstateflipped[0][2][l][1][bertiniblue]
		}\,.
	\end{equation}
	Using this representation we can depict the unitarity of the local gate with the following diagrammatic relations
	\begin{equation}
		\fineq[-0.8ex][.85][1]{
			\tsfmatV[0][1.25][r][1][][][bertinired][topright]
			\tsfmatV[0][0][r][1][][][bertiniblue][topright]
			\draw[ thick] (-0.5,1.5) -- (-0.5,1.75);
			\draw[ thick] (0.5,1.5) -- (0.5,1.75);
		}
		= 
		\fineq[-0.8ex][.85][1]{
			\tsfmatV[0][1.25][r][1][][][bertiniblue][topright]
			\tsfmatV[0][0][r][1][][][bertinired][topright]
			\draw[ thick] (-0.5,1.5) -- (-0.5,1.75);
			\draw[ thick] (0.5,1.5) -- (0.5,1.75);
		}
		=
		\fineq[-0.8ex][.85][1]{
			\draw[ thick] (-0.45,0.15) -- (-0.45,2.35);
			\draw[ thick] (0.45,0.15) -- (0.45,2.35);
		}
		\,.
		\label{eq:unitaritynonfolded}
	\end{equation}
	
	To simplify the diagrams it is convenient to fold them in two. In particular, folding the blue part of the circuit underneath the red one, we can represent the reduced density matrix in \eqref{eq:reducedAdiag} as follows
	\begin{align}
		\!\!\rho_{A}(t)\!=\! 
		\frac{1}{d^{|A|/2}}\fineq[-0.8ex][0.5][1]{
			\foreach \i in {0,...,-2}
			{
				\draw[thick, dashed] (-.5,2*\i+5.25)--(9.5,2*\i+5.25);
				\draw[thick, dashed] (-.5,2*\i+4.75)--(9.5,2*\i+4.75);
			}
			\foreach \i in {0,...,1}{
				\tsfmatV[2*\i][0][r][6][tr][init][bertiniorange][topright]}
			\foreach \i in {2,...,3}{
				\tsfmatV[2*\i][0][r][6][][init][bertiniorange][topright]}
			\tsfmatV[8][0][r][6][tr][init][bertiniorange][topright]
			\foreach \i in {2,4,6}{
				\hook[9][\i][t][l]
				\hook[9][\i-1][b][l]
				\hook[-1][\i][t][r]
				\hook[-1][\i-1][b][r]}
		}\,,
		\label{eq:DMfolded}
	\end{align}
	where we introduced the \emph{double gate} 
	\begin{equation}
		\fineq[-0.8ex][.85][1]{
			\tsfmatV[0][0][r][1][][]
		}:= W = U\otimes_r U^*\,,
		\label{eq:doublegate}
	\end{equation}
	the \emph{double initial-state matrix}
	\begin{equation}
		\fineq[-0.8ex][.85][1]{
			\drawinitstate[0][2][l][1]
		} = m \otimes_r m^*\,,
	\end{equation}
	the \emph{loop state} 
	\begin{equation}
		\fineq[-0.8ex][.85][1]{
			\draw[very thick] (-0.15,0.25) -- (-0.15,-0.251);
			\draw[thick, fill=white] (-.15,-0.25) circle (0.1cm); 
		} =
		\frac{1}{\sqrt{d}}\,\,
		\fineq[-0.8ex][.85][1]{
			\draw[thick] (-2,0.25) -- (-2,-0.251);
			\draw[thick] (-1.5,0.4) -- (-1.5,-0.101);
			\draw[thick] (-2,-0.25) to[out=-85,in=-80] ( (-1.505,-0.1);
		}:= \ket{\mcirc},
		\label{eq:foldedwire}
	\end{equation}
	and, finally, the shorthand notation  
	\begin{equation}
		|A|:= (\# \text{ of sites in $A$})\,.
	\end{equation} 
	In the above equations $\otimes_r$ denotes the tensor product between different copies or \emph{replicas} of the time sheet (different from $\otimes$ which is the one between different spatial sites in the same copy).

	In this folded representation, the unitarity relations \eqref{eq:unitaritynonfolded} are depicted as  
	\begin{equation}
		\fineq[-0.8ex][0.75][1]{
			\tsfmatV[0][0][r][1][][][bertiniorange][topright]
			\draw[thick, fill=white] (-.5,1.5) circle (0.1cm); 
			\draw[thick, fill=white] (.5,1.5) circle (0.1cm); 
		}
		=
		\fineq[-0.8ex][0.75][1]{
			\def\eps{0.5}
			\draw[ thick] (-4.25,0.5) -- (-4.25,-0.5);
			\draw[ thick] (-3.25,-0.5) -- (-3.25,0.5);
			\draw[thick, fill=white] (-4.25,0.5) circle (0.1cm); 
			\draw[thick, fill=white] (-3.25,0.5) circle (0.1cm); 
		}\,,
		\qquad
		\fineq[-0.8ex][0.75][.85]{
			\tsfmatV[0][0][r][1][][][bertiniorange][topright]
			\draw[thick, fill=white] (-.5,0.5) circle (0.1cm); 
			\draw[thick, fill=white] (.5,0.5) circle (0.1cm); 
		}
		=
		\fineq[-0.8ex][0.75][.85]{
			\def\eps{0.5}
			\draw[ thick] (-4.25,0.5) -- (-4.25,-0.5);
			\draw[ thick] (-3.25,-0.5) -- (-3.25,0.5);
			\draw[thick, fill=white] (-4.25,-0.5) circle (0.1cm); 
			\draw[thick, fill=white] (-3.25,-0.5) circle (0.1cm); 
		}\,. 
		\label{eq:unitarity}
	\end{equation}
	Moreover, since the double gate is itself unitary we also have  
	\begin{equation}
		\fineq[-0.8ex][.85][1]{
			\tsfmatV[0][1.25][r][1][][][bertiniorange][topright]
			\tsfmatV[0][0][r][1][][][bertinigreen][topright]
			\draw[ thick] (-0.5,1.5) -- (-0.5,1.75);
			\draw[ thick] (0.5,1.5) -- (0.5,1.75);
		}
		= 
		\fineq[-0.8ex][.85][1]{
			\tsfmatV[0][1.25][r][1][][][bertinigreen][topright]
			\tsfmatV[0][0][r][1][][][bertiniorange][topright]
			\draw[ thick] (-0.5,1.5) -- (-0.5,1.75);
			\draw[ thick] (0.5,1.5) -- (0.5,1.75);
		}
		=
		\fineq[-0.8ex][.85][1]{
			\draw[ thick] (-0.45,0.15) -- (-0.45,2.35);
			\draw[ thick] (0.45,0.15) -- (0.45,2.35);
		}\,,
		\label{eq:unitarityfolded}
	\end{equation}
	where we introduced
	\begin{equation}
		\fineq[-0.8ex][.85][1]{
			\tsfmatV[0][0][r][1][][][bertinigreen]
		}:= W^\dag = U^\dag\otimes_r U^T\,.\label{eq:doublegatedag}
	\end{equation}

	\section{Generalised folding algorithm and generalised temporal entanglement}
	\label{sec:folding}
	
	A standard class of observables in quantum circuits are correlation functions of local operators. In particular, let us focus on non-equilibrium dynamical two-point functions of the form 
	\begin{equation}
		C_{ab}(x_1,x_2,t_1,t_2)=\tr\left[\rho_0 a_{x_1}(t_1) b_{x_2}(t_2)\right],
		\label{eq:2bodycorrelator}
	\end{equation}
	where we took $t_2\geq t_1\geq 0$, ${a_x := a_x(0)}$ and ${b_x := b_x(0)}$ are local operators, and $\rho_0=\ketbra{\Psi_0}$ is the initial state (cf.~\eqref{eq:initialstate}). Note that \eqref{eq:2bodycorrelator} contains non-equilibrium one point functions as a special case that is obtained by setting $a_x=\1$.

	In fact, the upcoming discussion will also be applicable to the case where $\rho_0$ is the infinite-temperature state, which, in generic situations, is the only stationary state of the system. In this case the correlation takes the following equilibrium form 
	\begin{equation}
		C^{\rm eq}_{ab}(x_1,x_2,t)=\tr\left[a_{x_1} b_{x_2}(t)\right].
		\label{eq:2bodycorrelatorth}
	\end{equation}
	Because of the strict light cone structure of the quantum circuit, the correlation function \eqref{eq:2bodycorrelator} is non-trivial (i.e.\ causally connected) only if (see Fig.~\ref{fig:paths}(b))
	\begin{equation}
		|\lceil x_1 \rceil - \lceil x_2 \rceil | \leq t_1+t_2\,.
		\label{eq:timelikecondition}
	\end{equation}
	while \eqref{eq:2bodycorrelatorth} only if 
	\begin{equation}
		|\lceil x_1 \rceil - \lceil x_2 \rceil | \leq t\,.
		\label{eq:timelikeconditionth}
	\end{equation}
	For the sake of definiteness from now on we assume $x_2< x_1$ and $x_1,x_2 \in\mathbb Z_L$, while to lighten the notation we drop the dependence of the correlation on $x_1,x_2,t_1,t_2$.

	Considering this case we can represent \eqref{eq:2bodycorrelator} diagrammatically as 
	\begin{align}
		C_{ab}= \fineq[-0.8ex][0.5][1]{
			\foreach \i in {1,...,-2}
			{
				\draw[thick, dashed] (3.5,2*\i+5.25)--(15.5,2*\i+5.25);
				\draw[thick, dashed] (3.5,2*\i+4.75)--(15.5,2*\i+4.75);
			}
			\foreach \i in {2,...,7}{
				\tsfmatV[2*\i][0][r][8][tr][init][bertiniorange][topright]}
			\draw[thick, fill=black] (10.5,2.5) circle (0.15cm); 
			\node at (10.2,2.2) {\LARGE $a$};
			\draw[thick, fill=black] (6.5,8.5) circle (0.15cm); 
			\node at (6.2,8.2) {\LARGE $b$};
			\foreach \i in {2,4,...,8}{
				\hook[15][\i][t][l]
				\hook[15][\i-1][b][l]
				\hook[3][\i][t][r]
				\hook[3][\i-1][b][r]}
		}\,,
		\label{eq:Corrfolded}
	\end{align}
	where, for simplicity, we assumed that $a_x$ and $b_x$ act non-trivially only on one site, we depicted them as 
	\begin{equation}
		\fineq[-0.4 ex][.85][1]{
			\draw[very thick] (-0.15,0.25) -- (-0.15,-0.25);
			\draw[thick, fill=black] (-.15,0) circle (0.1cm); 
			\node at (-.5,0) {\large $a$};
		} = a\otimes_r \1,
		\qquad 
		\fineq[-0.5 ex][.85][1]{
			\draw[very thick] (-0.15,0.25) -- (-0.15,-0.25);
			\draw[thick, fill=black] (-.15,0) circle (0.1cm); 
			\node at (-.5,0) {\large $b$};
		} = b\otimes_r \1, 
	\end{equation}
	and set 
	\begin{equation}
		\fineq[-0.8ex][.85][1]{
			\draw[very thick] (-0.15,0.25) -- (-0.15,-0.251);
			\draw[thick, fill=black] (-.15,-0.25) circle (0.1cm); 
			\node at (-.5,-.25) {\large $a$};
		} = \fineq[-0.8ex][.85][1]{
			\draw[very thick] (-0.15,0.25) -- (-0.15,-0.251);
			\draw[thick, fill=black] (-.15,0) circle (0.1cm); 
			\node at (.125,0) {\large $a$};
			\draw[thick, fill=white] (-.15,-0.25) circle (0.1cm); 
		}. 
	\end{equation}
	Let us now illustrate how the diagram  \eqref{eq:Corrfolded} can be evaluated using the folding algorithm of Ref.~\cite{banuls2009matrix}. The starting point is to represent it in terms of transfer matrices ``in space''. Namely, one introduces three different transfer matrices 
	\begin{align}
		\mathcal T_{x}=\fineq[-0.8ex][0.5][0.5]{
			\tsfmatV[0][0][r][8][tr][init]
			\node[scale=2] at (1,-0.5) {$x$};},\quad 
		\mathcal T^{(a)}_{x}=\fineq[-0.8ex][0.5][0.5]{
			\tsfmatV[0][0][r][8][tr][init]
			\draw[thick, fill=black] (.5,2.5) circle (0.15cm); 
			\node at (.2,2.2) {\LARGE $a$};
			\node[scale=2] at (1,-0.5) {$x$};},
		\quad
		\mathcal T^{(b)}_{x}=\fineq[-0.8ex][0.5][0.5]{
			\tsfmatV[0][0][r][8][tr][init]
			\draw[thick, fill=black] (.5,8.5) circle (0.15cm); 
			\node at (.3,8.2) {\LARGE $b$};
			\node[scale=2] at (1,-0.5) {$x$};},
		\label{eq:spaceTM}
	\end{align}
	so that the diagram \eqref{eq:Corrfolded} can be written as 
	\begin{align}
		C_{ab} =&  {\rm tr}[{\mathcal T}_{1}\cdots {\mathcal T}_{x_1-1} {\mathcal T}^{(a)}_{x_1}{\mathcal T}_{x_1+1}\cdots \notag\\ 
		&\qquad {\mathcal T}_{x_2-1}{\mathcal T}^{(b)}_{x_2}{\mathcal T}_{x_2+1}\cdots {\mathcal T}_{L} ],
		\label{eq:correlationtransfer}
	\end{align}
	where we consider the generic case of non-translational invariant circuits. We remark that the space transfer matrices in \eqref{eq:spaceTM} are matrix product operators (MPO) with finite bond dimension $\chi =d^2$. 
	
	The next step is to note that unitarity can simplify products of transfer matrices. To illustrate this point, let us write down the product of $2t_2$ transfer matrices ($\mathcal{T}_x$ in Eq.~\eqref{eq:spaceTM}). In diagrams it takes the following form 
	\begin{equation}
		\!\!\fineq[-0.8ex][0.45][0.45]{      
			\foreach \x in {2,4,...,16}{
				\tsfmatV[\x][0][r][8][tr][init]
			}
			\draw[dashed, red, line width = 1pt] (6+4,0)--++(8,8);
			\draw[dashed, red, line width = 1pt] (6+4,0)--++(-8,8);
		}.
	\end{equation}
	Unitarity (cf. Eq.~\eqref{eq:unitarityfolded}) allows us to cancel all the gates above the red dashed lines and propagate the bullets to the legs crossing the dashed lines. We therefore have the following rank-1 decomposition
	\begin{equation}
		{\mathcal T}_{y_1}\cdots {\mathcal T}_{y_{2t_2}} = \ketbra*{{{R}_{y_1}}}{L_{y_{2t_2}}}, \qquad \forall y_j,
		\label{eq:lightcone}
	\end{equation}
	where we introduced the following vectors on the folded time lattice 
	\begin{equation}
		\ket*{{{R}_x}}= \fineq[-0.8ex][0.5][0.5]{
			\eigenVR[-1][0][r][7]
			\node[scale=2] at (0,-0.5) {$x$};  
		}\,,
		\label{eq:standardinfluenceR}
	\end{equation}
	\begin{equation}
		\bra*{L_x}= \fineq[-0.8ex][0.5][0.5]{
			\draw[thick] (-.75,7.5) -- (-.35,7.15);
			\draw[fill=white] (-.75,7.5) circle (0.15cm);
			\eigenVL[-1][0][l][6]
			\node[scale=2] at (-1,-0.5) {$x$};  
		}\,.
		\label{eq:standardinfluenceL}
	\end{equation}
	This means that, for $L>x_2-x_1+2t_2$, Eq.~\eqref{eq:correlationtransfer} can be written as
	\begin{equation}
		\!\!\!C_{ab} \!=\!\! \mel{L_{x_1-1}}{{\mathcal T}^{(a)}_{x_1}{\mathcal T}_{x_1+1}\cdots {\mathcal T}_{x_2-1}{\mathcal T}^{(b)}_{x_2}}{R_{x_2+1}}\!.
		\label{eq:CorrLargesystem}
	\end{equation}
	This representation sheds light on the physical interpretation of the two vectors $\bra*{L_x}$ and $\ket*{{{R}_x}}$. These objects encode the effect of the rest of the system on the subsystem of size $x_2-x_1$ where ${a_x}$ and ${b_x}$ act. Since their role is analogous to that of the influence functional of Feynman and Vernon~\cite{feynman1963the}, they have been dubbed ``influence matrices''~\cite{lerose2021influence}. Note that in the translational invariant case one can use \eqref{eq:lightcone} to show that $\bra*{L_x}$ and $\ket*{{{R}_x}}$ are the unique \emph{fixed points}, i.e.\ eigenvectors corresponding to eigenvalue one, of the space transfer matrix ${\mathcal T}$ (which is $x$-independent in translational invariant circuits). 
	
	The representation \eqref{eq:CorrLargesystem} is the main instrument of the folding algorithm. Assuming that one can find an efficient MPS representation for the influence matrices (see Sec.~\ref{sec:TE}), Eq.~\eqref{eq:CorrLargesystem} gives a way to compute two-point functions as matrix elements of an MPO --- the product of $x_2-x_1+1$ space transfer matrices --- between two MPS. Since the bond dimension of the MPO is bounded by $d^{2(x_2-x_1+1)}$ this operation can be performed efficiently for small distances $x_2-x_1$. On the other hand, the computation becomes rapidly unfeasible when the distance increases. This represents a serious limitation as, for instance, two point functions for arbitrary distances fulfilling \eqref{eq:timelikecondition} are needed to compute transport coefficients in linear response~\cite{mahan1981many, altland2010condensed,bertini2021finitetemperature}. { To circumvent this problem we propose an alternative method for contracting the diagram in Eq.~\eqref{eq:Corrfolded}: Instead of contracting it in the space direction, we contract it in a more general \emph{space-like} direction such that the two points lie on the same time-like surface, see  the macroscopic-scale illustration in Fig.~\ref{fig:TE_intro}. The only relevant macroscopic feature of the time-like surface is its the spacetime slope $v_{\gamma}$. A more precise lattice definition is given in the upcoming Sec.~\ref{subsec:gen_folding}. While in Sec.~\ref{sec:TE} we discuss the computational complexity of encoding the influence matrices in an MPS.}
	
	\subsection{Generalised Folding Algorithm}
	\label{subsec:gen_folding}
	
	For a precise definition of the generalised folding algorithm it is useful to distinguish between two different regimes: 
	\begin{itemize}
		\item[(I)] $0 \leq x_2-x_1\leq t_2-t_1$;
		\item[(II)] $t_2-t_1<x_2-x_1\leq t_2+t_1$; 
	\end{itemize}
	Note that (II) only arises out-of-equilibrium: the equilibrium correlation in Eq.~\eqref{eq:2bodycorrelatorth} exists only in the regime (I). Moreover, (I) is also the only regime arising for non-equilibrium one-point functions.

	\subsubsection{Regime $(I)$}
	
	In this regime there exists a path $\tilde\gamma$ connecting $a$ and $b$ that is entirely contained in the causal light cone emanating from $a$, i.e., it goes from $a$ to $b$ without ever ``turning back''. We call this kind of paths \emph{time-like paths}, because all the space-time points they reach are causally connected.  
	
\begin{figure*}
		
		\ifdef{0}{}{ \scalebox{0.875}{
				\begin{tikzpicture}
					[baseline={([yshift=-0.6ex]current bounding box.center)},scale=.6] 
					\node at (3,9) {\large $(a)$};  
					\clip (3,-.5) rectangle (16,10); 
					\foreach \i in {2,...,7}{
						\tsfmatV[2*\i][0][r][8][tr][init][bertiniorange][topright]}
					\draw[thick, fill=black] (21-10.5,2.5) circle (0.15cm); 
					\node at (20-10.2+0.35,2.2) {\large $a$};
					\draw[thick, fill=black] (20-12.5,8.5) circle (0.15cm); 
					\node at (20-12.2,8.2) {\large $b$};
					\foreach \i in {2,4,...,8}{
						\hook[15][\i][t][l]
						\hook[15][\i-1][b][l]
						\hook[3][\i][t][r]
						\hook[3][\i-1][b][r]}
					\draw[line width=0.75mm, red, rounded corners, stealth-] (20.5-10,2.5) -- (20-14,7) -- (20-13,8);
					\node at (20-13.4,7) {\large ${\color{red}\tilde \gamma}$};
					\draw[line width=0.75mm, blue, rounded corners, stealth-] (20-9,0) -- (20-10,1) -- (20-9,2) -- (20.5-10,2.5);
					\node at (20-9.3,1) {\large ${ \tilde{\tilde \gamma}}$};
					\foreach \i in {-7,-5,-3,-1,1,3,5,7,9}{
						\draw[line width=0.3mm, black, dashed, rounded corners] (20-9+\i,0) -- (20-10+\i,1) -- (20-9+\i,2) -- (20.5-10+\i,2.5) -- (20-14+\i,7) -- (20-13+\i,8);
					}                       
				\end{tikzpicture}
			}
			\qquad
			\scalebox{0.875}{
				\begin{tikzpicture}
					[baseline={([yshift=-0.6ex]current bounding box.center)},scale=.6] 
					\node at (3,9) {\large $(b)$};    
					\clip (3,-.5) rectangle (16,10);  
					\foreach \i in {2,...,7}{
						\tsfmatV[2*\i][0][r][8][tr][init][bertiniorange][topright]}
					\draw[thick, fill=black] (14.5,2.5) circle (0.15cm); 
					\node at (14.2,2.2) {\large $a$};
					\draw[thick, fill=black] (4.5,8.5) circle (0.15cm); 
					\node at (4.2,8.2) {\large $b$};
					\foreach \i in {2,4,...,8}{
						\hook[15][\i][t][l]
						\hook[15][\i-1][b][l]
						\hook[3][\i][t][r]
						\hook[3][\i-1][b][r]}
					\draw[line width=0.75mm, black, rounded corners, stealth-] (15,0) -- (14,1) -- (15,2) --(14,3) -- (9,8);
					\node at (12.2,4.2) {\large $\bar \gamma$};
					\foreach \i in {-11,-9,-7,-5,-3,-1,1,3,5,7,9}{
						\draw[line width=0.3mm, black, dashed, rounded corners] (15+\i,0) -- (14+\i,1) -- (15+\i,2) --(14+\i,3) -- (9+\i,8);
					}  
				\end{tikzpicture}
		}}
		\includegraphics[scale=0.875]{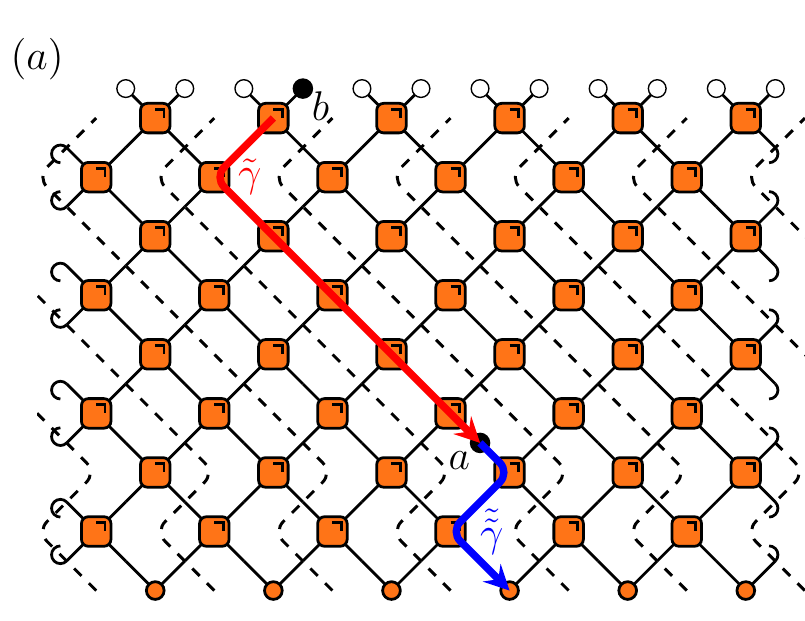}
		\includegraphics[scale=0.875]{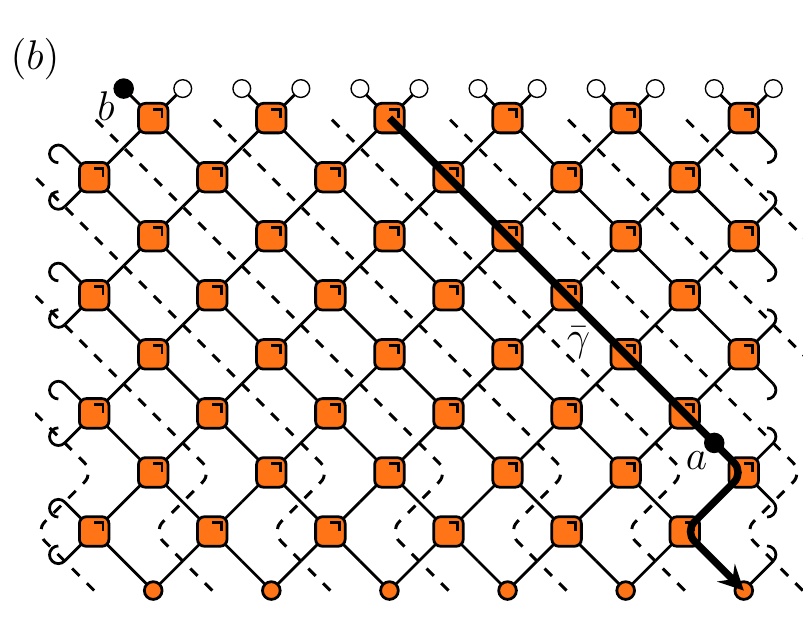}
		\caption{Diagrammatic representation of the path to contract correlation functions. Panels (a) and (b) depict examples of paths used in the contraction of the correlation functions in the regimes I and II respectively. Dashed lines helps to identify the transfer matrix and time slices to cut open the diagram. }
		\label{fig:paths}
	\end{figure*}

	To specify $\tilde\gamma$ we start from the gate below $b$ and move down in discrete jumps, see Fig.~\ref{fig:paths} (a). At each jump we reach one neighbouring gate: either the one at south east or the one at south west. Using the variable ${\tilde\gamma}_i=\pm$ to keep track of whether on the $i$-th step we jump on the left or on the right, we can represent the path by means of the following sequence
	\begin{equation}
		\tilde \gamma = \{{\tilde \gamma}_1, \ldots, {\tilde \gamma}_{N}\},
	\end{equation}
	where $N=2(t_2-t_1)-1$ is the length of the path. For instance  
	\begin{equation}
		\tilde\gamma = \{-,+,+,+,+\}, 
	\end{equation}
	is the path depicted in red in Fig.~\ref{fig:paths} (a). The path $\tilde \gamma$ can be extended to a path $\gamma$ that reaches the initial state by concatenating it with another time-like path $\tilde{\tilde \gamma}$ from $(x_1,t_1)$ to $(y,0)$ for some $y\in[x_1-t, x_1+t]$. The total length of the path $\gamma$, i.e.\ the total number of jumps, is then $2t_2$. For instance, in the example of Fig.~\ref{fig:paths} one can consider 
	\begin{equation}
		\gamma = \{-,+,+,+,+\} \circ \{+,-,+\},
		\label{eq:examplepath}
	\end{equation}
	where $\circ$ denotes the composition operator. The average slope of a given path $\gamma$ is given by 
	\begin{equation}
		v_{\gamma}=\frac{1}{|\gamma|}\sum_{i=1}^{|\gamma|} \gamma_i\,,
		\label{eq:vgamma}
	\end{equation}
	where $|\gamma|$ is the length of the path. {  As mentioned before, $v_{\gamma}$ is the only bit of information required for a coarse grained description of the path.}  
	
	Since the path $\gamma$ does not turn back, we can use it to ``slice'' the diagram of the correlation function. Namely, we subdivide it in a number of  time-like slices by cutting the bonds in a direction parallel to $\gamma$ (see the black dashed lines in Fig.~\ref{fig:paths}) and connect them with suitably defined transfer matrices. In particular, for the configuration in Fig.~\ref{fig:paths} (a) the transfer matrices are given by
	\begin{equation}
		\!\!\!\!{\mathcal T}_{\gamma,x}=\!\!\!\!\!\!\!\fineq[-.8ex][0.5][0.5]{
			\tsfmatV[-3][7][r][1][tr][][bertiniorange][topright]
			\tsfmatV[-4][6][r][1][][][bertiniorange][topright]
			\tsfmatV[-3][5][r][1][][][bertiniorange][topright]
			\tsfmatV[-2][4][r][1][][][bertiniorange][topright]
			\tsfmatV[-1][3][r][1][][][bertiniorange][topright]
			\tsfmatV[0][2][r][1][][][bertiniorange][topright]
			\tsfmatV[1][1][r][1][][][bertiniorange][topright]
			\tsfmatV[0][0][r][1][][init][bertiniorange][topright]
			\node[scale=2] at (1,-0.5) {$x$};
		}\!\!,
		\quad
		{\mathcal T}^{(ab)}_{\gamma,x}=\!\!\!\!\!\!\!\!\!\fineq[-.8ex][0.5][0.5]{
			\tsfmatV[-3][7][r][1][tr][][bertiniorange][topright]
			\tsfmatV[-4][6][r][1][][][bertiniorange][topright]
			\tsfmatV[-3][5][r][1][][][bertiniorange][topright]
			\tsfmatV[-2][4][r][1][][][bertiniorange][topright]
			\tsfmatV[-1][3][r][1][][][bertiniorange][topright]
			\tsfmatV[0][2][r][1][][][bertiniorange][topright]
			\tsfmatV[1][1][r][1][][][bertiniorange][topright]
			\tsfmatV[0][0][r][1][][init][bertiniorange][topright]
			\draw[thick, fill=black] (-2.5,8.5) circle (0.15cm); 
			\node at (-2.2,8.2) {\LARGE $b$};
			\draw[thick, fill=black] (.5-.1,2.5+.1) circle (0.15cm); 
			\node at (.2,2.2) {\LARGE $a$};
			\node[scale=2] at (1,-0.5) {$x$};}\!\!.
	\end{equation}
	In this way we can write Eq.~\eqref{eq:Corrfolded} as
	\begin{align}
		C_{ab} =&  {\rm tr}[{\mathcal T}_{\gamma,1}\cdots {\mathcal T}_{\gamma,x_1-1} {\mathcal T}^{(ab)}_{\gamma, x_1}{\mathcal T}_{\gamma,x_1+1}\cdots {\mathcal T}_{\gamma,L} ].
		\label{eq:correlationtransfer}
	\end{align}
	This expression can again be simplified using the unitarity of the gates. In particular, we again have
	\begin{equation}
		{\mathcal T}_{\gamma,y_1}\cdots {\mathcal T}_{\gamma,y_{2t_2}} = \ketbra*{{{R}_{\gamma,y_1}}}{L_{\gamma,y_{2t_2}}}, \qquad \forall y_j,
		\label{eq:lightconegamma}
	\end{equation}
	where we introduced the generalised or ``boosted'' influence matrices 
		\begin{equation}
		\bra{L_{\gamma,x}}=\!\!\!\!\!\!\!\fineq[-.8ex][0.5][0.5]{
			\tsfmatV[-4][5][r][1][][][bertiniorange][topright]
			\tsfmatV[-3][4][r][1][][][bertiniorange][topright]
			\tsfmatV[-2][3][r][1][][][bertiniorange][topright]
			\tsfmatV[-1][2][r][1][][][bertiniorange][topright]
			\tsfmatV[0][1][r][1][][][bertiniorange][topright]
			\tsfmatV[-1][0][r][1][][init][bertiniorange][topright]
			\node[scale=2] at (0,-0.5) {$x$};
			\eigenVL[-5][0][l][5]
			\tsfmatV[-4][0][l][4][][init]
			\tsfmatV[-2][0][l][2][][init]
			\draw[thick] (-.75-2-.9+0.1,7.5+1-.9-0.1) -- (-.35-2-.9,7.15+1-.9);
			\draw[fill=white] (-4.5,6.5) circle (0.15cm); 
			\draw[fill=white] (-4.5+1,6.5+1) circle (0.15cm); 
		}, 
		\label{eq:Lgamma}
	\end{equation}
	\begin{equation}
		\ket{R_{\gamma,x}}	=\fineq[-0.8ex][0.5][0.5]{
			\tsfmatV[-4][6][r][1][][][bertiniorange][topright]
			\tsfmatV[-3][5][r][1][][][bertiniorange][topright]
			\tsfmatV[-2][4][r][1][][][bertiniorange][topright]
			\tsfmatV[-1][3][r][1][][][bertiniorange][topright]
			\tsfmatV[0][2][r][1][][][bertiniorange][topright]
			\tsfmatV[1][1][r][1][][][bertiniorange][topright]
			\tsfmatV[2][0][r][1][][init][bertiniorange][topright]
			\node[scale=2] at (1,-0.5) {$x$};
			\eigenVR[0][0][l][1][0]
			\draw[thick, fill=white] (1.5,2.5) circle (0.15cm); 
			\draw[thick, fill=white] (1.5-1,2.5+1) circle (0.15cm); 
			\draw[thick, fill=white] (1.5-2,2.5+2) circle (0.15cm); 
			\draw[thick, fill=white] (1.5-3,2.5+3) circle (0.15cm); 
			\draw[thick, fill=white] (1.5-4,2.5+4) circle (0.15cm); 
			\draw[thick, fill=white] (1.5-5,2.5+5) circle (0.15cm); 
			\draw[thick, fill=white] (1.5+1,2.5-1) circle (0.15cm); 
			\draw[thick, fill=white] (1.5+2,2.5-2) circle (0.15cm); 
		}\,.
		\label{eq:Rgamma}
	\end{equation}
	Therefore, for $L>2t_2$, we find  
	\begin{equation}
		\!\!\!C_{ab} \!=\!\! \mel{L_{\gamma, x_1-1}}{{\mathcal T}^{(ab)}_{\gamma, x_1}}{R_{\gamma, x_1+1}}\!.
		\label{eq:CorrLargesystemgamma}
	\end{equation}
	As opposed to Eq.~\eqref{eq:CorrLargesystem}, this expression can be always efficiently contracted if $\bra{L_{\gamma, x}}$ and $\ket{R_{\gamma, x}}$ admit an efficient MPS representation.

	\subsubsection{Regime $(II)$}
	In this regime there is no time-like path connecting $a$ and $b$. This means that we cannot embed both $a$ and $b$ in the same ``thin'' transfer matrix as done in Eq.~\eqref{eq:CorrLargesystemgamma}. The best strategy in this case is to slice the diagram \eqref{eq:Corrfolded} using transfer matrices corresponding to the path $\bar \gamma = \gamma_{\rm lc}\circ \tilde{\tilde \gamma}$, where 
	\begin{equation}
		\gamma_{\rm lc} = \{+,...,+\},
		\label{eq:lightconepath}
	\end{equation}
	{ is the fastest path allowed by causality} and $\tilde{\tilde \gamma}$ is an arbitrary time-like path between the initial state and $a$ (cf. Fig.~\ref{fig:paths} (b)). Repeating the above analysis we find that for $L>x_2-x_1+t_2+t_1$ the correlations can be expressed as 
	\begin{align}
		&\!\!\!C_{ab} \!=\!\! \bra{L_{\bar \gamma, x_1-1}} {\mathcal T}^{(a)}_{\bar \gamma, x_1} {\mathcal T}_{\bar \gamma, x_1+1} \cdots \notag\\
		&\qquad\qquad{\mathcal T}_{\bar \gamma, x_2-t_2+t_1-1} {\mathcal T}^{(b)}_{\bar \gamma, x_2-t_2+t_1}\ket{R_{\bar \gamma, x_2-t_2+t_1+1}}\!,
		\label{eq:CorrLargesystemgamma2}
	\end{align}
	where we introduced 
	\begin{equation}
		\!\!\!\!{\mathcal T}^{(a)}_{\gamma,x}=\!\!\!\!\!\!\!\fineq[-.8ex][0.5][0.5]{
			\tsfmatV[-3][7][r][1][tr][][bertiniorange][topright]
			\tsfmatV[-4][6][r][1][][][bertiniorange][topright]
			\tsfmatV[-3][5][r][1][][][bertiniorange][topright]
			\tsfmatV[-2][4][r][1][][][bertiniorange][topright]
			\tsfmatV[-1][3][r][1][][][bertiniorange][topright]
			\tsfmatV[0][2][r][1][][][bertiniorange][topright]
			\tsfmatV[1][1][r][1][][][bertiniorange][topright]
			\tsfmatV[0][0][r][1][][init][bertiniorange][topright]
			\node[scale=2] at (1,-0.5) {$x$};
			\draw[thick, fill=black] (.5-.1,2.5+.1) circle (0.15cm); 
			\node at (.2,2.2) {\LARGE $a$};
		}\!\!,
		\quad
		{\mathcal T}^{(b)}_{\gamma,x}=\!\!\!\!\!\!\!\!\!\fineq[-.8ex][0.5][0.5]{
			\tsfmatV[-3][7][r][1][tr][][bertiniorange][topright]
			\tsfmatV[-4][6][r][1][][][bertiniorange][topright]
			\tsfmatV[-3][5][r][1][][][bertiniorange][topright]
			\tsfmatV[-2][4][r][1][][][bertiniorange][topright]
			\tsfmatV[-1][3][r][1][][][bertiniorange][topright]
			\tsfmatV[0][2][r][1][][][bertiniorange][topright]
			\tsfmatV[1][1][r][1][][][bertiniorange][topright]
			\tsfmatV[0][0][r][1][][init][bertiniorange][topright]
			\draw[thick, fill=black] (-3.5,8.5) circle (0.15cm); 
			\node at (-3.7,8.2) {\LARGE $b$};
			\node[scale=2] at (1,-0.5) {$x$};
		}\!\!.
	\end{equation}
	We see that the expression \eqref{eq:CorrLargesystemgamma2} involves the product of 
	\begin{equation}
		n=x_2-x_1-t_2+t_1+1
	\end{equation}
	transfer matrices, which means $0\leq n\leq 2 t_1$. This has two immediate implications: First, the representation \eqref{eq:CorrLargesystemgamma2} gives an advantage over \eqref{eq:CorrLargesystem} because it involves less transfer matrices. Second, when both $x_2-x_1$ and $t_1$ are large the contraction of \eqref{eq:CorrLargesystemgamma2} becomes inefficient.

	\subsection{Generalised Temporal Entanglement}
	\label{sec:TE}
	
	{ 
		In extreme summary, the upshot of the previous subsection is that an efficient representation of the generalised influence matrices does indeed lead to an efficient computational scheme for the calculation of correlation functions in many physically relevant cases~\footnote{This is always true for two-point functions on equilibrium states (cf.~\eqref{eq:2bodycorrelatorth}) (which encode linear response coefficients~\cite{bertini2021finitetemperature}) and one-point functions. For non-equilibrium two-point functions this is the case away from the regime $t_2-t_1<x_2-x_1\leq t_2+t_1$}. This motivates us to investigate whether an efficient representation of the generalised influence matrices is possible. In particular, here we assess whether these objects admit an efficient MPS representation by computing their entanglement. This is the fundamental question to which the rest of this paper is devoted.}
	
	The entanglement of the influence matrices is computed in three steps: 
	
	\noindent (i) We define reduced density matrices corresponding to an arbitrary non-disjoint bipartition $A\bar A$ of the lattice along the path $\gamma$  
	\begin{align}
		&\rho_{H,\gamma,A}=\tr_{\bar A} \frac{\ketbra{H_{\gamma,x}}{H_{\gamma,x}}}{\|\!\ket{H_{\gamma,x}}\|^2}, &
		& H=L,R\,.
		\label{eq:defrhogamma}
	\end{align}
	(ii) We compute their R\'enyi entropies 
	\begin{equation}
		S^{(\alpha)}_{H,A}(\gamma) := S^{(\alpha)}({\rho_{H,\gamma,A}}),\qquad H=L,R\,,\quad \alpha\in\mathbb R\,,
		\label{eq:maxRenyi}
	\end{equation}
	where we introduced the function 
	\begin{equation}
		S^{(\alpha)}(\rho):=\frac{1}{1-\alpha} \log {\rm tr}[{\rho^{\alpha}}]\,.
		\label{eq:Sfun}
	\end{equation}
	(iii) We maximise them over all possible bipartitions $A\bar A$ where $A$ is a contiguous region.

	Before proceeding we note that 
	\begin{equation}
		\rho_{L,\gamma,A}(W) = \rho_{R,\bar\gamma,A}(W'),
		\label{eq:rhoLrhoR}
	\end{equation}
	where we highlighted the dependence on the double gate \eqref{eq:doublegate}, introduced 
	\begin{equation}
		W' = \fineq[-0.8ex][.85][1]{
			\tsfmatV[0][0][r][1][][][bertinigreen][bottomleft]
		},
	\end{equation}
	and denoted by $\bar \gamma=\{-\gamma_1, \ldots, -\gamma_{2t_2}\}$ the mirror image of the path $\gamma$ with respect to the vertical line passing through $b$. 
	
	In the following we will use this relation to focus only on the entanglement properties of one of $\bra{L_{\gamma,x}}$ and $\ket{R_{\gamma,x}}$: the ones of the other are easily inferred from \eqref{eq:rhoLrhoR} upon replacing $W$ with $W'$. Therefore, from now on we will only look at the entanglement of $\bra{L_{\gamma,x}}$, and, to lighten the notation, we set 
	\begin{equation}
		\rho_{L,\gamma,A} \mapsto \rho_{\gamma,A}, \qquad  S^{(\alpha)}_{L,A}(\gamma)\mapsto  S_A^{(\alpha)}(\gamma)\,.
		\label{eq:rhogamma}
	\end{equation}
	Moreover, we also drop the dependence of $\bra{L_{\gamma,x}}$ on the point $x$ at which it is computed, i.e., 
	\begin{equation}
		\bra{L_{\gamma,x}} \mapsto \bra{L_{\gamma}}\,.
	\end{equation}

	\section{Temporal Entanglement in Generic Unitary Circuits}
	\label{sec:TEinRU}

	{ In this section we specify the unitary gates in Eq.~\eqref{eq:Uloc} to be (independent) Haar random matrices. We consider the temporal entanglement of the state in \eqref{eq:Lgamma} for a typical realisation of the disorder and in the long time limit. In our analysis we focus on initial states in product form, i.e.\ we take $m$ as in Eq.~\eqref{eq:mprod}. Indeed, we expect that the choice of the initial state, as long as it is short-range entangled, does not affect the general scaling of entanglement in a random circuit.} 
	
		The use of the Haar random unitaries follows from the philosophy of random matrix theory. By dispensing with all system-specific details, these strongly chaotic gates allow for analytic calculations while retaining the universal properties of entanglement in strongly interacting systems. Recently, there have been various applications of random unitary circuits to explain aspects of quantum chaos and other non-equilibrium features of generic quantum systems, see for instance Refs.~\cite{vonKeyserlingk2018operator,znidaric_exact_2008,harrow_random_2009,emerson_pseudo-random_2003,skinner2019measurement,nahum2017quantum,li2019measurement,chan2018spectral,chan_unitary-projective_2019}  and the review~\cite{fisher_random_2022} for a more comprehensive list of references. 
	
	The (R\'enyi) entanglement in a random unitary circuit is described by a statistical mechanical model written in terms of permutation degrees of freedom~\cite{zhou2019emergent,nahum_quantum_2017,jonay2018coarsegrained}. The von Neumman entropy is at the replica limit of the model. The permutations originate from pairings of the unitary evolution with its time reversal. To be more specific, let us consider the example of the $n$-th R\'enyi entropy, with $\mathbb N\ni n\geq 2$, of $\rho_A(t)$: the regular density matrix reduced to a subregion $A$ (cf.~\eqref{eq:reducedAdiag}). In each copy of the time evolved reduced density matrix $\rho_A(t)$, there is one forward and one backward time sheet (cf. Eq.~\eqref{eq:DMfolded}). Therefore, in total, there are $n$ forward and $n$ backward time sheets. When performing random averaging over the gates, each copy of a given gate and its hermitian conjugate are paired in a fashion similar to the Wick theorem of the free fields. The boundary conditions for the R\'enyi entropies are domain walls between different types of pairings. If we view the pairings as spin degrees of freedom, the effective statistical mechanical model describing the entanglement is in the ordered phase, and the domain wall continues to exist in the bulk, possibly splitting in a cascade of more elementary domain walls. The (R\'enyi) entanglement entropy is given by the free energy of these generically interacting domain walls. All these microscopic details can be encoded in a coarse grained line tension of the domain wall, which gives rise to the growth rate of entanglement in the long time limit. 
	
	{ 
		In the upcoming subsections we obtain the general scaling of temporal entanglement in three steps 
		\begin{itemize}
			\item[1.] We show that the boundary conditions to evaluate the purity of $\bra{L_{\gamma}} $ correspond to domain walls in the statistical problem (Sec.~\ref{subsec:bd_t_ent}).
			\item[2.] Averaging over the random unitary gates, we show that the minimal-energy configurations are those where the domains penetrate in the bulk. Minimising the free energies by means of the line tension formalism we find a linear growth of temporal entanglement (Sec.~\ref{subsec:ev_analysis}). 
			\item[3.] Recalling the arguments of Ref.~\cite{zhou2020entanglement} we infer that the domain wall picture can be applied also to a single realisation of random circuit (without averaging) or, equivalently, to systems without randomness (Sec.~\ref{sec:typicalcircuit}). 
		\end{itemize}
		A technical note: in the upcoming calculations we consider the state $\bra{L_{\gamma}}$, with the (anti-)slope $v \ge 0$, see Fig.~\ref{fig:dw_config}. Indeed, the $v \ge 0$ condition gives rise to non-trivial domain wall configurations. In the case $v < 0$ our analysis can be applied to $\ket{R_{\gamma}}$.}

	\subsection{The boundary conditions for temporal entanglement}
	\label{subsec:bd_t_ent}
	
	The expression of the $n$-th R\'enyi entropy contains $n$-copies of the forward and backward evolution by the circuit. The pairings emerge naturally on the boundary when contracting copies of these circuits to evaluate R\'enyi entropies, with or without random averaging.
	
	\begin{figure}[h]
		\centering
		\includegraphics[width=\columnwidth]{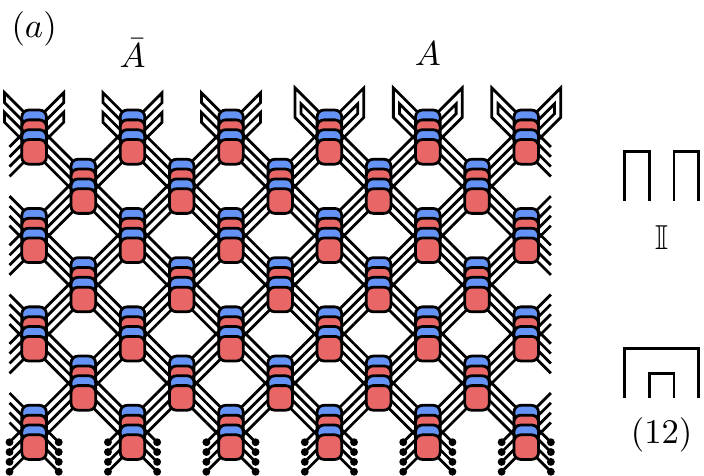}
		\includegraphics[width=\columnwidth]{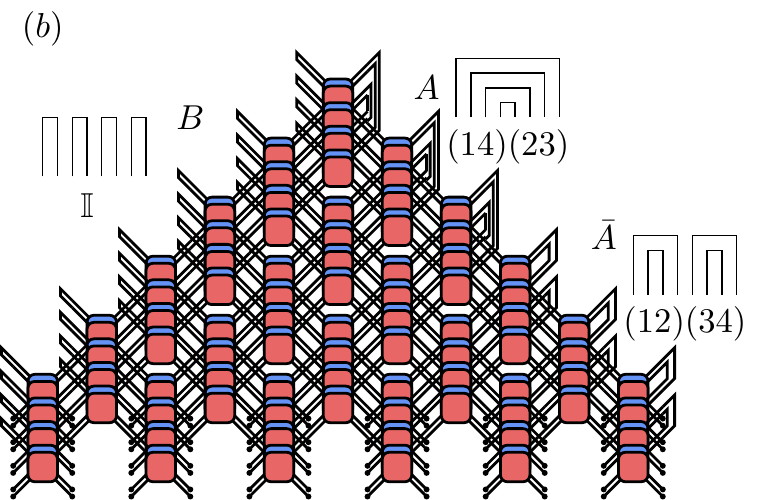}
		\caption{Permutation boundary conditions for (a) the state purity $\tr(\rho_{A}^2(t))$ and (b) the operator purity $\tr_{A}( \tr_{\bar{A}}( \ketbra{L_{\gamma}} )^2)$. The dots at the bottom represent a generic product initial state. Red gates are forward evolution $u$; blue gates are backward evolution $u^*$. (a) The top boundary conditions implements the partial trace, the matrix multiplication of $\rho_{A}$ with itself and the trace in $\bar A$. (Right) Region $A$ has boundary condition $\I$, region $\bar{A}$ has boundary condition $(12)$. (b) The top boundary conditions are permutation elements in $S_4$. They are respectively given by $\I$, $(12)(34)$ and $(14)(23)$. We consider a general contiguous partition. The ratio of size $|A|$ and the total size $|A|+ |\bar{A}|$ is set to be $r \in [0,1]$.}
		\label{fig:bd_cond}
	\end{figure}
	
	Let us illustrate this idea in the example of the purity $\tr(\rho_{A}^2(t))$ of the quantum state $\rho_{A}(t)$, see Fig.~\ref{fig:bd_cond}(a). To form the reduced density matrix $\rho_{A}(t)$, we take partial trace in each copy of $\rho(t)$. The partial trace operation is denoted as a contraction of the corresponding indices from the forward and backward copies of the circuit. Multiplying two copies of $\rho_A(t)$ and taking the trace, we obtain the swap contraction in region $A$. In this quantity, there are two copies of the unitary gate $U$ and two copies of $U^*$. There are two ways to contract them, which we denote as $\I$ and $(12)$ permutations
	\begin{align}
		\label{eq:pairing}
		\I\!:\,\, &\contraction[1.5ex]{}{U}{\otimes}{U^*}
		\contraction[1.5ex]{U \otimes U^*\otimes}{U}{\otimes}{U^*}
		U \otimes U^* \otimes U \otimes U^*,
		&
		(12)\!:\,\,&\contraction[2ex]{}{U}{\otimes U^* \otimes U}{\otimes U^*}
		\contraction{U \otimes }{U^*}{\otimes}{U}
		U \otimes U^* \otimes U \otimes U^*.
	\end{align}
	The top boundary thus has a domain wall boundary condition between $\I$ and $(12)$ permutations. 
	
	Temporal entanglement is defined for an ``operator state'', namely a state in the folded space. Therefore, the ket itself involves a forward and a backward evolution: see, e.g., Eq.~\eqref{eq:Lgamma}. The permutation boundary conditions are the same if we were to consider an operator state on a spatial slice, which have been computed explicitly in Ref.~\cite{wang2019barrier}. For completeness, we repeat the derivation for the second R\'enyi entropy for the operator state $\bra{L_{\gamma}}$. The boundary conditions involve permutations in the symmetric group $S_4$. The purity of a sub-region on the temporal slice is
	\begin{equation}
		\frac{\tr_{A}  ( \tr_{\bar A}( \ketbra{L_{\gamma}} )^2) }{ \braket{L_{\gamma}}^2 }.
	\end{equation}
	Here we choose the initial product state in the diagrams to be normalised to $1$, which is different from the $m$ state in \eqref{eq:Uloc}. To be consistent with the random circuit literature, we choose to normalise the boundary condition as shown in Fig.~\ref{fig:bd_cond} for upward pointing legs. For downward pointing legs, we use permutations normalised as the loop state as in Eq.~\eqref{eq:foldedwire}. The temporal R\'enyi-2 entropy thus has two terms
	\begin{equation}
		\label{eq:S_op_rho_A}
		\!\!\!\!S^{(2)}_{A}(\gamma) =\!\! - \log(\tr_{A}  ( \tr_{\bar A}( \ketbra{L_{\gamma}} )^2))\!+\! 2 \log (\braket{L_{\gamma}})
	\end{equation}
	where the second term is twice the R\'enyi-2 entropy of $A\bar{A}$. The boundary conditions for the first term is shown in Fig.~\ref{fig:bd_cond}(b).

	\subsection{Entanglement in terms of domain wall line tension: disorder average}
	\label{subsec:ev_analysis}
	
	In Fig.~\ref{fig:bd_cond}, we see that pairings (permutations) emerge naturally as boundary conditions when evaluating entanglement-related quantities. In fact, in quantum chaotic systems these pairings are also the dominant degrees of freedom in the interior of the multi-layer unitary evolution. One simple way to introduce pairings in the bulk is through random averaging of the gate over Haar ensemble. Indeed, the latter are the only degrees of freedom surviving the average. 
	
	Taking the purity diagram in Fig.~\ref{fig:bd_cond}(a) as an example, each four-layer gate after random averaging can only give a tensor of $\I$ or $(12)$ as the output at its bottom legs. Hence we can label the gate with ``spin'' variables taking values in $\I$ or $(12)$.  The $\I$ and $(12)$ can form contiguous domains connecting to the $\I$ and $(12)$ on the boundary.  We label a general domain wall between a pairing $\sigma$ on the left and a pairing $\mu$ on the right as $\sigma^{-1} \mu$. The domain wall between $\I$ and $(12)$ is thus $\I^{-1} (12) = (12)$. Due to constraints from unitarity and locality of the interactions, this $(12)$ domain wall can only wander within the light cone and can not branch. The entanglement is the free energy, or tension of the domain wall. Since disorder fluctuations are negligible over large enough scales, the system is asymptotically translationally invariant and the domain wall macroscopically should be a straight line. Using $v$ to denote the inverse of the domain wall's slope, we can write the R\'enyi entropy at leading order as
	\begin{equation}
		\label{eq:state_purity_min}
		-\log \overline{\tr[\rho_{A}^2(t)]} \simeq s_{\rm eq}  \min_{v} \mathcal{E}_H(v) t, 
	\end{equation}
	where $\overline{\cdots}$ denotes the average over Haar random gates, { $\mathcal{E}_H(v)$ is the line tension of a domain wall in the Haar random circuit~\footnote{Here we use $\mathcal{E}_H$ to denote the line tension for the average purity decay (annealed average). The expression is exact. The line tension for the average second R\'enyi entanglement (quenched average) has $\mathcal{O}(d^{-4})$ correction on top of this, see Ref.~\cite{zhou2019emergent}.} 
		\begin{equation}
			\mathcal{E}_H(v) = \frac{\log \frac{d^2+1}{d}+ \frac{1+v}{2} \log \frac{1+v}{2} + \frac{1-v}{2} \log \frac{1-v}{2} }{\log d},
			\label{eq:linetensionHaar}
		\end{equation}
		and $s_{\rm eq}$ is the infinite temperature equilibrium entropy $\log d$ (we recall that $d$ is the local Hilbert space dimension). For a generic product initial state, the domain wall end point at the bottom is not fixed and the R\'enyi entropy is obtained by minimising over different slopes.}
	
\begin{figure}[h]
		\centering
		\includegraphics[width=\columnwidth]{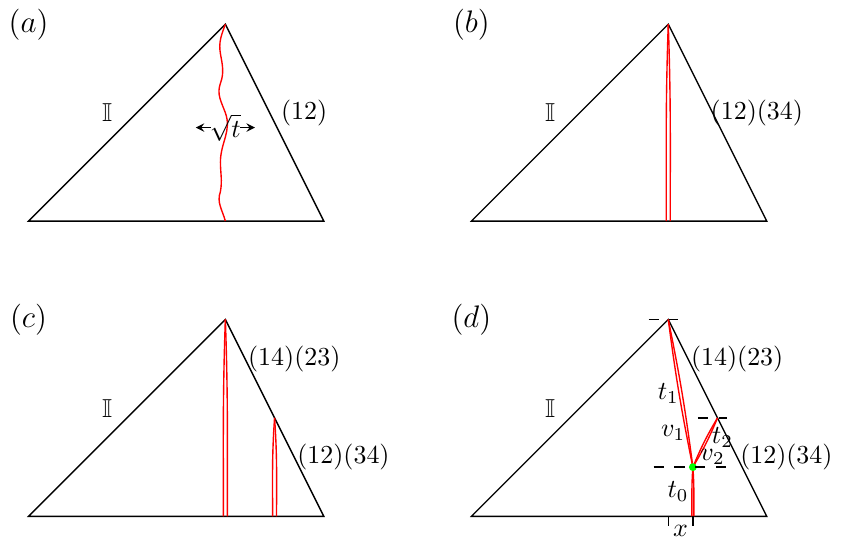}
		\caption{Domain wall configurations for purity-like diagrams. (a) Domain wall for purity $\tr(\rho_A^2)$. The single domain wall (a transposition $(12)$) undergoes a random walk down to the bottom boundary. The size fluctuation is $\sqrt{t}$, which is subleading to $t$ and ignored in other sub-figures. (b) Domain walls for $(\tr\rho^2_A)^2 $. The two transpositions are independent. (c) Configurations when the two sets of domain wall dot no interact. (d) The two sets of domain wall meet in a middle time slice. The three segments have time duration $t_1$, $t_2$ and $t_0$, and anti-slopes $v_1$, $v_2$ and $0$ receptively. The anti-slope of the right edge of the triangle is $v$. The green vertex represents the fusion. It can have a non-trivial weight, but only affects the free energy by an $\mathcal{O}(1)$ amount. }
		\label{fig:dw_config}
	\end{figure}

	Since the random circuit is left-right symmetric after disorder averaging, the minimum in \eqref{eq:state_purity_min} is taken at $v = 0$, i.e. a vertical line, see Fig.~\ref{fig:dw_config}(a). This gives a linear growth where the line tension is the entanglement growth rate, which is called entanglement velocity~\cite{kim2013ballistic}. 
	
	Let us now consider the entanglement of the operator state $\bra{L_\gamma}$. In particular, we consider the following averaged version of it 
	\begin{equation}
		\!\!\!\bar S^{(2)}_{A}(\gamma) : =\!\! - \log{\overline{\tr_{A}  ( \tr_{\bar A}( \ketbra{L_{\gamma}} )^2)}}\!+\!\log\overline{(\braket{L_{\gamma}})^2}\!.
		\label{eq:S_op_rho_Aave}
	\end{equation}
	As noted above, this quantity involves permutations in $S_4$. In this case the leading contribution is again given by suitable domain walls, however, differently from before these domain walls will involve multiple elementary transpositions. 
	
	For instance, let us consider the second term on the right hand side of Eq.~\eqref{eq:S_op_rho_Aave}  
	\begin{equation}
		F_2(t) :=- \log\overline{(\braket{L_{\gamma}}^2)}.
	\end{equation}
	The boundary condition is $\I$ on subsystem $B$ and $(12)(34)$ on subsystem $A\bar{A}$ (cf. boundary conditions of the first term in Fig.~\ref{fig:bd_cond}(b)). Hence it has two commutative domain walls at the intersection of $B$ and $A\bar{A}$. These two domain walls are independent, each of them go vertically down to the bottom (Fig.~\ref{fig:dw_config}(b)) and give a contribution of $\mathcal{E}_H( 0 ) t$. And thus the second term corresponds to  twice the state R\'enyi entropy
	\begin{equation}
		\label{eq:F_2_t}
		F_{2}(t) \simeq 2 s_{\rm eq} \mathcal{E}_H(0) t.
	\end{equation}

	The boundary conditions of the first term in Eq.~\eqref{eq:S_op_rho_A} contains four domain walls. There are two commutative transpositions $(14)(23)$ at the tip of the diagram (cut between $B$ and $A\bar{A}$), and other two transpositions $(13)(24)$ at the entanglement cut between $A$ and $\bar{A}$. The two transpositions $(14)(23)$ {\it alone} are independent, so are $(13)(24)$. If the two sets do not meet at an intermediate time slice before reaching the bottom boundary, we can separately minimise their free energies. The equilibrium configuration is that all the domain walls go down vertically (Fig.~\ref{fig:dw_config}(c)), giving total free energy 
	\begin{align}
		\label{eq:F1t_naive}
		F_{1,\text{vert}}(t)&:= - \log \overline{(\tr_{A}  ( \tr_{\bar A}( \ketbra{L_{\gamma}} )^2))}\notag\\
		&\simeq 2 s_{\rm eq} \mathcal{E}_H(0) t + 2 s_{\rm eq} \mathcal{E}_H(0) (1-r) t,
	\end{align}
	where $r \in [0, 1]$ is the ratio of size $|A|$ and total size $|A| + |\bar{A}|$. 
	The difference with $F_2(t)$ in Eq.~\eqref{eq:F_2_t} is then $\mathcal{E}_H(0) t$: This gives us the following upper bound
	\begin{equation}
		\label{eq:TE_upper_bound_1}
		{\bar S_A^{(2)}(\gamma)}\lesssim  2(1-r) \mathcal{E}(0) t.
	\end{equation}
	
	If the two sets meet in the middle, the domain walls can fuse to different permutations according to the 
	group multiplication rules. For example, if $(13)(24)$ and $(14)(23)$ completely fuse together, the domain wall becomes $(13)(24) \times (14)(23) = (12)(34)$. We see that the number of domain walls reduces to two, thus reducing the energy cost. For this reason a configuration like the one displayed in Fig.~\ref{fig:dw_config}(d) can compete for the minimal free energy. 
	
	We now set up the minimisation problem assuming the two sets to have (anti-)slope $v_1$ and $v_2$ before they meet in the middle. After the fusion, the resulting domain wall $(12)(34)$ is composed by two independent components. They cost free energy $2\mathcal{E}_H(0) t_0$ for the remaining duration of $t_0$. The vertex can have a non-trivial weight but, as long as it is not zero, it only brings in a $\mathcal{O}(1)$ correction and can be neglected when considering the leading order free energy in the long time limit. The geometry is depicted in Fig.~\ref{fig:dw_config}(d). Therefore, we can write the free energy as 
	\begin{equation}
		\label{eq:F1_t_t0}
		\!\!\!F_{1,Y}(t, t_0) \simeq 2 s_{\rm eq} \left( \mathcal{E}_H( v_1  ) t_1  + \mathcal{E}_H( v_2 ) t_2 + \mathcal{E}_H(0 ) t_0\right)\!, 
	\end{equation}
	where $t_0$, $t_1$ and $t_2$ are the duration of the two sets of domain walls and satisfy the geometric relations. The subscript $Y$ denotes the merging configuration.
	\begin{equation}
		\begin{aligned}
			v_1 &= \frac{x}{t_1}, \quad v_2 = \frac{x - r v_{\gamma} t}{t_2}.
		\end{aligned}
	\end{equation}
	To parameterize time, we set
	\begin{equation}
		\label{eq:def_r_r0}
		t_0 = r_0 t \qquad t_1 - t_2 = rt \quad t_1 + t_0 = t
	\end{equation}
	where $r \in [0,1]$ is the ratio of size $|A|$ with respect to $|A| + |\bar{A}|$, and $r_0 \in [0,1-r]$ is the portion of the merged domain wall. $r_0 = 0$ corresponds to merge at the bottom; $r_0  = 1-r$ corresponds to taking $v_1 = v_{\gamma}$, i.e. merge immediately 
	when available. 
	To minimize $F_1(t, t_0)$, we first fix $t_0$ and vary $x$. The implicit $x$ derivative gives
	\begin{equation}
		\mathcal{E}_H' \left( \frac{x}{ t_1}\right) + \mathcal{E}_H'\left( x-\frac{v_{\gamma} r t}{ t_2 }\right)
		= \mathcal{E}_H'(v_1 ) + \mathcal{E}_H'(v_2 ) = 0.
	\end{equation}
	For a reflection symmetric system the physical solution is 
	\begin{equation}
		v_1 = -v_2,
	\end{equation}
	namely the two sets of domain walls in Fig.~\ref{fig:dw_config}(d) meet symmetrically from left and right towards each other. 
	
	{ The minimisation with respect to $t_0$ depends on the explicit form of the line tension function. In particular, an explicit calculation for the case of the random circuit line tension $\mathcal{E}_H$ is carried out Appendix~\ref{app:f1min_haar}. The resulting expression of $\min_{t_0} F_{1,Y}(t,t_0)$ is piece-wise continuous in $v$ and depends on $d$. Nevertheless it is still linear in $t$ (cf.\ Eq.~\eqref{eq:F1_F2_min}). Combining with the linear bound in Eq.~\eqref{eq:TE_upper_bound_1}, we conclude that in a Haar random circuit
		\begin{equation}
			\bar S^{(2)}_{A}(\gamma) \simeq  s_{\rm eq} v_{\rm TE, H}^{(2)} t.
		\end{equation}
		where the temporal R\'enyi entanglement velocity ${v_{\rm TE, H}^{(2)} > 0}$. The explicit expressions of $ v_{\rm TE, H}^{(2)}$ for Haar random circuit can be found in Eq.~\eqref{eq:vTEH}}

	\subsection{Typical circuit without averaging}
	\label{sec:typicalcircuit}
	
	In this subsection, we argue that the line tension formalism discussed above can be applied to the calculation of temporal entanglement in generic chaotic circuits without introducing disorder averaging. The recipe is to replace ${\mathcal{E}}_H(v)$ with a ``dressed'' line tension function ${\mathcal{E}}(v)$ characterised by the following general properties~\cite{jonay2018coarsegrained}
	\begin{equation}
		{\mathcal{E}}(v)\geq |v|,\qquad {\mathcal{E}}''(v)\geq 0,\qquad {\mathcal{E}}(v)= {\mathcal{E}}(-v),
		\label{eq:genpropEps}
	\end{equation}
	where the last one follows from the parity symmetry of the system.

	The basic arguments follow Ref.~\cite{zhou2020entanglement}, where the concept of line tension function is generalised to non-random circuits and we will briefly recall them here. The key observation is that the pairings between a unitary gate and its hermitian conjugate continues to dominate the configuration sum, or ``path integral'' that determines the R\'enyi entropies. Indeed, these are real positive quantities that do not suffer from phase cancellation. A single domain wall, such as $(12)$, will be dressed by non-pairing degrees of freedom, but only perturbatively to have an $\mathcal{O}(1)$ width. 
	
	Our problem is slightly more complicated than the one discussed in Ref.~\cite{zhou2020entanglement}  because the contributions generating temporal entanglement involve composite domain walls (for instance a domain wall $(123)$ can appear). Nevertheless, the two sets of domain walls $(14)(23)$, and $(13)(24)$ have the same dressed line tension function ${\mathcal{E}}(v)$ when they do not interact. Indeed, $(14)(23)$ can be mapped to $(12)(34)$ by relabelling the third and forth copies of the unitary and its complex conjugate --- it is a symmetry of the multi-replica dynamics if we look at the patch of $(14)(23)$ alone. The symmetry no longer holds when the two sets of domain walls meet each other and interact. This process, however, only dresses the interaction vertex in Fig.~\ref{fig:dw_config}(d), which introduces an order $\mathcal{O}(1)$ correction to the free energy. Below the interaction vertex, the two sets of domain walls fuse to $(12)(34)$, which again has the same line tension function ${\mathcal{E}}(v)$. This justifies the use of a single line tension function, ${\mathcal{E}}(v)$, to characterise the scaling of the temporal R\'enyi-2 in Eq.~\eqref{eq:S_op_rho_A}. 
	
	Specifically, following the steps discussed in the previous section we obtain 
	\begin{equation}
		S^{(2)}_A(\gamma)  \simeq s_{\rm eq}  v_{\rm TE}^{(2)} t, 
	\end{equation}
	where the temporal R\'enyi entanglement velocity is determined by
	\begin{equation}
		\label{eq:v_TE_full}
		v_{\rm TE}^{(2)} = \min( \min_{r_0\in[0,r]} {\mathcal F}(r_0), 2(1-r) \mathcal{E}(0) ). 
	\end{equation}
	The parameters $r_0$ and $r$ are defined in Eq.~\eqref{eq:def_r_r0} and 
	\begin{align}
		\!\!\!\!{\mathcal F}(r_0) \!=\! &2 [\mathcal{E}({v_1(r_0)}) \!-\!  \mathcal{E}( 0 ) ](1\!-\!r_0)  \label{eq:gfun1} \\
		&+ 2\mathcal{E}( v_1(r_0)) (1\!-\!r_0-r).  \label{eq:gfun2}
	\end{align}
	In this case, the remaining minimisation over $r_0$ cannot be performed explicitly as the minimum depends on the precise form of the line tension. However, we can show that for $\mathcal{E} (0)>0$ 
	\begin{equation}
		\min_{r_0\in[0,r]} {\mathcal F}(r_0)= 0 \quad \Leftrightarrow \quad \mathcal{E} ( v_\gamma )  = \mathcal{E} ( 0 ), 
		\label{eq:maincondition}
	\end{equation}
	which implies generic linear growth of the temporal R\'enyi-2 entropy apart from marginal cases. 
	
	We begin to prove this property by noting that the two terms in \eqref{eq:gfun1} and \eqref{eq:gfun2} are both non-negative
	\begin{equation}
		\begin{aligned}
			[\mathcal{E}({v_1(r_0)}) \!-\!  \mathcal{E}( 0 )] (1\!-\!r_0) &\ge 0, \\
			\mathcal{E}( v_1(r_0)) (1\!-\!r_0 - r)  &\ge 0,
		\end{aligned}
		\label{eq:terms}
	\end{equation}
	in the relevant range $r_0 \in [0, r]$, because of the convexity and parity of line tension function. Indeed, these two properties imply that the function is either constant or has a unique local minimum in $v=0$, i.e. 
	\be
	\mathcal{E}( v)\geq \mathcal{E}(0)>0, \quad \forall v\,.
	\ee
	In fact, the above inequality indicates that \eqref{eq:terms} can both be zero for generic $r$ only if $r_0=r$ and

	\begin{equation}
		\label{eq:ev_e0}
		\mathcal{E} ( v_\gamma )  = \mathcal{E} ( 0 ).
	\end{equation}
	Noting that the reversed implication is obvious, we conclude the proof. 
	
	In fact, due to the general properties~\eqref{eq:genpropEps} of the line-tension function, Eq.~\eqref{eq:ev_e0} admits solution only in two cases: either constant line tension function (\eqref{eq:margin_case_1}) or $v_\gamma=0$ (\eqref{eq:margin_case_2}). These are the two marginal cases anticipated in Sec.~\ref{sec:mainresults}.

	\section{Temporal Entanglement In Chaotic Dual-Unitary Circuits}
	
	\label{sec:TEinDU}
	
	{ In this section we consider the first of the two marginal cases identified in Eq~\eqref{eq:ev_e0}: the one in which the line tension function is constant. As discussed in Sec.~\ref{sec:mainresults} (cf.\ the discussion around Eq.~\ref{eq:margin_case_1}), this situation can only be realised when the local gates forming the time evolution operator in Eq.~\eqref{eq:U1U2} are dual unitary~\cite{bertini2019exact}.} In terms of our diagrammatic representation the dual-unitarity condition means that the gates fulfil 
	\be
	\fineq[-0.8ex][.85][1]{
		\tsfmatV[0][0][r][1][][][bertinired][topright]
		\tsfmatV[1.25][0][r][1][][][bertiniblue][bottomleft]
		\draw[ thick] (.5,1.5) -- (.75,1.5);
		\draw[ thick] (.5,0.5) -- (.75,0.5);
	}
	= 
	\fineq[-0.8ex][.85][1]{
		\tsfmatV[0][0][r][1][][][bertiniblue][bottomleft]
		\tsfmatV[1.25][0][r][1][][][bertinired][topright]
		\draw[ thick] (.5,1.5) -- (.75,1.5);
		\draw[ thick] (.5,0.5) -- (.75,0.5);
	}
	=
	\fineq[-0.8ex][.85][1]{
		\draw[ thick] (0,1.5) -- (1,1.5);
		\draw[ thick] (0,0.5) -- (1,0.5);
	}\,,
	\label{eq:dualunitaritynonfolded}
\end{equation}
in addition to the standard unitarity conditions \eqref{eq:unitaritynonfolded}. Without additional fine tuning, the gates fulfilling \eqref{eq:unitaritynonfolded} and \eqref{eq:dualunitaritynonfolded} are quantum chaotic~\cite{bertini2019exact, bertini2020operator, bertini2021random}. 

Imposing the condition~\eqref{eq:dualunitaritynonfolded} enables one to make a number of exact statements concerning dynamics and spectral properties of the quantum circuit~\cite{bertini2019entanglement, gopalakrishnan2019unitary, piroli2020exact, bertini2020operator, bertini2020operator2, claeys2020maximum, jonay2021triunitary, chan2018solution, claeys2020maximum, zhou2022maximal, reid2021entanglement, bertini2018exact, bertini2021random, fritzsch2021eigenstate, claeys2021ergodic, suzuki2022computational, giudice2022temporal}. In particular, dual-unitary circuits have been shown to admit a class of ``solvable'' initial states for which one can compute exactly the full time evolution of any local subsystem~\cite{piroli2020exact}. For solvable initial states the generalised influence matrices $\bra{L_{\gamma}}$ and $\ket{R_{\gamma}}$ (cf.~\eqref{eq:Rgamma} and \eqref{eq:Lgamma}) take the following product form
\begin{equation}
	\bra{L_{\gamma}}=\bra{\mcirc}^{\otimes |\gamma|},\qquad \ket{R_{\gamma}}=\ket{\mcirc}^{\otimes |\gamma|},
\end{equation}
where $\ket{\mcirc}$ is the loop state of Eq.~\eqref{eq:foldedwire}. This form immediately implies a strictly vanishing temporal entanglement.

Here, however, we are interested in the behaviour of temporal entanglement for \emph{generic}, non-solvable, initial states. Specifically --- recalling that for the family of initial states \eqref{eq:initialstate} considered here the solvable instances correspond to the cases where the matrix $m$ is unitary~\cite{foligno2022growth} --- we consider the case where $m$ is \emph{not} unitary. 

Since the time evolution in a chaotic system should not depend on the initial configuration, one might expect that the behaviour of solvable states is somewhat representative of the generic situation. Namely, that the temporal entanglement is aways small for dual-unitary circuits. In fact, as we now discuss, this intuition turns out to be incorrect: Even though higher temporal R\'enyi entropies are bounded by a sub-linear function of time (in agreement with our entanglement-membrane analysis of the previous section) the von Neumann temporal entanglement entropy is always linear for non-fine-tuned dual unitary circuits. In the upcoming subsections we show these facts by analysing separately the cases of higher R\'enyi entropies and von Neumann entropy.   

{ For simplicity, in the main text we consider paths ${\gamma=A\bar A}$ with \emph{constant} slope $v_\gamma$. This means that the slope is the same in both $A$ and $\bar A$. This assumption is lifted in Appendices~\ref{app:secIVAcalc} and~\ref{app:secIVBcalc} where we present the most general form of our results.}

\subsection{Bound on temporal higher R\'enyi entropies}
\label{sec:boundRenyis}

{ 
	In this subsection we show that higher temporal R\'enyi entropies are \emph{sub-linear} in time, in agreement with the entanglement membrane analysis of the previous section. More precisely, we prove the following bound   
	\begin{align}
		S_A^{(\alpha>1)}(\gamma) \leq	  \frac{\alpha}{\alpha-1} \log\left[\frac{d^{\tA}\mathcal P(\ttot/2)}{\mathcal P(\tAb/2)}\right], 
		\label{eq:DUleftminent}
	\end{align}
	where $A$, $\bar{A}$ correspond to a contiguous bipartition of the $2t$ legs of the influence matrix
	\begin{align}
		\bra{L_{\gamma}}=\!\!\!\!\!\!\!\fineq[-.8ex][0.5][0.5]{
			\tsfmatV[-4][5][r][1][][][bertiniorange][topright]
			\tsfmatV[-3][4][r][1][][][bertiniorange][topright]
			\tsfmatV[-2][3][r][1][][][bertiniorange][topright]
			\tsfmatV[-1][2][r][1][][][bertiniorange][topright]
			\tsfmatV[0][1][r][1][][][bertiniorange][topright]
			\tsfmatV[-1][0][r][1][][init][bertiniorange][topright]
			\eigenVL[-5][0][l][5]
			\tsfmatV[-4][0][l][4][][init]
			\tsfmatV[-2][0][l][2][][init]
			\draw[thick] (-.75-2-.9+0.1,7.5+1-.9-0.1) -- (-.35-2-.9,7.15+1-.9);
			\tsfmatD[0][4][l][3][][no]
			\draw[fill=white] (-4.5,6.5) circle (0.15cm); 
			\draw[fill=white] (-4.5+1,6.5+1) circle (0.15cm);
			\draw [decorate, decoration = {brace}, thick,shift={(-7.,2.)}]   (8,2.75+3) -- (8,1.3);
			\draw [decorate, decoration = {brace}, thick,shift={(-7.,-2.)}]   (8,1.75+3) -- (8,1.5);
			\node[scale=2]  at (1.7,5.3) {$A$};
			\node[scale=2]  at (1.7,1.4) {$\bar A$};
		},
	\end{align}
	$\tA$ ($\tAb$) is the number of up-pointing legs in $A$ ($\bar A$) fulfilling
	\be
	\tA+\tAb=(1+v_\gamma)t\equiv \ttot\,,
	\ee
	while 
	\begin{equation}
		\mathcal P(t) = \tr[\rho_{[0,\infty)}(t)^2],
		\label{eq:purity}
	\end{equation}
	is the purity of the ``regular'' reduced density matrix $\rho_{[0,\infty)}(t)$ (cf.~\eqref{eq:DMfolded}) corresponding to a half-infinite sub-system $[0,\infty)$ with open boundary conditions.
	
	{To prove Eq.~\eqref{eq:DUleftminent} we proceed in two steps, which are detailed in Appendix~\ref{app:secIVAcalc}.\\
		
		\noindent {\bf Step 1}: General bound on higher R\'enyi entropies. We take advantage of the unitarity of the gates and of Eckart-Young Theorem~\cite{eckart1936approximation} to bound the temporal R\'enyi entropies in terms of the norm of the state $\bra{L_\gamma}$

		\begin{align}
			\max_A S_A^{(\alpha)}(\gamma) \le \frac{\alpha}{\alpha-1} \log (\frac{\braket{L_{\gamma_{\bar A}}} }{\braket{L_{\gamma}}}). 
			\label{eq:boundRenyigeneral}
		\end{align}
		\noindent {\bf Step 2}: Dual-unitary case. Specialising the treatment to the dual-unitary case we can relate $\braket{L_{\gamma}}$ to the spatial purity (cf.~\eqref{eq:purity}). In particular we find 
		\be
		\braket{L_{\gamma}}= d^{\tau} \mathcal P(\tau/2).
		\ee
		Plugging it into \eqref{eq:boundRenyigeneral} we obtain \eqref{eq:DUleftminent}.}

	The physical interpretation of \eqref{eq:DUleftminent} is immediate: in dual-unitary circuits the growth of higher temporal R\'enyi entropies is controlled by that of spatial purity. If the initial state is solvable, then the purity is minimised to $d^{-2 t}$ and the temporal entanglement is zero. For more general, non-solvable states the purity is no longer strictly $d^{-2 t}$, but --- since dual-unitary circuits maintain a maximal entanglement velocity~\cite{foligno2022growth} --- it can only acquire sub-exponential corrections. This implies that all higher temporal R\'enyis are \emph{sub-linear} in time. 
	
	To make further progress we introduce the following assumption
	\begin{asp}
		\label{asp:1}
		For any generic dual-unitary circuit evolving from a non solvable state we have 
		\be
		\mathcal P(t) \simeq \frac{C t}{d^{2t}},
		\label{eq:scalingass}
		\ee
		where $\simeq$ denotes the leading order in the asymptotic expansion for large times and $C>0$ a time independent constant. 
	\end{asp}
	The scaling in Eq.~\eqref{eq:scalingass} can be proven by averaging each dual unitary gate of the circuit over random single qubit rotations of its legs (see Ref.~\cite{foligno2022growth} and Appendix \ref{app:Masymptotics}). Thus, we expect it to hold for typical dual unitary circuits: This is in agreement with our numerical investigations, as shown in Fig.~\ref{plot:correction_maximal_growth} for some representative examples. 
	
	As discussed in Appendix~\ref{app:Masymptotics} we expect the case $C=0$ to hold only for solvable initial states, for which $\mathcal P(t) = {d^{-2t}}$ and the temporal entanglement is identically 0 for any bipartition \cite{piroli2020exact}. The discussion in Appendix~\ref{app:Masymptotics}, however, also shows that these states are ``unstable'' from the point of view of purity scaling: for any arbitrary small perturbation of a solvable state one has Eq.~\eqref{eq:scalingass} with ${C>0}$ and the behaviour of temporal entanglement is the one we discuss here.
	
	\begin{figure}
		\ifdef{0}{}{	
			\begin{tikzpicture}[scale=1,remember picture]
				\begin{axis}[grid=major,
					legend columns=2,
					legend style={at={(0.6,0.7)},anchor=south east,font=\scriptsize,draw= none, fill=none},
					xtick distance=1,
					mark size=1.3pt,	
					xlabel=$t$,
					ylabel=$d^{2t}\mathcal P(t)-d^{2t-1}\mathcal P(t-1/2)$,
					y label style={at={(axis description cs:.1,.5)},anchor=south,font=\normalsize	},		
					tick label style={font=\normalsize	},	
					x label style={font=\normalsize	},	
					]
					\addplot[
					smooth,
					thick,
					mark=*,
					blue,
					dashed
					] table[x expr=(\thisrowno{0})/2, y expr=(\thisrowno{1})] {plots_data/Mt2.dat};
					\addlegendentry{p=0.27}
					\addplot[
					smooth,
					thick,
					mark=*,
					orange,
					dashed
					] table[x expr=(\thisrowno{0})/2, y expr=(\thisrowno{2})] {plots_data/Mt2.dat};
					\addlegendentry{p=0.33}	
					\addplot[
					smooth,
					thick,
					mark=*,
					green,
					dashed
					] table[x expr=(\thisrowno{0})/2, y expr=(\thisrowno{3})] {plots_data/Mt2.dat};
					\addlegendentry{p=0.40}	
					\addplot[
					smooth,
					thick,
					mark=*,
					red,
					dashed
					] table[x expr=(\thisrowno{0})/2, y expr=(\thisrowno{4})] {plots_data/Mt2.dat};
					\addlegendentry{p=0.47}	
					\addplot[
					smooth,
					thick,
					mark=*,
					purple,
					dashed
					] table[x expr=(\thisrowno{0})/2, y expr=(\thisrowno{5})] {plots_data/Mt2.dat};
					\addlegendentry{p=0.53}	\addplot[
					smooth,
					thick,
					mark=*,
					brown,
					dashed
					] table[x expr=(\thisrowno{0})/2, y expr=(\thisrowno{6})] {plots_data/Mt2.dat};
					\addlegendentry{p=0.60}
					\addplot[
					smooth,
					thick,
					mark=*,
					pink,
					dashed
					] table[x expr=(\thisrowno{0})/2, y expr=(\thisrowno{7})] {plots_data/Mt2.dat};
					\addlegendentry{p=0.67}
				\end{axis}	
			\end{tikzpicture}
			}
		\includegraphics[scale=1]{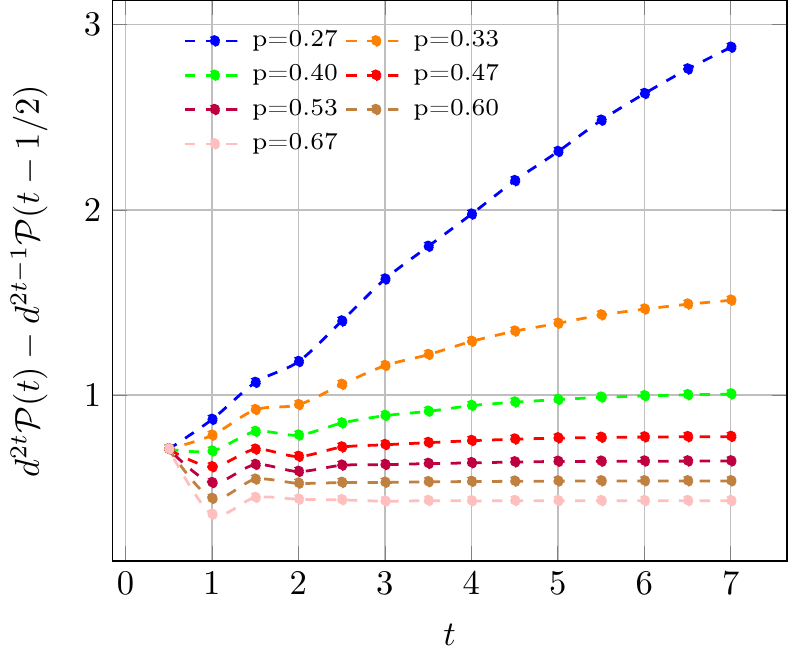}
		\caption{Evolution of $d^{2t}\mathcal P(t)-d^{2(t-1)}\mathcal P(t-1)$ as a function of $t$ for homogeneous dual-unitary circuits with different entangling power $p$ (cf. Eq.~\eqref{eq:entanglingpowerdef}). If the scaling \eqref{eq:scalingass} holds this quantity should saturate to the constant ${C/2}$. We considered $d=2$ and parameterised the gates as described in Appendix~\ref{sec:parameterisation}. Note that $C$ grows upon decreasing $p$: this is consistent with the fact that at the non-interacting point $p=0$ the purity decays with an exponent $\lambda< 2 \log d$. }
		\label{plot:correction_maximal_growth}
	\end{figure}

	Using Assumption~\ref{asp:1} and considering an appropriate scaling of the bipartition
	\begin{align}
		r\equiv \frac{\abs{A}}{2t}<1,
		\label{eq:rdef1}
	\end{align}
	it is immediate to show that the higher R\'enyi entropies saturate to a constant~\footnote{This holds for $r$ strictly smaller than 1. For $r=1-A/t^\alpha$, logarithmic violations may occur.}
	\begin{equation}
		S_A^{(\alpha)}(\gamma) \simeq \frac{\alpha}{\alpha-1}\log(1-r).
		\label{eq:epsbound}
	\end{equation}

	A direct numerical test of Eq.~\eqref{eq:epsbound} is not straightforward as we have only access to short times. Therefore, we can only consider gates for which the asymptotic form Eq.~\eqref{eq:scalingass} is attained early. With this restriction the bound appears convincingly obeyed. For instance, in Fig.~\ref{plot:higherrenyibound} we consider a comparison between Eq.~\eqref{eq:epsbound} and the four gates of Fig.~\ref{plot:correction_maximal_growth} with higher entangling power. 
	\begin{figure}\ifdef{0}{}{
		\begin{tikzpicture}[scale=1,remember picture]
			\begin{axis}[grid=major,
				legend columns=2,
				legend style={at={(0.75,0.67)},anchor=south east,font=\scriptsize,draw= none, fill=none},
				xtick distance=1,
				mark size=1.3pt,	
				xlabel=$t$,
				ylabel=$S^{(\infty)}\left(\bra{L_\gamma}\right)$,
				y label style={at={(axis description cs:.1,.5)},anchor=south,font=\normalsize	},		
				tick label style={font=\normalsize	},	
				x label style={font=\normalsize	},	
				]
				\addplot[
				thick,
				mark=*,
				blue,
				dashed
				] table[x expr=(\thisrowno{0}), y expr=(\thisrowno{2})] {plots_data/renyisaturation4.dat};
				\addlegendentry{$p=0.46$}\addplot[
				thick,
				mark=*,
				red,
				dashed
				] table[x expr=(\thisrowno{0}), y expr=(\thisrowno{3})] {plots_data/renyisaturation4.dat};
				\addlegendentry{$p=0.53$}\addplot[
				thick,
				mark=*,
				green,
				dashed
				] table[x expr=(\thisrowno{0}), y expr=(\thisrowno{4})] {plots_data/renyisaturation4.dat};
				\addlegendentry{$p=0.6$}\addplot[
				thick,
				mark=*,
				orange,
				dashed
				] table[x expr=(\thisrowno{0}), y expr=(\thisrowno{5})] {plots_data/renyisaturation4.dat};
				\addlegendentry{$p=0.66$}
				\addplot[
				thick,
				mark=*,
				black,
				dashed
				] table[x expr=(\thisrowno{0}), y expr=(\thisrowno{1})]
				{plots_data/renyisaturation4.dat};		
				\addlegendentry{Asymptotic bound}		\end{axis}	
		\end{tikzpicture}}\includegraphics{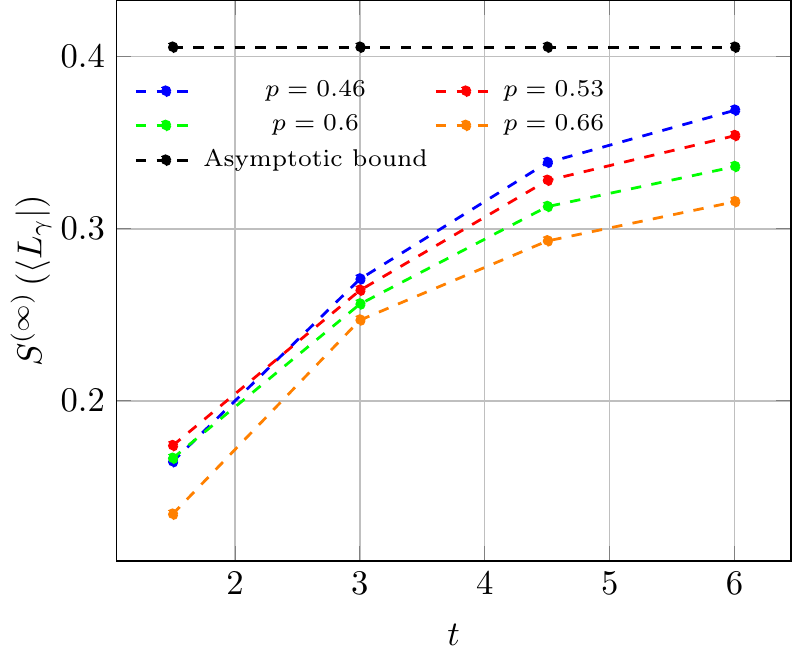}
		\caption{
			Growth of $S^{(\infty)}$ for the state $L_\gamma$, taken with $v_{\gamma}=1$, for gates of various entangling power $p$, compared with the asymptotic bound given by Eq. \eqref{eq:epsbound}, for a fixed value of $r=\frac{1}{3}$.	
		}\label{plot:higherrenyibound}
	\end{figure}
}

\subsection{Linear Growth of Temporal Entanglement Entropy}
\label{sec:VNentropy}

{ 
	In the previous Sec.~\ref{sec:boundRenyis}, we showed that the higher R\'enyi entropies are bounded by a constant for any partition with ratio $r<1$. However, since R\'enyi entropies are non-increasing functions of the R\'enyi index, this  result only provides a lower bound for the temporal entanglement entropy, i.e.,  
	\begin{equation}
		S_A(\gamma) = \lim_{\alpha\to1} S^{(\alpha)}_A(\gamma)\,.
	\end{equation}
	
	In this subsection, we show that $S_A(\gamma)$ in a typical dual unitary circuit grows linearly in time for non-solvable initial states.
	
	Denoting again by $r$ the ratio between the number of legs in region $A$ and the total one (cf. Eq.~\eqref{eq:rdef1}), we can bound $S_A(\gamma)$ from above and below 
	\begin{equation}
		\label{eq:SAgammabound}
		s(r) t \log d \le S_A(\gamma ) \le s(r) t \log d + \mathcal{O}(\log t),
	\end{equation}
	with the same function
	\begin{equation}
		\!\!\!\!\!\!\!\!s(r) = \left\lbrace
		\begin{aligned}
			&  {(1+v_\gamma)}r^2 & \quad  r \in [0, \frac{2}{v_\gamma+3}] \\
			& \frac{4(1-r)[(2+v_\gamma)r - 1]}{1+v_\gamma} & r\in (\frac{2}{v_\gamma+3}, 1].\\
		\end{aligned} \right.
		\label{eq:finalresultDU}
	\end{equation}
	The $\log(t)$ margin in Eq.~\eqref{eq:SAgammabound} is sub-leading with respect to the linear scaling of $s(r) t \log d$ and therefore the latter determines the long time scaling of $S_A(\gamma)$.

	In the derivation of the upper bound, we only use Assumption~\ref{asp:1}. For the lower bound, we additionally employ
	\begin{asp}
		\label{asp:2}
		The membrane picture of entanglement holds for the second R\'enyi entropy of the state in Eq.~\eqref{eq:k-state}. 
	\end{asp}
	This assumption is in line with general expectations from the membrane theory~\cite{zhou2020entanglement} and can be verified numerically. A representative example is reported in Fig.~\ref{plot:sigmaentropy2}. We see that, even though there are strong deviations for short times, the numerical results seems to approach the membrane theory predictions as time increases(see Appendix~\ref{app:renyigrowth} for a more thorough discussion of the validity of this assumption).\\
	\begin{figure}\ifdef{0}{}{
		\begin{tikzpicture}[scale=1,remember picture]
			\begin{axis}[
				grid=major,
				legend columns=1,
				legend style={at={(0.25,0.6)},anchor=south east,font=\scriptsize,draw= none, fill=none},
				mark size=1.3pt,	
				xlabel=$r$,
				ylabel=$S^{(1)}({\rho_{\tau_A}})$,
				y label style={at={(axis description cs:.1,.5)},anchor=south,font=\normalsize	},		
				tick label style={font=\normalsize	},	
				x label style={font=\normalsize	},	
				]
				\addplot[
				smooth,
				thick,
				mark=*,
				orange,
				dashed
				]  table[x expr=(\thisrowno{0}), y expr=(\thisrowno{1})] {plots_data/VonNeumannsigma2t14.dat};
				\addlegendentry{$t=7$}
				\addplot[
				smooth,
				thick,
				mark=*,
				blue,
				dashed
				]  table[x expr=(\thisrowno{0}), y expr=(\thisrowno{1})] {plots_data/VonNeumannsigma2t12.dat};
				\addlegendentry{$t=6$}
				
				\addplot[
				smooth,
				thick,
				mark=*,
				green,
				dashed
				]  table[x expr=(\thisrowno{0}), y expr=(\thisrowno{1})] {plots_data/VonNeumannsigma2t10.dat};
				\addlegendentry{$t=5$}
				\addplot[
				smooth,
				thick,
				mark=*,
				cyan,
				dashed
				]  table[x expr=(\thisrowno{0}), y expr=(\thisrowno{1})] {plots_data/VonNeumannsigma2t8.dat};
				\addlegendentry{$t=4$}			
				\addplot[
				smooth,
				thick,
				mark=*,
				red,
				dashed
				]  table[x expr=(\thisrowno{0}), y expr=(\thisrowno{1})] {plots_data/VonNeumannsigma2t6.dat};
				\addlegendentry{$t=3$}
				\addplot[dashed,black,domain=0:1,very thick]{1-2*abs(x-0.5)};\end{axis}	
		\end{tikzpicture}}\includegraphics{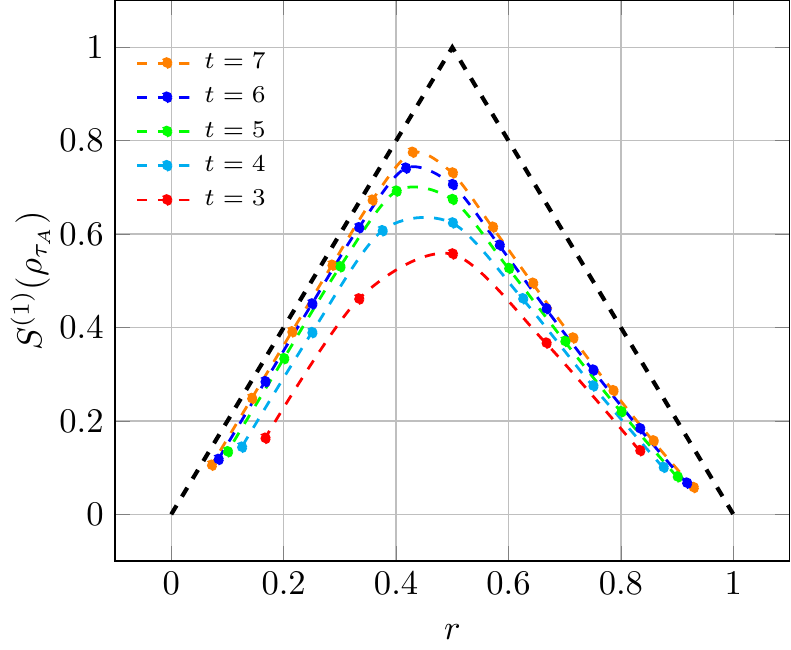}
		\caption{Slope of the entanglement entropy of the matrices $\rho_{\tA}$ as a function of the ratio $r$, for various values of $t$ accessible numerically. The dashed line represent the asymptotic profile of the curve according to Eq. \eqref{eq:membranepicturelowerb} by Assumption~\ref{asp:2}.}
		\label{plot:sigmaentropy2}
	\end{figure}

	We sketch the proof in three steps.\\
	
	{\bf Step 1}: Reduction. First we observe that the entanglement of a state, for a given bipartition, is unaffected by the action of unitary matrices acting locally on the two separate bipartitions.
	Thanks to this observation and the dual unitarity of the gates we can consider a simplified version of our state $\bra{L_\gamma}\mapsto \bra{L_{{\gamma'}}}$ where the new path ${\gamma'}$ corresponds to the edges of the light cones of the two bipartitions (see the detailed derivation in App.~\ref{eq:proofboundLstar}). This is easily understood graphically by looking at the diagram below and noting that  the area in red corresponds to matrices that, when viewed horizontally, are unitary. Therefore, removing them will not affect its entanglement
	\begin{equation}
		\!\!\bra{L_{\gamma}} = \!\!\!\!
		\fineq[-0.8ex][1][1]{
			\fill[red!20] (2.5,0)--++(-1.25/2,1.25/2)--++(1-2.5/4,1-2.5/4) coordinate (A);
			\fill[red!20] (A)--++(-1.25/2,1.25/2)--(2,2)--cycle;
			\draw (0,0)--(2,2)--(2.5,0)--cycle;
			\foreach \x in {0,0.5,2,2.5}{
				\draw (\x-0.15,0) arc (180:360:0.15);
				\draw[fill] (\x,-0.15) circle (0.03);
			}
			\foreach \x in {1,1.5}{
				\node () at (\x,-0.15) {$\cdots$};	
			}
			\foreach \x in {0.05,0.25,1.65,1.85}{
				\draw (\x,\x)--++(0,0.2);
				\draw (\x,\x+0.25) circle (0.05);
			}
			\draw (-0.15,0)--++(0,0.05);
			\draw (-0.15,-0.15+0.25) circle (0.05);
			\foreach \x in {0.52,0.82,...,1.6}{
				\node () at (\x,1.03*\x+0.15) {$\udots$};	
			}
			\draw[dashed] (2.5,0)--++(-1.25/2,1.25/2)--++(1-2.5/4,1-2.5/4)--++(-1.25/2,1.25/2);
			\node[right] () at (2.125,1.5+0.2) {$A$};
			\node[right] () at (2.375,0.5+0.2) {$\bar{A}$};
			\node () at (1.9,1.6) {$u$};
			\node () at (2.15,0.6) {$u$};
		}\mapsto \fineq[-0.8ex][1][1]{
			
			\draw (2.5,0)--(0,0)--(1.625,1.625);
			\foreach \x in {0,0.5,2,2.5}{
				\draw (\x-0.15,0) arc (180:360:0.15);
				\draw[fill] (\x,-0.15) circle (0.03);
			}
			\foreach \x in {1,1.5}{
				\node () at (\x,-0.15) {$\cdots$};	
			}
			\foreach \x in {0.05,0.25,1.65-.375,1.85-.375,1.65,1.85}{
				\draw (\x,\x)--++(0,0.2);
				\draw (\x,\x+0.25) circle (0.05);
			}
			\draw (-0.15,0)--++(0,0.05);
			\draw (-0.15,-0.15+0.25) circle (0.05);
			\foreach \x in {0.52,0.82,...,1.3}{
				\node () at (\x,1.03*\x+0.15) {$\udots$};	
			}
			\draw (2.5,0)--++(-1.25/2,1.25/2)--++(1-2.5/4,1-2.5/4)--++(-1.25/2,1.25/2);
		}\!\!\!.
	\end{equation}
	After this operation we end up with some bullet states on the top of region $A$, which do not entangle with any other part of the system: They can also be removed without affecting the result. 
	
	To sum up, as far as the entanglement is concerned, we can reduce $\bra{L_{\gamma} }$ to the following state
	\begin{equation}
		\bra{L_{\gamma'}} 
		= 
		\fineq[-0.8ex]{
			\draw (0,0)--(2.5,0)--++(-1.25/2,1.25/2)--++(1-2.5/4,1-2.5/4)--++(-1.25/2,1.25/2) --cycle;
			\foreach \x in {0,0.5,2,2.5}{
				\draw (\x-0.15,0) arc (180:360:0.15);
				\draw[fill] (\x,-0.15) circle (0.03);
			}
			\foreach \x in {1,1.5}{
				\node () at (\x,-0.15) {$\cdots$};	
			}
			\foreach \x in {0.05,0.25,1.3,1.5}{
				\draw (\x,\x)--++(0,0.2);
				\draw (\x,\x+0.25) circle (0.05);
			}
			\draw (-0.15,0)--++(0,0.05);
			\draw (-0.15,-0.15+0.25) circle (0.05);
			\foreach \x in {0.5,0.8,...,1.3}{
				\node () at (\x,1.03*\x+0.15) {$\udots$};	
			}
			\node[right] () at (2,0.5+0.2) {$\bar{A}$};
			\draw [decorate,decoration={brace,amplitude=5pt,raise=0pt,mirror},yshift=0pt]
			(2.5-1.25/2+1-2.5/4,1.25/2+1-2.5/4) -- ++(-1.25/2,1.25/2) node [black,midway,xshift=10pt,yshift=10pt] {$\tA$};
		}\,,
	\end{equation}
	in which region $A$ has $\tA = {(1+v_\gamma)} |A|/2$ sites and region $\overline{A}$ has $|\overline{A}|$ sites. 
	
	{\bf Step 2}: Lower and upper bounds. We define $\tA+1$ orthogonal projectors in region $A$
	\begin{equation}
		\begin{aligned}
			&P_0 = \ketbra{\mcirc} \otimes \ketbra{\mcirc} \otimes \cdots \ketbra{\mcirc} \otimes \I_{\bar{A}}\\
			&P_1 = \ketbra{\mcirc} \otimes \ketbra{\mcirc} \otimes \cdots (\I_{d^2} - \ketbra{\mcirc} ) \otimes \I_{\bar{A}} \\
			&P_k = \ketbra{\mcirc}^{\otimes \tA -k } \otimes (\I_{d^2} - \ketbra{\mcirc} ) \I_{d^2}^{\otimes k-1} \otimes \I_{\bar{A}} \\
			&P_{\tA} = (\I_{d^2} - \ketbra{\mcirc} ) \otimes \I_{d^2} \otimes \cdots \I_{d^2}  \otimes \I_{\bar{A}}. 
		\end{aligned}\label{eq:projdefinition}
	\end{equation}
	In words, the projector has three different actions, which we highlight by different colours in the in the following graphical equation
	\begin{equation}
		\bra{L_{\gamma'}}P_k = 
		\fineq[-0.8ex]{
			\draw (0,0)--(2.5,0)--++(-1.25/2,1.25/2)--++(1-2.5/4,1-2.5/4)--++(-1.25/2,1.25/2) --cycle;
			\foreach \x in {0,0.5,2,2.5}{
				\draw (\x-0.15,0) arc (180:360:0.15);
				\draw[fill] (\x,-0.15) circle (0.03);
			}
			\foreach \x in {1,1.5}{
				\node () at (\x,-0.15) {$\cdots$};	
			}
			\foreach \x in {0.05,0.25,1.3,1.5}{
				\draw (\x,\x)--++(0,0.2);
				\draw (\x,\x+0.25) circle (0.05);
			}
			\draw (-0.15,0)--++(0,0.05);
			\draw (-0.15,-0.15+0.25) circle (0.05);
			\foreach \x in {0.5,0.8,...,1.3}{
				\node () at (\x,1.03*\x+0.15) {$\udots$};	
			}
			\node[right] () at (2,0.5+0.2) {$\bar{A}$};
			\draw [decorate,decoration={brace,amplitude=5pt,raise=0pt},yshift=0pt]
			(2.5-1.25/2+1-2.5/4,1.25/2+1-2.5/4) -- ++(-1.25/2,1.25/2) node [black,midway,xshift=-10pt,yshift=-10pt] {$\tA$};
			\fill[red!50] (2.5-1.25/2+1-2.5/4-1.25/4,1.25/2+1-2.5/4+1.25/4)-- ++(-1.25/4,1.25/4) --++(0.1,0.1) --++ (1.25/4,-1.25/4) --cycle;
			\fill[green!50](2.5-1.25/2+1-2.5/4-1.25/4,1.25/2+1-2.5/4+1.25/4)-- ++(0.1,-0.1) --++ (0.1,0.1) --++(-0.1,0.1) --cycle;
			\fill[blue!50](2.5-1.25/2+1-2.5/4,1.25/2+1-2.5/4)--++(-1.25/4+0.1,1.25/4-0.1)--++(0.1,0.1)--++(1.25/4-0.1,-1.25/4+0.1)--cycle;
			\draw [decorate,decoration={brace,amplitude=5pt,raise=0pt,mirror},yshift=0pt]
			(2.5-1.25/2+1-2.5/4+0.1,1.25/2+1-2.5/4+0.1) -- ++(-1.25/4,1.25/4) node [black,midway,xshift=10pt,yshift=10pt] {$k$};
		}\,.
	\end{equation}
	The $k$-th projector $P_k$ keeps the bottom $k-1$ sites (blue) intact, projects each of the top $\tA - k$ sites (red) to a bullet state and the $k$-th site (green) to the orthogonal complement of the bullet state. 
	
	One can easily verify that the projectors are orthogonal and form a complete basis, i.e.,  
	\be
	P_i P_j = \delta_{ij} P_i, \qquad \sum_{k=0}^{\tA} P_k= \I_A \otimes \I_{\bar{A}}\,.
	\ee 
	These projectors decompose $|L_{\gamma'} \rangle $ into $\tA +1$ states
	, which are orthogonal in $A$. Namely, 
	\be
	\tr_A ( P_i |L_{\gamma'} \rangle \langle L_{\gamma'} |P_j) = 0, \qquad i\neq j\,. 
	\ee 
	The reduced density matrix 
	\be
	\rho_{\bar{A}} = \frac{1}{\langle L_{\gamma'} | L_{\gamma'} \rangle } \tr_{A} (|L_{\gamma'} \rangle \langle L_{\gamma'} |),
	\ee
	is then written as a classical mixture of $\tA +1$ reduced density matrices
	\begin{equation}
		\begin{aligned}
			\rho_{\bar{A}} = \sum_{k=0}^{\tA} p_k \rho_k, 
		\end{aligned}
	\end{equation}
	where the classical probability is
	\begin{equation}
		p_k =  \frac{\tr (P_k |L_{\gamma'} \rangle \langle L_{\gamma'} | )}{\langle L_{\gamma'} | L_{\gamma'} \rangle },
		\label{eq:pkdef}
	\end{equation}
	and the reduced density matrices are 
	\begin{equation}
		\rho_k = \frac{\tr_A (P_k |L'_{\gamma} \rangle \langle L'_{\gamma} | )}{\tr (P_k |L'_{\gamma} \rangle \langle L'_{\gamma} | )}.
		\label{eq:rdms}
	\end{equation}
	The concavity lower bound and mixing upper bound of $S(\rho_{\bar{A}})$ confine the von Neumann entropy to the following interval 
	\begin{equation}
		\!\!\!\!\!\sum_{k=0}^{\tA } p_k S(\rho_k )  \le S_{\gamma}(A) \le \!\!\sum_{k=0}^{\tA } p_k S(\rho_k ) \!-\! \sum_{k=0}^{\tA } p_k \log p_k\,.  
		\label{eq:bounds}
	\end{equation}
	The Shannon entropy of the classical probability $p_k$ is at most ${\log (\tA + 1) \sim \mathcal{O}(\log t)}$. Therefore, we conclude that 
	\begin{equation}
		\sum_{k=1}^{\tA } p_k S(\rho_k )  \le S_{\gamma}(A) \le \sum_{k=1}^{\tA } p_k S(\rho_k ) + \mathcal{O}(\log t)\,, 
		\label{eq:bounds2}
	\end{equation}
	where we also removed $k = 0$ from the summation since ${S(\rho_0) = 0}$.
	
	{\bf Step 3}: Evaluation of $\sum_{k=1}^{\tA} p_k S(\rho_k )$. We evaluate $p_k$ in App.~\ref{app:pkevaluation} using Assumption \ref{asp:1}. The asymptotic expression reads as 
	\begin{equation}
		p_k = 
		\left\lbrace
		\begin{aligned}
			& \frac{1}{(1+v_\gamma)t} & \quad k \ne 0 \\
			& \frac{|\bar{A}|}{2 t} & \quad k = 0 \\
		\end{aligned} \right. .
		\label{eq:pkexpr}
	\end{equation}
	\begin{figure}
		\ifdef{0}{}{	
			\begin{tikzpicture}[scale=1,remember picture]
				\begin{axis}[grid=major,
					legend columns=2,
					legend style={at={(0.65,0.75)},anchor=south east,font=\scriptsize,draw= none, fill=none},
					xtick distance=1,
					mark size=1.3pt,	
					xlabel=$k$,
					ylabel=$p_k$,
					y label style={at={(axis description cs:.1,.5)},anchor=south,font=\normalsize	},		
					tick label style={font=\normalsize	},	
					x label style={font=\normalsize	},	
					]
					\addplot[
					thick,
					mark=*,
					red,
					dashed
					] table[x expr=(\thisrowno{0}), y expr=(\thisrowno{4})/10.269] {plots_data/pk.dat};
					\addlegendentry{p=0.47}	
					\addplot[
					thick,
					mark=*,
					purple,
					dashed
					] table[x expr=(\thisrowno{0}), y expr=(\thisrowno{3})/8.81] {plots_data/pk.dat};
					\addlegendentry{p=0.53}	\addplot[
					thick,
					mark=*,
					brown,
					dashed
					] table[x expr=(\thisrowno{0}), y expr=(\thisrowno{2})/7.55] {plots_data/pk.dat};
					\addlegendentry{p=0.60}
					\addplot[
					thick,
					mark=*,
					pink,
					dashed
					] table[x expr=(\thisrowno{0}), y expr=(\thisrowno{1})/6.24] {plots_data/pk.dat};
					\addlegendentry{p=0.67}
					\addplot[dashed,black,domain=1:7,very thick]{1/14};
					\addplot[dashed,black,domain=0:1,very thick]{1/2-(1/2-1/14)*(x)};
				\end{axis}	
			\end{tikzpicture}}\includegraphics{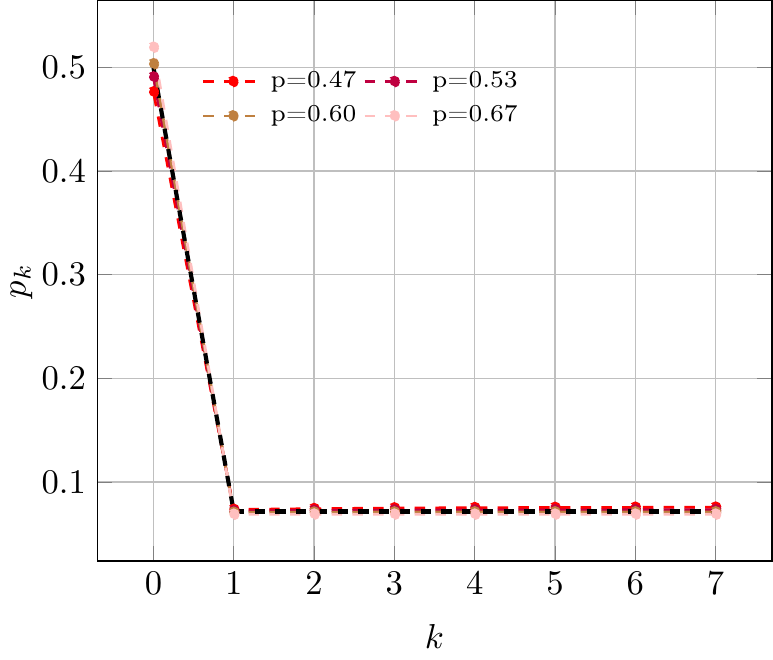}
			\caption{$p_k$ computed for various gates correspoding to different entangling powers $p$, for a state $\bra{L_\gamma}$ with $v_\gamma=1$, $t=7$ and a bipartition corresponding to $r=\frac{1}{2}$. We show the asymptotic behavior in black, according to Eq. \eqref{eq:epsbound} derived from Assumption~\ref{asp:1}.  }\label{fig:pk}
		\end{figure}
		In Fig.~\ref{fig:pk}  we compare this expression with the exact numerical evaluation of $p_k$ for short times. We see that for the cases where Eq.~\eqref{eq:scalingass} holds at short times the agreement is excellent.

		According to \eqref{eq:rdms}, $\rho_k$ can be viewed as  the reduced density matrix of the pure state 
		\begin{equation}
			\label{eq:k-state}
			\bra{L_k}=  \bra{L_{\gamma'}}P_k=
			\fineq[-0.8ex][1][1]{
				\draw (0,0)--(2.5,0)--++(-1.25/2,1.25/2)--++(1-2.5/4,1-2.5/4)--++(-1.25/2,1.25/2) --cycle;
				\foreach \x in {1,1.5}{
					\node () at (\x,-0.15) {$\cdots$};	
				}
				\foreach \x in {0,0.5,2,2.5}{
					\draw (\x-0.15,0) arc (180:360:0.15);
					\draw[fill] (\x,-0.15) circle (0.03);
				}
				\foreach \x in {0.05,0.25,1.3,1.5}{
					\draw (\x,\x)--++(0,0.2);
					\draw (\x,\x+0.25) circle (0.05);
				}
				\draw (-0.15,0)--++(0,0.05);
				\draw (-0.15,-0.15+0.25) circle (0.05);
				\foreach \x in {0.5,0.8,...,1.3}{
					\node () at (\x,1.03*\x+0.15) {$\udots$};	
				}
				\node[right] () at (2.65,0.5) {$\bar{A}$};
				\draw [decorate, decoration = {brace},shift={(0.15,0.2)}]   (2.5,0.5+0.2)  --++ (0,-0.8);
				\draw [decorate,decoration={brace,amplitude=5pt,raise=0pt,mirror},shift={(0.4,0.4)}]
				(2.5-1.25/2+1-2.5/4,1.25/2+1-2.5/4) -- ++(-1.25/2,1.25/2) node [black,midway,xshift=10pt,yshift=10pt] {$k$};
				\foreach \x in {1.3,1.1}{
					\draw (\x+0.05/1.414,3.25-\x+0.05/1.414)--++(0.2/1.414,0.2/1.414);
					\draw (\x,3.25-\x) circle (0.05);
				}
				\foreach \x in {1.9,2.1}{
					\draw (\x+0.05/1.414-0.08,3.25-\x+0.05/1.414)--++(0.2/1.414,0.2/1.414);
				}
				
				\draw (1.7+0.05/1.414-0.08,3.25-1.7+0.05/1.414)--++(0.2*1.414,0.2*1.414);
				\draw[fill=black,rotate around={45:(1.7+0.05/1.414-0.08+0.1*1.414,3.25-1.7+0.05/1.414+0.1*1.414)}]
				(1.7+0.05/1.414-0.08+0.05*1.414,3.25-1.7+0.05/1.414+0.05*1) rectangle (1.7+0.05/1.414-0.08+0.15*1.414,3.25-1.7+0.05/1.414+0.15*1.414);
				
				\foreach \x in {2.9, 3.05,3.2}{
					\draw (\x+0.05/1.414-0.8,3.25-\x+0.05/1.414)--++(0.2/1.414,0.2/1.414);
				}
				
				\foreach \x in {0,0.15,0.3}{
					\draw (2.6+\x+0.05/1.414-0.7,3.25-2.6+\x+0.05/1.414)--++(0.2/1.414,-0.2/1.414);
				}
				
			}
		\end{equation}
		where the first $\tA - k$ sites are  (projected by $P_k$ to be) bullet states, and we inserted on the $k$-th site (counted from the bottom of $A$) a projector indicated by a black square
		\begin{align}
			\fineq[-0.8ex][0.5][1]{
				\draw (-.5,0)--(1.5,0);
				\draw[fill=black] (.25,-.25) rectangle (.75,.25);
			}=\I_{d^2} - \ketbra{\mcirc} .
		\end{align}

		By using the bound of Hilbert space dimension, we can write
		\begin{align}
			S(\rho_k ) &\le \min( 2 (k-1)\log d+\log(d^2-1) , 2|\bar{A}|\log(d) )\notag\\
			&\simeq 2 \min( k, |\bar{A}|)\log(d) ,\label{eq:geometricupperbound}
		\end{align}
		where we considered a scaling limit where $k, \abs{A}, t $ are taken to infinity and the relative ratios kept constant. Plugging back into \eqref{eq:bounds2} we find the following upper bound
		\begin{align}
			&\sum_{k=1}^{\tA} p_k S(\rho_k ) \lesssim \frac{ 2 \log (d)}{(1+v_\gamma) t} \sum_{k=1}^{\tA} \min(k, |\bar{A}| ) \\
			&\simeq \frac{8 t \log(d)}{1 + v_\gamma} \int_0^{r (1+v_\gamma)/2} \!\!\!\!\!\!\!\!\!\!\!\!\!\!\!\!\!\!\!\!\!\!\!{\rm d}x\,\, \min( x, 1-r) = s(r) t\log(d)\,.\notag
		\end{align}
		
		Using Assumption \ref{asp:2}, we can evaluate $S(\rho_k)$ using the membrane picture, as detailed in Appendix~\ref{app:renyigrowth}, to find
		\begin{equation}
			S( \rho_k ) \simeq \min( k, |\bar{A}|) 2 \log d. 
			\label{eq:membranepicturelowerb}
		\end{equation}
		This means that $S(\rho_k)$ saturates the trivial bound in Eq. \eqref{eq:geometricupperbound}, leading to Eq.~\eqref{eq:SAgammabound}.
		
		\begin{figure}\ifdef{0}{}{
			\begin{tikzpicture}[scale=1,remember picture]
				\begin{axis}[
					grid=major,
					legend columns=1,
					legend style={at={(0.4,0.7)},anchor=south east,font=\scriptsize,draw= none, fill=none},
					mark size=1.3pt,	
					xlabel=$r$,
					ymin=0,
					ylabel=$ {S_A(\gamma )}/{t}$,
					y label style={at={(axis description cs:.1,.5)},anchor=south,font=\normalsize	},		
					tick label style={font=\normalsize	},	
					x label style={font=\normalsize	}
					]
					\addplot[
					smooth,
					thick,
					mark=*,
					blue,
					dashed
					] table[x expr=(\thisrowno{0}), y expr=(\thisrowno{1})/6.5] {plots_data/VonNeumannbounds.dat};	
					\addlegendentry{Lower bound}
					\addplot[
					smooth,
					thick,
					mark=*,
					red,
					dashed
					] table[x expr=(\thisrowno{0}), y expr=(\thisrowno{2})/6.5] {plots_data/VonNeumannbounds.dat};	
					\addlegendentry{Exact value}
					\addplot[
					smooth,
					thick,
					mark=*,
					green,
					dashed
					] table[x expr=(\thisrowno{0}), y expr=(\thisrowno{3})/6.5] {plots_data/VonNeumannbounds.dat};	
					\addlegendentry{Upper bound}
					
				\end{axis}	
			\end{tikzpicture}}\includegraphics{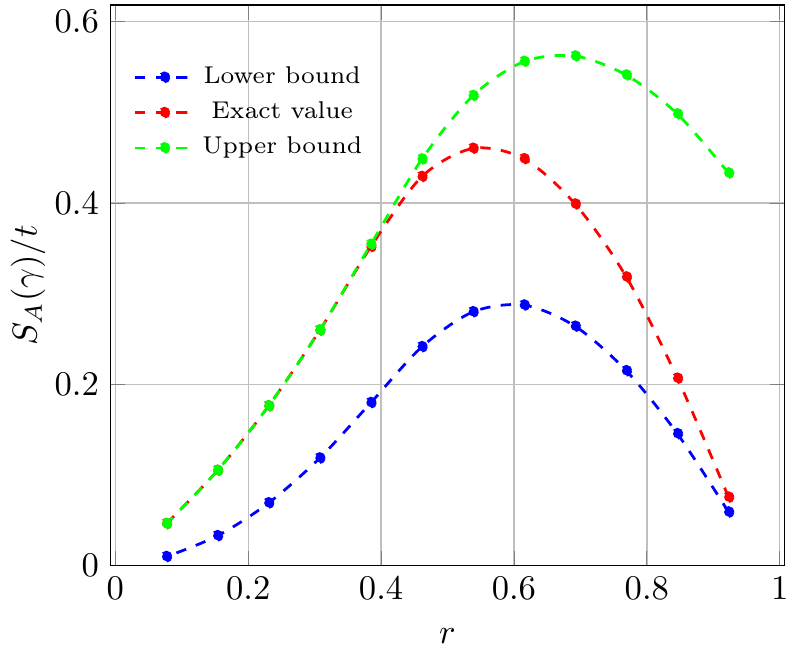}
			\caption{Slope of the entanglement entropy  as a function of the ratio $r$, for various values of $t$ accessible numerically, in the case $v_{\gamma}=1$. The lower  and upper bounds are found evaluating numerically Eq. \eqref{eq:bounds}. }
			\label{plot:Von_neumann_growth_r}
		\end{figure}
		\begin{figure}\ifdef{0}{}{
			\begin{tikzpicture}[scale=1,remember picture]
				\begin{axis}[
					grid=major,
					legend columns=1,
					legend style={at={(0.5,0.55)},anchor=south east,font=\scriptsize,draw= none, fill=none},
					mark size=1.3pt,	
					xlabel=$r$,
					ymin=0,
					ymax=0.55,
					ylabel=${\Delta_t S_A(\gamma)}$,
					y label style={at={(axis description cs:.1,.5)},anchor=south,font=\normalsize	},		
					tick label style={font=\normalsize	},	
					x label style={font=\normalsize	}
					]
					\addplot[dashed,black,domain=0.:0.5,very thick]{2*(x)^2*ln(2)};
					\addlegendentry{Analytic prediction}
					\addplot[
					only marks,
					thick,
					mark=*,
					blue,
					dashed
					] table[x expr=(\thisrowno{0}), y expr=(\thisrowno{1})] {plots_data/fitteddata.dat};	
					\addlegendentry{$t=\infty$ extrapolation}
					
					\addplot[
					red,smooth,
					thick,
					no markers
					] table[x expr=(\thisrowno{0}), y expr=(\thisrowno{1})] {plots_data/fulldatainterpolation.dat};	
					\addlegendentry{$t=5$}
					\addplot[
					green,smooth,
					thick,
					no markers
					] table[x expr=(\thisrowno{0}), y expr=(\thisrowno{2})] {plots_data/fulldatainterpolation.dat};	
					\addlegendentry{$t=5.5$}
					\addplot[
					violet,smooth,
					thick,
					no markers
					] table[x expr=(\thisrowno{0}), y expr=(\thisrowno{3})] {plots_data/fulldatainterpolation.dat};	
					\addlegendentry{$t=6$}\addplot[
					brown,smooth,
					thick,
					no markers
					] table[x expr=(\thisrowno{0}), y expr=(\thisrowno{4})] {plots_data/fulldatainterpolation.dat};	
					\addlegendentry{$t=6.5$}\addplot[
					orange,smooth,
					thick,
					no markers
					] table[x expr=(\thisrowno{0}), y expr=(\thisrowno{5})] {plots_data/fulldatainterpolation.dat};	
					\addlegendentry{$t=7$}
					
					\addplot[dashed,black,domain=0.5:1,very thick]{2*(1-(x))*(3*(x)-1)*ln(2)};
				\end{axis}
			\end{tikzpicture}} \includegraphics{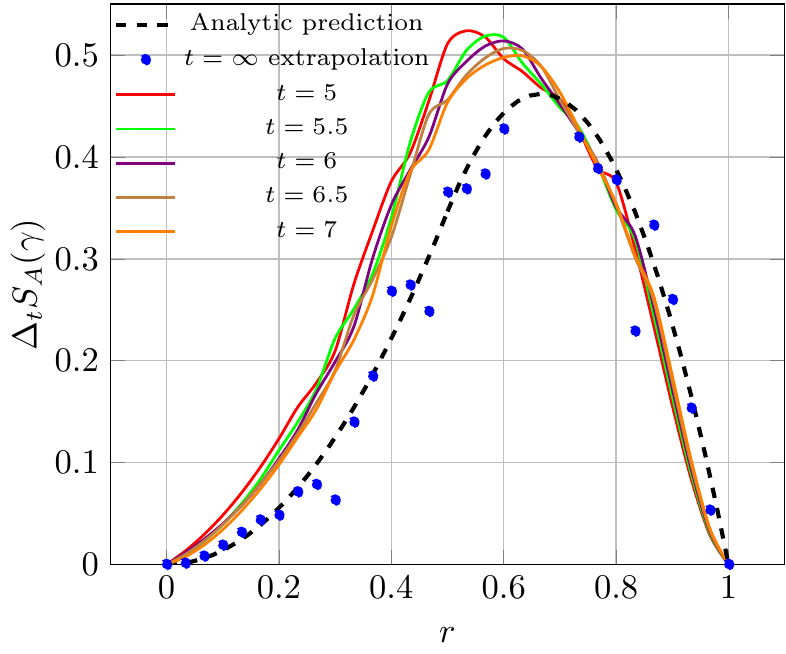}\ifdef{0}{}{
				\begin{tikzpicture}
					\begin{axis}[
						grid=major,
						legend columns=1,
						legend style={at={(0.62,0.85)},anchor=south east,font=\scriptsize,draw= none, fill=none},
						mark size=1.3pt,	
						xlabel=$r$,
						ymin=0,
						ymax=0.5,
						ylabel=${S_A(\gamma_t)-S_A(\gamma_{t-1/2})}$,
						y label style={at={(axis description cs:.1,.5)},anchor=south,font=\normalsize	},		
						tick label style={font=\normalsize	},	
						x label style={font=\normalsize	}
						]
						\addplot[	   only marks,blue,
						error bars/.cd, 
						y dir=both, x dir=both,
						x explicit, y explicit]table [x index=0, y index=1,y error index=2] {plots_data/fitteddataerror.dat};	
						\addlegendentry{Extrapolated data at $t=\infty$}
						
						\addplot[dashed,black,domain=0.:0.5,very thick]{2*(x)^2*ln(2)};
						\addplot[dashed,black,domain=0.5:1,very thick]{2*(1-(x))*(3*(x)-1)*ln(2)};
						\addlegendentry{Analytic prediction}
					\end{axis}	
				\end{tikzpicture}
			}
			\caption{Slope of the entanglement entropy for the state $\bra{L_\gamma}$, obtained by taking finite differences of $S_A(\gamma)$ for two subsequent time-steps. Given the discrete nature of the states, only some rational values of $r$ are allowed at each time, so we interpolated between those in order to take the difference. In blue, we show an extrapolation of this data in the limit $t\rightarrow\infty$, which we ultimately compare with the asymptotic prediction, in black (obtained from Eq. \eqref{eq:finalresultDU}). The extrapolation is attained by observing that, due to the logarithmic form of the corrections to ${S_A(\gamma)}$, for large enough $t$ we have ${\Delta_t S_A(\gamma)}\simeq A + B/t$, where $A$ is the desired asymptotic value. Then, we performed a linear fit of the data in $1/t$ to estimate $A$.}
			\label{plot:fitteddata}
		\end{figure}

		\begin{figure}
\ifdef{0}{}{\begin{tikzpicture}[scale=1,remember picture]
					\begin{axis}[
						grid=major,
						legend columns=2,
						legend style={at={(0.6,0.7)},anchor=south east,font=\scriptsize,draw= none, fill=none},
						mark size=1.3pt,	
						xlabel=$t$,
						ylabel=$\max_A S_A(\gamma)$,
						y label style={at={(axis description cs:.1,.5)},anchor=south,font=\normalsize	},		
						tick label style={font=\normalsize	},	
						x label style={font=\normalsize	},	
						]
						\addplot[
						smooth,
						thick,
						mark=*,
						blue,
						dashed
						] table[x expr=(\thisrowno{0})/2, y expr=(\thisrowno{1})] {plots_data/VonNeumann.dat};
						\addlegendentry{p=0.27}
						\addplot[
						smooth,
						thick,
						mark=*,
						orange,
						dashed
						] table[x expr=(\thisrowno{0})/2, y expr=(\thisrowno{2})] {plots_data/VonNeumann.dat};
						\addlegendentry{p=0.33}	
						\addplot[
						smooth,
						thick,
						mark=*,
						green,
						dashed
						] table[x expr=(\thisrowno{0})/2, y expr=(\thisrowno{3})] {plots_data/VonNeumann.dat};
						\addlegendentry{p=0.40}	
						\addplot[
						smooth,
						thick,
						mark=*,
						red,
						dashed
						] table[x expr=(\thisrowno{0})/2, y expr=(\thisrowno{4})] {plots_data/VonNeumann.dat};
						\addlegendentry{p=0.47}	
						\addplot[
						smooth,
						thick,
						mark=*,
						purple,
						dashed
						] table[x expr=(\thisrowno{0})/2, y expr=(\thisrowno{5})] {plots_data/VonNeumann.dat};
						\addlegendentry{p=0.53}	\addplot[
						smooth,
						thick,
						mark=*,
						brown,
						dashed
						] table[x expr=(\thisrowno{0})/2, y expr=(\thisrowno{6})] {plots_data/VonNeumann.dat};
						\addlegendentry{p=0.60}
						\addplot[
						smooth,
						thick,
						mark=*,
						pink,
						dashed
						] table[x expr=(\thisrowno{0})/2, y expr=(\thisrowno{7})] {plots_data/VonNeumann.dat};
						\addlegendentry{p=0.67}
						\addplot[dashed,domain=0:6,very thick]{2*x/3 * ln(2)+0.35};
					\end{axis}	
			\end{tikzpicture}}\includegraphics{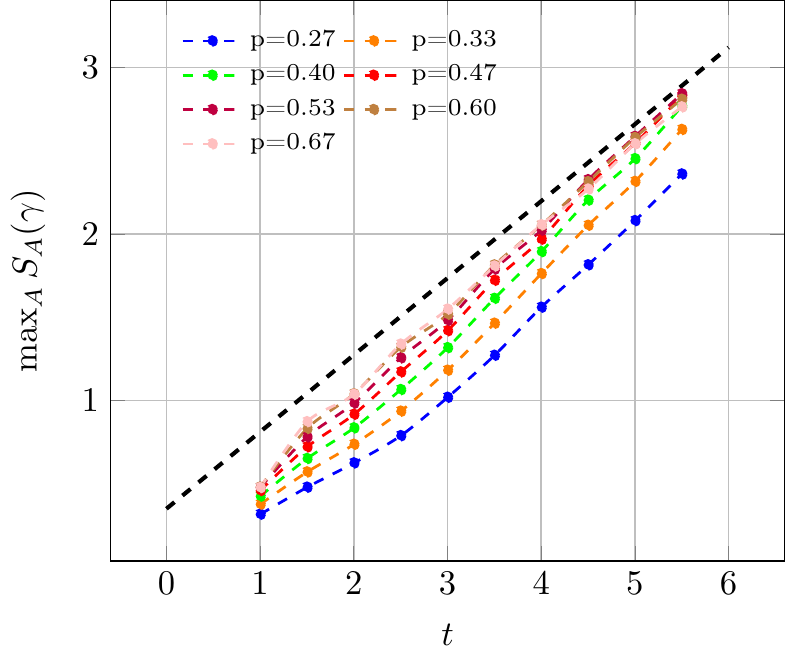}
			\caption{Maximal entanglement of the influence matrix on the diagonal path $\gamma_{\rm lc}$ (cf.~\eqref{eq:lightconepath}) versus the length of the path $|\gamma_{\rm lc}|$ for different values of the entangling power $p$. The entanglement reported is the maximum attained among all the possible non-disjoint bipartitions. The gates are parameterised as explained in Appendix~\ref{sec:parameterisation}. The asymptotic growth seems to be independent of the entangling power. The initial $p$-dependent transient is larger for smaller values of $p$. This is consistent with the fact that for $p=0$ the gates are SWAPs and $\bra{L_{\gamma_{\rm lc}}}$ is a product state for all initial states~\eqref{eq:initialstate}. The black dashed line corresponds to the theoretical prediction of the growth, as in Eq. \eqref{eq:finalresultDU}, plus an arbitrary constant chosen for convenience.}
			\label{plot:Von_neumann_growth}
		\end{figure}

		As for the bound in the previous subsection, an independent numerical test of Eq.~\eqref{eq:finalresultDU} is hampered by the fact that our numerical investigations are restricted to short times. At the accessible times $S(\rho_k )$ are typically far from their asymptotic form and the Shannon entropy of $p_k$ is non-negligible. This is demonstrated in Fig.~\ref{plot:Von_neumann_growth_r} where we plot the exact numerical evaluations of $S_A(\gamma)$ and the lower and upper bound in Eq.~\eqref{eq:bounds}: the two bounds should collapse for large times but are still rather far at the maximal accessible times. To circumvent this complication, we plot the the difference of $S_A(\gamma)$ at two subsequent time steps from finite-time numerics and extrapolate to $t \rightarrow \infty$. We find a fair agreement with Eq.~\eqref{eq:finalresultDU}, see Fig.~\ref{plot:fitteddata}. Interestingly, the finite-time effects seem not to affect the maximal slope of the temporal entanglement entropy, which is in good agreement with Eq.~\eqref{eq:finalresultDU} even at short times, see Fig.~\ref{plot:Von_neumann_growth}. 
		
	}

	\section{Temporal Entanglement of the Vertical State}
	\label{sec:TEinRUvertical}
	
	In this section, we consider the second marginal case of Eq.~\eqref{eq:ev_e0} in which the time-like surface is vertical $(v_\gamma=0)$ and the chaotic quantum circuit is arbitrary. Namely, we look at the scaling in $|\gamma|=2t$ of the temporal entanglement of the original influence matrix for generic circuits. 
	
	We find that higher R\'enyi entropies grow logarithmically in time
	\begin{equation}
		S^{(\alpha)}_A(\gamma) \sim \log(t), \qquad \alpha>1\,.
		\label{eq:logscalingvertical}
	\end{equation}
	We begin by showing the sub-linear growth via a direct application of the Eckart-Young strategy employed  App.~\ref{app:upperboundRenyi}. Specifically, we use the upper bound 
	\begin{align}
		\!\!\!\!S^{(\alpha)}_A(\gamma) &\!\leq\! \frac{\alpha}{1-\alpha}\log\! \frac{\left({\bra{L_\gamma}(\ket{\Psi_A}\otimes\ket{\Psi_{\bar{A}}})}\right)^2}{{\braket{L_\gamma}}{\braket{\Psi_A}}{\braket{\Psi_{\bar A}}}}\notag\\
		&:=  \frac{2\alpha}{1-\alpha} \log r_t,
		\label{eq:EYboundv=0}
	\end{align}
	by means of the overlap of the state $\bra{L_{\gamma}}$ and a factorised state $\bra{\Psi_A}\otimes \bra{\Psi_{\bar{A}}}$. 
	
\begin{figure}\begin{subfigure}[]{	 
				\includegraphics[scale=1]{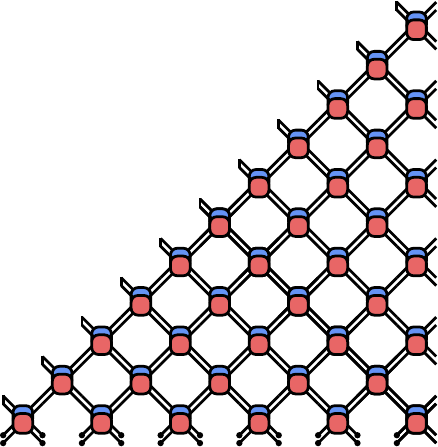}
				\ifdef{0}{}{\fineq[-0.8ex][0.4][0.4]{
						\foreach \x in {0,1,...,5}{
							\foreach \y in {0,...,\x}{	
								\roundgate[2*\x-\y][\y+0.20][1][tr][bertiniblue]
								\roundgate[2*\x-\y][\y+0.00][1][tr][bertinired]
								\roundgate[10-\y][\y+10-2*\x+0.20][1][tr][bertiniblue]
								\roundgate[10-\y][\y+10-2*\x+0.00][1][tr][bertinired]
							}
						}
						\foreach \x in {0,...,10}{
							\draw[thick](\x-.5,.48+\x)--(\x-.5,\x+.72);
						}
						\foreach \x in {0,1,...,10}{
							\foreach \y in {0,0.2}{
								\fill (\x-0.5,-0.5+\y) circle (0.08);
							}
						}
		}}}	\end{subfigure} \begin{subfigure}[]{
				\includegraphics[scale=1]{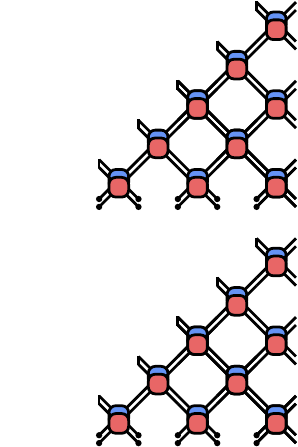}
				\ifdef{0}{}{\fineq[-0.8ex][0.4][0.4]{
						\foreach \x in {0,1,...,2}{
							\foreach \y in {0,...,\x}{	
								\roundgate[2*\x-\y][\y+0.20][1][tr][bertiniblue]
								\roundgate[2*\x-\y][\y+0.00][1][tr][bertinired]
								\roundgate[4-\y][\y+4-2*\x+0.20][1][tr][bertiniblue]
								\roundgate[4-\y][\y+4-2*\x+0.00][1][tr][bertinired]
							}
						}
						\foreach \x in {0,...,4}{
							\draw[thick](\x-.5,.48+\x)--(\x-.5,\x+.72);
						}
						\foreach \x in {0,1,...,4}{
							\foreach \y in {0,0.2}{
								\fill (\x-0.5,-0.5+\y) circle (0.08);
							}
						}
						\foreach \x in {0,1,...,2}{
							\foreach \y in {0,...,\x}{	
								\roundgate[2*\x-\y][-6+\y+0.20][1][tr][bertiniblue]
								\roundgate[2*\x-\y][-6+\y+0.00][1][tr][bertinired]
								\roundgate[4-\y][\y-2-2*\x+0.20][1][tr][bertiniblue]
								\roundgate[4-\y][\y-2-2*\x+0.00][1][tr][bertinired]
							}
						}
						\foreach \x in {0,...,4}{
							\draw[thick](\x-.5,.48+\x-6)--(\x-.5,\x+.72-6);
						}
						\foreach \x in {0,1,...,4}{
							\foreach \y in {0,0.2}{
								\fill (\x-0.5,-0.5+\y-6) circle (0.08);
							}
						}\draw[fill = white,white,shift={(-3,-4)},opacity=0] (0,0) rectangle (1,8);
		}	}}\end{subfigure} 
		\caption{(a) The vertical state; (b) A tensor product state on $A$ and $\bar A$. The state is written as a tensor product of two vertical states defined on a time lattice with half of the sites.}
		\label{fig:L_overlap}
	\end{figure}

\begin{figure}
		\ifdef{0}{}{\begin{tikzpicture}[scale=1,remember picture]
				\begin{loglogaxis}[grid=major,
					legend columns=1,
					legend style={at={(0.27,0.1)},anchor=south east,font=\scriptsize,draw= none, fill=none},
					mark size=0.5pt,	
					xlabel=$t$,
					ylabel=$\bar r_t$,
					y label style={at={(axis description cs:.15,.6)},anchor=south,font=\normalsize	},		
					tick label style={font=\normalsize	},	
					x label style={font=\normalsize	},	
					]
					\addplot[
					thick,
					no markers,
					blue,
					] table[x index=0,y index=1] {plots_data/dataratio2.dat}	;
					\addlegendentry{$d=2$}
					\addplot[
					thick,
					violet,
					no markers,
					] table[x index=0,y index=2] {plots_data/dataratio2.dat}	;
					\addlegendentry{$d=3$}
					\addplot[
					thick,
					brown,
					no markers,
					] table[x index=0,y index=3] {plots_data/dataratio2.dat}	;
					\addlegendentry{$d=4$}
					\addplot[
					thick,
					cyan,
					no markers,
					] table[x index=0,y index=4] {plots_data/dataratio2.dat}	;
					\addlegendentry{$d=5$}
					\addplot[domain=2:600,dashed] { x^(-1.25)*10};
					\addlegendentry{ $10{t^{-\frac{5}{4}}}{}$}
				\end{loglogaxis}	
		\end{tikzpicture}}
		\includegraphics{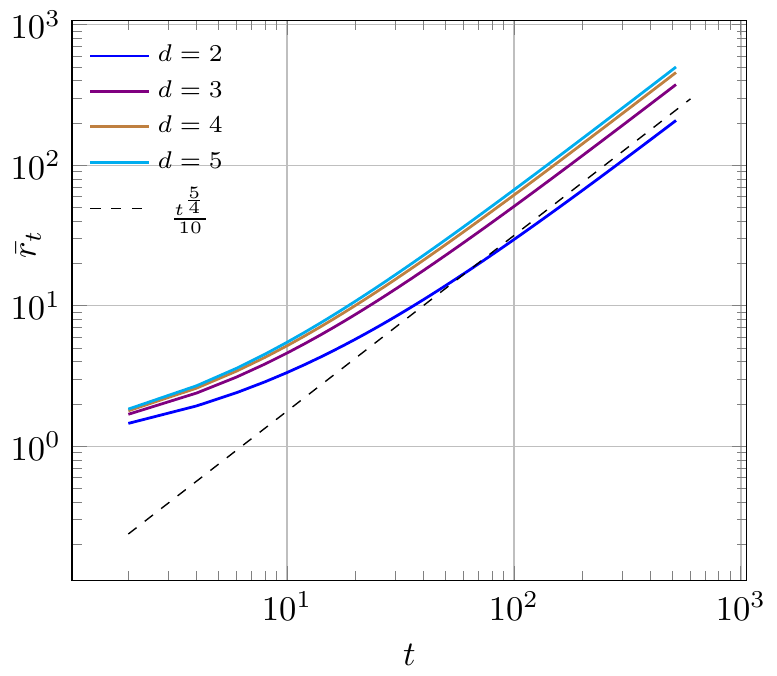}
		\caption{Polynomial decay of $\bar r_t$. According to \eqref{eq:eqfinal} we show the asymptotic expected behavior $\propto t^{-\frac{5}{4}}$, as a dotted line. In the calculation we considered sites of local dimension $d$ and an initial product state.
		}
		\label{plot:ratiodecay}
	\end{figure}

	To find a product state with large overlap we employ the membrane theory. Specifically, we consider the state 
	\begin{equation}
		\bra{\Psi_A}\otimes \bra{\Psi_{\bar{A}}}
		= \bra{L_{\gamma/2}}\otimes \bra{L_{\gamma/2}},
	\end{equation}
	which is depicted in Fig.~\ref{fig:L_overlap}(b). 
	\begin{figure}
		\includegraphics[width=\columnwidth]{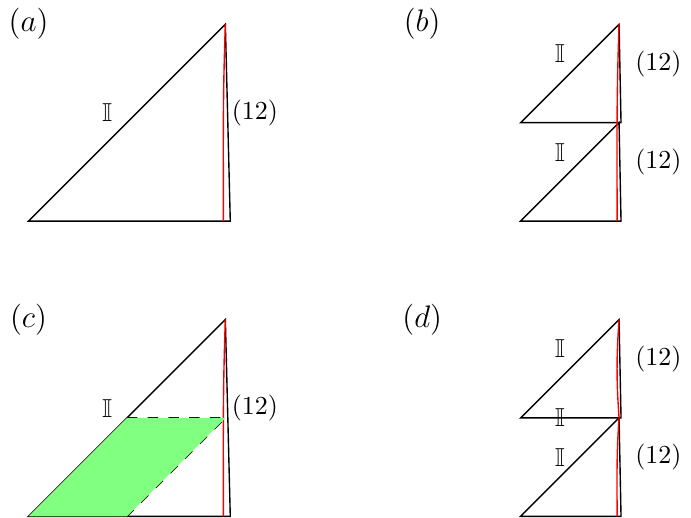}
		\caption{The domain wall analysis that produces \eqref{eq:normssmallT} and \eqref{eq:overlapsmallT}. (a) $\overline{\braket{L_\gamma}}$: the dominant configuration is one domain wall going down vertically at equilibrium. (b) $\overline{(\bra{\Psi_A}\otimes\bra{\Psi_{\bar{A}}})(\ket{\Psi_{\bar{A}}}\otimes\ket{\Psi_A})}$: the two parts of the product states factorise so there are two independent vertical domain walls with half the size of (a). (c) $\overline{\bra{L_\gamma} (\ket{\Psi_A}\otimes\ket{\Psi_{\bar{A}}})}$: the green region is where $\bra{L_\gamma}$ and $\bra{\Psi_{\bar{A}}}\otimes\bra{\Psi_A}$ differ. It has only one copy of $u \times u^*$, which, upon averaging, generates a patch of $\I$. So the domain wall will avoid the green region and goes down vertically. (d) The green region after average provides the $\I$ boundary conditions for the two dashed lines in (c) (see main text for the bottom rim of the upper triangle, also see App. \ref{app:averagedunitaries}). The top triangle has a pinned domain wall ending at the right most point, the bottom triangle still host a free random walk.}
		\label{fig:dw_overlap}
	\end{figure}
	Assuming the circuit to be Haar random, the norm of these states are determined by the line tension at $v =0$ as follows
	\begin{equation}
		\begin{aligned}
			\overline{\braket{L_\gamma}} &\sim \exp( -  \mathcal{E}_H(0) \log(d)  t ),\\
			\overline{\braket{\Psi_A}} &\sim \exp( -  \mathcal{E}_H(0) \log(d)  t/2 ),\\
			\overline{\braket{\Psi_{\bar{A}}}} &\sim \exp( -  \mathcal{E}_H(0) \log(d)  t/2 ),
		\end{aligned}
		\label{eq:normssmallT}
	\end{equation}
	where we recall (cf. Eq.~\eqref{eq:linetensionHaar})
	\begin{equation}
		\mathcal{E}_H(0)= \frac{\log ({d^2+1})-\log {2d}}{\log d}\,.
	\end{equation}
	The average of the overlap is
	\begin{equation}
		\overline{\bra{L_\gamma} (\ket{\Psi_A}\otimes\ket{\Psi_{\bar{A}}})} \sim \exp( -  \mathcal{E}_H(0) \log(d)  t ). 
		\label{eq:overlapsmallT}
	\end{equation}
	The estimation in Eq.~\eqref{eq:overlapsmallT} relies upon evaluating the random averaging in Fig.~\ref{fig:dw_overlap}(c), where the region in which the two states differ is only populated by the permutation $\I$. Thus, the minimal free energy configuration continues to have a domain wall going vertically down. Combining \eqref{eq:normssmallT} and \eqref{eq:overlapsmallT} we find 
	\begin{equation}
		\bar r_t =   \frac{\left|\overline{\bra{L_\gamma}(\ket{\Psi_A}\otimes\ket{\Psi_{\bar{A}}})}\right|}{\sqrt{\overline{\braket{L_\gamma}}\,\, \overline{\braket{\Psi_A}}\,\, \overline{\braket{\Psi_{\bar A}}}}} = O(t^\alpha),
		\label{eq:ratio}
	\end{equation}
	The exponent $\alpha$ can be found by studying the subleading contributions from the random walk of the domain wall. All the domain walls in Fig.~\ref{fig:dw_overlap} are subject to the non-crossing condition at the right boundary. If we view from bottom to top, this is the random walk that first hit $x = 0$ (the coordinate of the right boundary) for $t=0$ (the final time when viewing from bottom to top). The probability distribution for this process is known as the Levy-Smirnov distribution and reads as
	\begin{equation}
		p(x)=\frac{x}{t^{\frac{3}{2}}}e^{- x^2 / t}.
	\end{equation}
	The three independent averages inside the square root in the numerator of Eq.~\eqref{eq:ratio} (one in Fig.~\ref{fig:dw_overlap}(a), the other two in Fig.~\ref{fig:dw_overlap}(b)) correspond to a free boundary condition at the bottom, each of which contributes a polynomial factor $t^{-\frac{1}{2}}$ (integrate the Levy-Smirnov distribution in $x$). For the average of the overlap, the green region in Fig.~\ref{fig:dw_overlap}(c) represents the missing part in $\bra{\psi_A } \bra{\psi_{\bar{A}} }$ compared with $\bra{L_\gamma }$. It can only produce $\I$, which becomes the boundary condition of the lower triangle and bottom rim of the top triangle in Fig.~\ref{fig:dw_overlap}(d) (the boundary condition for the bottom rim after random averaging is $\ket{\mcirc}$ tensor the density matrix of single site state. But its overlap with the $\I$ and $(12)$ permutation is the same as $\ket{\mcirc \mcirc}$, so for random averaging we can have the replacement). The lower triangle contributes a $t^{-\frac{1}{2}}$ factor as we argued above. Instead, because of the $\I$ boundary condition at the bottom rim, the domain wall in the top triangle is penalised by a factor of ${1}/{d}$ when it further moves to the left. Thus the domain wall is pinned to a slope of $v = 0$. We end up with a $t^{-\frac{3}{2}}$ factor in the Levy-Smirnov distribution for the pinned domain wall. Putting all together, we have 
	\begin{equation}
		\bar r_t \simeq \frac{t^{-\frac{3}{2}} t^{-\frac{1}{2}}}{ \sqrt{t^{-\frac{1}{2}} t^{-\frac{1}{2}} t^{-\frac{1}{2}}}} = t^{-\frac{5}{4}}.
		\label{eq:predrt}
	\end{equation}
	These random wall arguments can be made more precise by solving a set of recursive relations of the averaged terms in \eqref{eq:ratio}, see Appendix \ref{app:averagedunitaries}. The prediction \eqref{eq:predrt} is compared with exact solution of the recursive relations in Fig.~\ref{plot:ratiodecay}. The power-law decay of $\bar r_t$ suggests that also $r_t$ in Eq.~\eqref{eq:EYboundv=0} should decay as a power law, leading to Eq.~\eqref{eq:logscalingvertical}.

\begin{figure}
		\ifdef{0}{}{\begin{tikzpicture}[scale=1,remember picture]
				\begin{axis}[grid=major,
					legend columns=2,
					legend style={at={(0.65,0.7)},anchor=south east,font=\scriptsize,draw= none, fill=none},
					mark size=1.3pt,	
					xlabel=$t$,
					ylabel=$\max S^{(1)}_A(\gamma)$,
					y label style={at={(axis description cs:.1,.5)},anchor=south,font=\normalsize	},		
					tick label style={font=\normalsize	},	
					x label style={font=\normalsize	},	
					]
					\addplot[
					thick,
					mark=*,
					blue,
					dashed
					] table[x index=0,y index=1] {plots_data/unitary_verticalstate.dat};
					\addlegendentry{p=0.37}
					\addplot[
					thick,
					mark=*,
					red,
					dashed
					] table[x index=0,y index=2] {plots_data/unitary_verticalstate.dat};\addlegendentry{p=0.58}
					\addplot[
					thick,
					mark=*,
					green,
					dashed
					] table[x index=0,y index=3] {plots_data/unitary_verticalstate.dat};\addlegendentry{p=0.56}
					\addplot[
					thick,
					mark=*,
					brown,
					dashed
					] table[x index=0,y index=4] {plots_data/unitary_verticalstate.dat};\addlegendentry{p=0.32	}\addplot[
					thick,
					mark=*,
					orange,
					dashed
					] table[x index=0,y index=5	] {plots_data/unitary_verticalstate.dat};\addlegendentry{p=0.42}
					\addplot[
					thick,
					mark=*,
					violet,
					dashed
					] table[x index=0,y index=6	] {plots_data/unitary_verticalstate.dat};\addlegendentry{p=0.62}
					\addplot[domain=2:7,dashed,very thick] { x * ln(2)  /3+1 };
					\addlegendentry{Dual unitary}
				\end{axis}	
		\end{tikzpicture}}
		\includegraphics{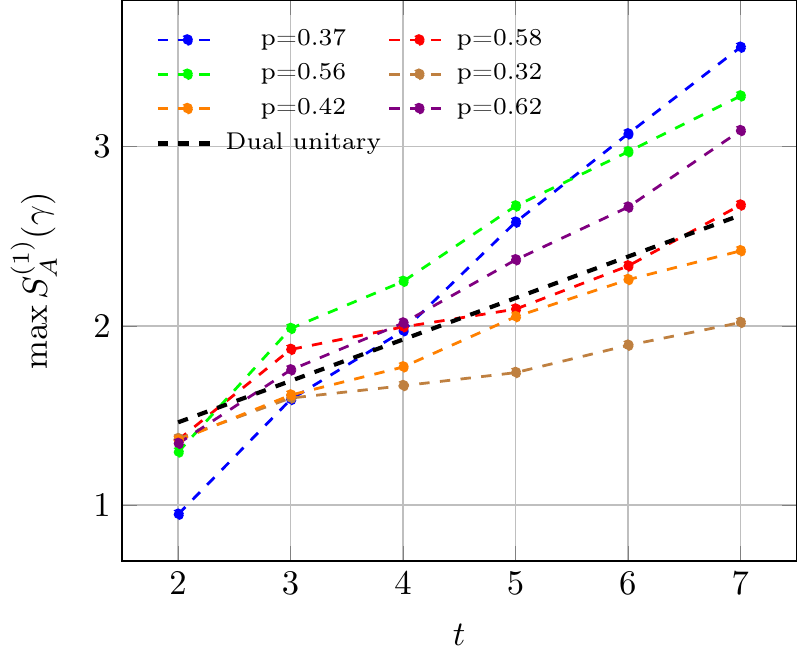}
		\caption{Growth of the entanglement entropy for the vertical cut state $\bra{L_\gamma}$, given random choices of the dual-unitary gate (kept constant in space and time) with entangling power $p$. We considered the contiguous bipartition $A\bar A$ of $\gamma$ yielding the maximum entanglement. The entangling power is computed according to Eq~5 in \cite{entanglingpoweroriginal}, which has been normalized by a factor ${(d+1)}/{(d-1)}$ in order to have $p\in [0,1]$.
			The black line represents the the growth of $S^{(1)}_A(\gamma)$ in dual-unitary circuits (Eq. \eqref{eq:finalresultDU}). 
		}
		\label{plot:verticalstateentropygrowth}
	\end{figure}

	On the other hand, a direct numerical evaluation is still compatible with a linear growth in time of $S^{(1)}(\gamma)$, see Fig.~\ref{plot:verticalstateentropygrowth}. Interestingly, we see that for certain choices of gates the growth of temporal entanglement entropy is slower than the lower bound for dual unitary circuits (see  Sec.~\ref{sec:VNentropy}). This indicates that dual-unitary circuits do not produce an extremal temporal entanglement growth.
	
	\section{{Temporal vs Spatial Entanglement}}
	\label{sec:temporalvsspatial}
	
\begin{figure}
		\ifdef{0}{}{\begin{tikzpicture}[scale=1.,remember picture]
				\begin{axis}[grid=major,
					legend columns=2,
					legend style={at={(0.6,0.75)},anchor=south east,font=\scriptsize,draw= none, fill=none},
					mark size=1.3pt,	
					xlabel=$t$,
					ylabel=$\max S^{(1)}_A(\gamma)$,
					y label style={at={(axis description cs:.1,.5)},anchor=south,font=\normalsize	},		
					tick label style={font=\normalsize	},	
					x label style={font=\normalsize	},	
					]
					\addplot[
					thick,
					mark=*,
					blue,
					] table[x index=0,y index=1] {plots_data/unitary_verticalstate.dat};
					\addlegendentry{p=0.37}
					\addplot[
					thick,
					mark=*,
					red,
					] table[x index=0,y index=2] {plots_data/unitary_verticalstate.dat};\addlegendentry{p=0.58}
					\addplot[
					thick,
					mark=*,
					green,
					] table[x index=0,y index=3] {plots_data/unitary_verticalstate.dat};\addlegendentry{p=0.56}
					\addplot[
					thick,
					mark=*,
					brown,
					] table[x index=0,y index=4] {plots_data/unitary_verticalstate.dat};\addlegendentry{p=0.32	}
					\addplot[
					thick,
					mark=*,
					blue,
					dashed
					] table[x index=0,y index=1	] {plots_data/unitary_spatialentanglement.dat};
					\addplot[
					thick,
					mark=*,
					red,
					dashed
					] table[x index=0,y index=2	] {plots_data/unitary_spatialentanglement.dat};
					\addplot[
					thick,
					mark=*,
					green,
					dashed
					] table[x index=0,y index=3	] {plots_data/unitary_spatialentanglement.dat};
					\addplot[
					thick,
					mark=*,
					brown,
					dashed
					] table[x index=0,y index=4	] {plots_data/unitary_spatialentanglement.dat};
				\end{axis}	
		\end{tikzpicture}}
		\includegraphics{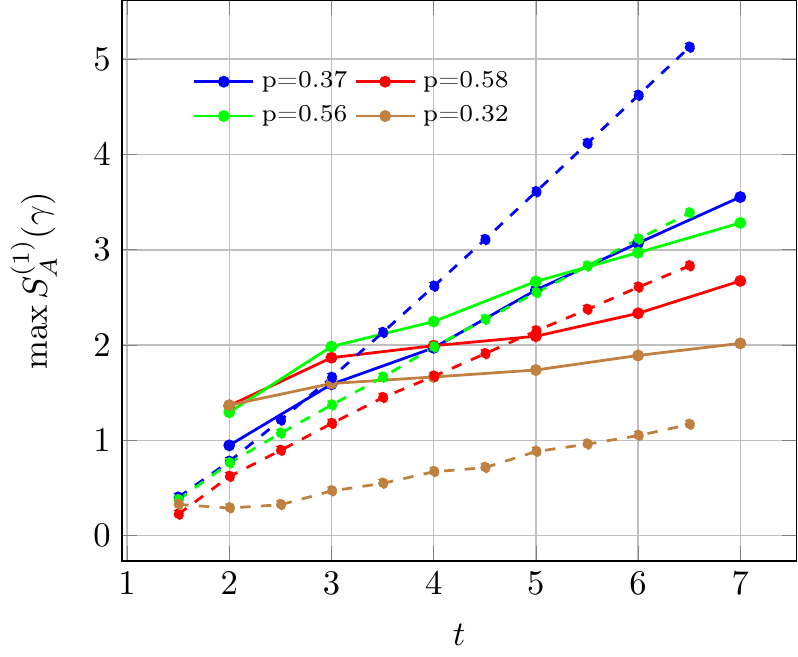}
		\caption{Comparison between the growth of spatial and temporal entanglement along the vertical path for generic unitary gates. The continuous lines report temporal entanglement of unitary gates with different entangling power (normalised such that $p\in[0,1]$) while the dashed ones the corresponding state entanglement. The plot suggests a faster asymptotic growth for the spatial entanglement.}
		\label{fig:spatialvstemporalverticalstate}
	\end{figure}

\begin{figure}\ifdef{0}{}{
			\begin{tikzpicture}[scale=1.,remember picture]
				\begin{axis}[grid=major,
					legend columns=2,
					legend style={at={(0.6,0.75)},anchor=south east,font=\scriptsize,draw= none, fill=none},
					mark size=1.3pt,	
					xlabel=$t$,
					ylabel=$\max S^{(1)}_A(\gamma)$,
					y label style={at={(axis description cs:.1,.5)},anchor=south,font=\normalsize	},		
					tick label style={font=\normalsize	},	
					x label style={font=\normalsize	},	
					]
					\addplot[
					thick,
					mark=*,
					red,
					] table[x index=0,y index=1] {plots_data/unitary_temporalentanglement_diag.dat};\addlegendentry{p=0.58}
					\addplot[
					thick,
					mark=*,
					green,
					] table[x index=0,y index=2] {plots_data/unitary_temporalentanglement_diag.dat};\addlegendentry{p=0.56}
					\addplot[
					thick,
					mark=*,
					brown,
					] table[x index=0,y index=3] {plots_data/unitary_temporalentanglement_diag.dat};\addlegendentry{p=0.32	}
					\addplot[
					thick,
					mark=*,
					orange,
					] table[x index=0,y index=4] {plots_data/unitary_temporalentanglement_diag.dat};\addlegendentry{p=0.42}
					\addplot[
					thick,
					mark=*,
					red,
					dashed
					] table[x index=0,y index=2	] {plots_data/unitary_spatialentanglement.dat};
					\addplot[
					thick,
					mark=*,
					green,
					dashed
					] table[x index=0,y index=3	] {plots_data/unitary_spatialentanglement.dat};
					\addplot[
					thick,
					mark=*,
					brown,
					dashed
					] table[x index=0,y index=4	] {plots_data/unitary_spatialentanglement.dat};
					\addplot[
					thick,
					mark=*,
					orange,
					dashed
					] table[x index=0,y index=5	] {plots_data/unitary_spatialentanglement.dat};
				\end{axis}	
		\end{tikzpicture}}
		\includegraphics{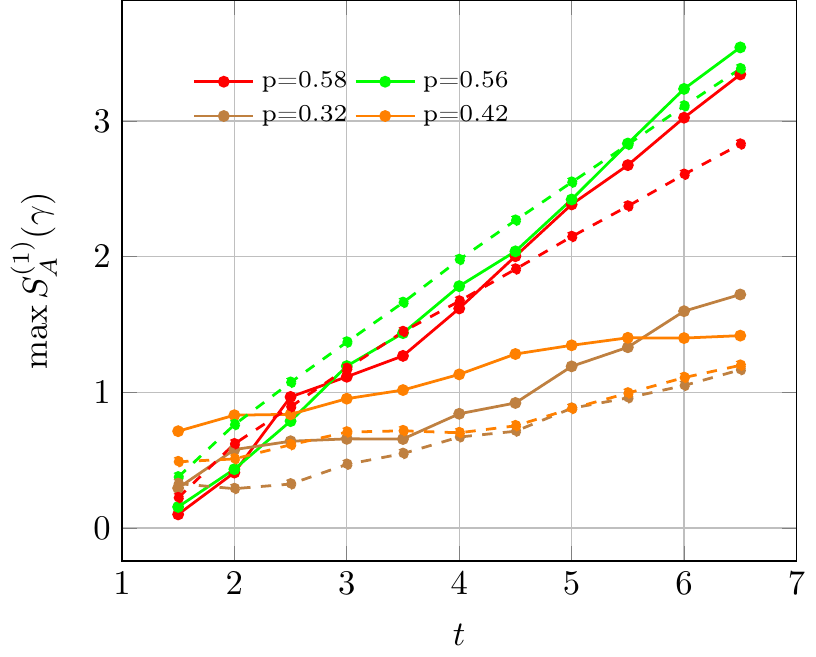}
		\caption{Comparison between the growth of spatial and temporal entanglement along the diagonal path for generic unitary gates. The continuous lines report temporal entanglement of unitary gates with different entangling power (normalised such that $p\in[0,1]$) while the dashed ones the corresponding state entanglement.}
		\label{fig:spatialvstemporaldiagonalstate}
	\end{figure}

	{ Having argued that temporal entanglement grows linearly after a quench in generic quantum circuits, the natural question is whether its growth is faster or slower than that of ``spatial entanglement'', i.e., regular state entanglement. 
		
		This question can be addressed precisely in the case of dual-unitary circuits. Indeed, for these circuits we have that state entanglement grows at the maximal possible speed for generic initial states~\cite{foligno2022growth}, i.e.
		\begin{equation}
			S_{{\rm sp}}(t) \simeq 2 t \log d.
			\label{eq:vNgrowth}
		\end{equation}
		On the other hand we can use our asymptotic result of Sec.~\ref{sec:VNentropy} to see that 
		\be
		\!\!\! S_A(\gamma ) \lesssim \max_r S(r ) t=  \frac{(1+v_{\gamma}) t}{(2+v_{\gamma})} \log d \leq  \frac{2t}{3} \log d\,,
		\label{eq:Sgammabound}
		\ee 
		where in the first step we computed the maximum of Eq.~\eqref{eq:finalresultDU} and in the second we used that it is monotonic in $v_\gamma$. 
		
		Comparing \eqref{eq:vNgrowth} and \eqref{eq:Sgammabound} we see that the temporal entanglement is lower than the spatial entanglement for every path $\gamma$. Our numerical investigations suggest that, for small enough $v_\gamma$, temporal entanglement grows slower than spatial entanglement also in generic quantum circuits. For instance, in Fig.~\ref{fig:spatialvstemporalverticalstate} we report a comparison between the entanglement of the vertical state ($v_\gamma=0$) and that of the regular time-evolving state for different times: We see that the former has a consistently smaller growth rate for all the gates considered. When the slope of the path is increased, however, the growth of temporal entanglement appears to match that of state entanglement. See for instance the comparison between spatial entanglement and temporal entanglement of the diagonal path ($v_\gamma=1$) reported in Fig.~\ref{fig:spatialvstemporaldiagonalstate}. }

	\section{Discussion}
	\label{sec:conclusions}

	In this work we studied space-like propagation approaches to quantum non-equilibrium dynamics. The main idea is to compute the time-evolution of relevant observables by exchanging the roles of space and time. For large enough volumes, the ``dual system'' living in the time direction --- also known as space-time swapped system~\cite{bertini2022growth, bertini2022nonequilibrium} --- reaches (left and right) stationary states dubbed ``influence matrices''~\cite{lerose2021influence}. Since in many-body systems characterising a stationary state is often easier and more efficient than characterising a time-evolving state at intermediate times, space-like propagation approaches are particularly promising and are attracting an increasing amount of attention~\cite{banuls2009matrix,muellerhermes2012tensor,hastings2015connecting, bertini2018exact,bertini2019exact,bertini2022entanglement, bertini2022growth, ippoliti2021postselectionfree, ippoliti2021fractal, muellerhermes2012tensor, sonner2022characterizing, frias2022light, bertini2019entanglement, piroli2020exact, klobas2021exact, klobas2021exact2, lerose2021scaling, giudice2022temporal, frias2022light, klobas2021entanglement, klobas2020space, thoenniss2022an, bertini2022nonequilibrium}.  
	
	Here we studied whether these ideas can be used to devise an efficient computational scheme to access correlation functions of local operators in generic systems. Our starting point has been the ``folding algorithm'' of Ref.~\cite{banuls2009matrix}, which uses the above idea to compute autocorrelation functions of local operators in one-dimensional quantum systems. The algorithm represents the time-evolving expectation value as a two-dimensional tensor network and proceeds by embedding the local operator in a system on the vertical time lattice which is then evolved in the space direction. To also access two-point functions between causally connected operators we generalised the folding algorithm by considering propagation in a generic space-like direction, i.e., in any direction in the two-dimensional space-time forming an angle $\alpha$ smaller than $\pi/4$ with the space direction. The idea is to consider the system on the lattice along on the time-like slice, or path, connecting the two points and evolve it in the orthogonal space-like direction.  
	
	We then investigated the efficiency of the generalised folding algorithm by computing the scaling in time of the temporal entanglement, i.e., the entanglement of the influence matrices~\cite{hastings2015connecting}. Performing a comprehensive investigation in chaotic quantum circuits we showed that for generic space-like evolutions (or states on time-like slice) the entanglement of the influence matrices grows linearly in time, preventing an efficient classical storing. { However, we also showed that the volume law scaling of temporal entanglement is much more subtle than one might expect due to the non-trivial structure of the temporal entanglement spectrum. Indeed, we found physically relevant cases where it separates into a few large Schmidt values (decaying at most polynomially in time) and many small ones (decaying exponentially). This means that the growth of temporal entanglement cannot be characterised via a replica trick.} 
	
	More specifically, we identified two cases where all temporal R\'enyi entropies with index larger than one grow sub-linearly in time: (i) standard space evolution (the one of the original folding algorithm of Ref.~\cite{banuls2009matrix}, where the time-like surface is vertical) in \emph{generic quantum circuits}; (ii) any space-like evolution in dual-unitary circuits. This phenomenon is very similar in nature to the sub-ballistic scaling of spatial R\'enyi entropies observed in circuits with diffusive conservation laws~\cite{rakovszky2019sub,huang2020dynamics}. As in the latter case the time evolving state has large overlap with a product state over a spatial bipartition of the system, in our case the influence matrices have large overlap with a product state (which we identified) over a temporal bipartition of the system. This means that the reduced density matrix has a small number of slowly decaying eigenvalues controlling the scaling of higher R\'enyi entropies.

	{ On the other hand, we showed that the von-Neumann temporal entanglement entropy grows {\it linearly} in time in both the cases (i) and (ii) but it has a strictly smaller rate of growth compared to regular state entanglement. Specifically, while for generic circuits we argued for a linear growth based on the absence of physical constraints and we characterised it numerically, for dual-unitary circuits we were able to provide a closed form expression for the slope of growth. This expression is always non-zero and smaller than the slope of growth of state entanglement. We stress that a strictly positive growth rate of temporal entanglement in dual-unitary circuits is particularly surprising because --- due to their maximally fast dephasing~\cite{piroli2020exact, kos2021thermalization} --- these systems are expected to be the chaotic system generating the lowest temporal entanglement~\cite{lerose2021influence}. We also emphasise that, to the best of our knowledge, this is the first analytical account of the non-commutativity of replica and large-time limit generating different scalings of R\'enyi entropies. } 
	
	Combined with the results of Ref.~\cite{giudice2022temporal}, our findings suggest that the behaviour of temporal entanglement after a quantum quench is a dynamical chaos indicator, i.e., it discriminates between integrable and chaotic dynamics. Indeed, while Ref.~\cite{giudice2022temporal} provided evidence for a generic sub-linear scaling of temporal entanglement in integrable models, here we showed that it grows linearly in chaotic systems (modulo some generiticity assumption on the initial state). This scenario is in agreement with the characterisation put forward in Ref.~\cite{dowling2022quantum}, which proposed volume-law spatio-temporal entanglement as the defining feature of quantum chaotic systems. From this point of view, temporal entanglement seems to behave similarly to the operator space entanglement of local operators~\cite{prosen2007is, prosen2007operator, pizorn2009operator, dubail2017entanglement} --- another conjectured dynamical chaos indicator~\cite{alba2019operator, bertini2020operator, bertini2020operator2, alba2020diffusion}.

	Our work opens several directions for future research. An obvious one is to understand whether it is possible to exploit our findings on the structure of the temporal entanglement spectrum to devise efficient computational schemes. In particular, the fact that influence matrices have a large product-state component might be used to extract information on the large time dynamics of certain special observables. 
	
	Another compelling question is to confirm our numerical observation that the von-Neumann entropy of the standard influence matrix grows linearly in time for generic circuits, but its growth is slower than that of regular state entanglement. Because of the non-commutativity of large time and replica limits this cannot be achieved by a direct application of the entanglement membrane approach. Indeed, in this case the membrane approach can only describe higher R\'enyi entropies and not von-Neumann: one cannot perform the analytic continuation. One possible strategy is to use the approach developed here for dual-unitary circuits: decompose the reduced density matrix as a convex combination and use data processing inequality and convexity of the von Neumann entropy to bound it.

	Finally, a further avenue for future research is to assess the performance of our generalised folding algorithm in non-ergodic systems, like nearly integrable ones, where the temporal entanglement grows slowly. This could provide a very efficient way to extract numerically linear transport coefficients and, more generally, characterise non-linear transport in such systems. For instance, it could be applied to the characterisation of anomalous transport in integrable systems with non-abelian charges~\cite{bulchandani2021superdiffusion}.

	\acknowledgements
	
	We thank Lorenzo Piroli, Pavel Kos, and Alessio Lerose for helpful discussions and valuable comments on the manuscript. TZ acknowledge discussion with and comments from Adam Nahum and Dmitry Abanin on the initial arxiv posting. This work has been supported by the Royal Society through the University Research Fellowship No.\ 201101 (AF and BB) and by the National Science Foundation under Grant No. NSF PHY-1748958 (BB and TZ). TZ is currently supported as a postdoctoral researcher from NTT Research Award AGMT DTD 9.24.20 and the Massachusetts Institute of Technology. We acknowledge the accommodation of the KITP program ``Quantum Many-Body Dynamics and Noisy Intermediate-Scale Quantum Systems'' in which part of the work took place. 
	
	\twocolumngrid
	\appendix
	
	\section{Minimization of the free energy term for Haar random circuit}
	\label{app:f1min_haar}
	
	{ 
		
		Here we explicitly carry out the minimisation of $F_{1,Y}(t, t_0)$ with respect to $t_0$ using the random circuit line tension in Eq.~\eqref{eq:linetensionHaar}. We recall that, setting $v_1 = -v_2$ in Eq.~\eqref{eq:F1_t_t0}, we have 
		\begin{equation}
			\!\!\!\!\!\!\!\!\!F_{1,Y}(t, r_0) \!\simeq\! 4 s_{\rm eq} t \!\left[ \mathcal{E}_H( v_1(r_0)  ) \!\!\left[\!1\!-\!r_0\!-\!\frac{r}{2}\right]\! +\! \mathcal{E}_H(0 ) \frac{r_0}{2}  \right]\!\!,
			\label{eq:F1}
		\end{equation}
		where 
		\begin{equation}
			v_1(r_0) =\frac{r v_\gamma }{2-r-2r_0} \in \left[\frac{r}{2-r-2r_0}{v_\gamma}, v_\gamma\right],
		\end{equation}
		and we use $r_0=t_0/t\in[0,r]$. The expression of the free energy in this case is
		\begin{align}
			F_{1,Y}(t,r_0) - F_2(t) &\simeq 2 s_{\rm eq} t  \biggl [( \mathcal{E}_H( v_1(r_0)  )\! -\!  \mathcal{E}_H( 0 ) ) (1- r_0) \notag \\
			&\quad+\mathcal{E}_H( v_1(r_0) )\!\!\left(1-r- \!r_0\!\right) \biggr]\!. 
			\label{eq:F1-F2}
		\end{align}
		We now solve this final minimisation using the explicit random circuit line tension $\mathcal{E}_H$. We set the $r_0$ derivative of the above expression to zero
		\begin{equation}
			\mathcal{E}_H( 0 ) - 2\mathcal{E}_H( v_1 ) + 2 v_1 \mathcal{E}'_H( v_1 ) = 0,
		\end{equation}
		and find that the equation is solved for $v_1 = v_d$ where 
		\begin{equation}
			v_d:=\frac{d-1}{\sqrt{d^2+1}}.
		\end{equation}
		The derivative $\partial_{r_0} F_{1,Y}(t,r_0)$ is negative for $v_1 \in [0,v_d)$ and positive for $v_1 \in (v_d, 1]$. So $v_1 = v_d$ is the minimal for $v_1 \in [0,1]$. However, since $v_1 \in [\frac{r}{2-r} v_{\gamma}, v_{\gamma}]$, depending on the choice of $r$, the free energy falls into three cases:  
		\begin{align}
			\label{eq:F1_F2_min}
			&\min_{r_0\in[0,r]} F_{1,Y}(t,r_0) - F_2(t) \simeq  2 s_{\rm eq}t  \\
			&\times
			\begin{cases}
				r [\mathcal{E}_H( v_\gamma) - \mathcal{E}_H(0)],  
				& v_{\gamma} \in [0,v_d),\\
				r[\mathcal{E}_H( v_d)\frac{v_\gamma}{v_d}-\mathcal{E}_H(0)\frac{v_d+v_\gamma}{2v_d}],  
				& v_{\gamma} \in [v_d , \frac{2-r}{r} v_d] \\
				(2-r)\mathcal{E}_H( \frac{r}{2-r} v_\gamma) - \mathcal{E}_H(0), &  v_{\gamma} \in (\frac{2-r}{r}v_d, 1 ]\\
			\end{cases}\,.\notag
		\end{align}
		To obtain $\bar S^{(2)}_{A}(\gamma)$, we further compare this minimum with the free energy of decoupled configurations in Fig.~\ref{fig:dw_config}, i.e., $2 s_{\rm eq} t (1-r)\mathcal{E}(0)$ from Eq.~\eqref{eq:TE_upper_bound_1}. 
		
		We first note that 
		\begin{equation}
			(2-r)\mathcal{E}_H\left(\frac{r}{2-r} v_\gamma\right) - \mathcal{E}_H(0) \ge  (1-r) \mathcal{E}_H(0), 
		\end{equation}
		due to $\mathcal{E}(v) \ge \mathcal{E}(0)$. So for the case of $v_{\gamma} \in (\frac{2-r}{r}v_d, 1 ]$ the $Y$-shaped configuration can never dominate. Physically the minimal for the $Y$ shape here corresponds to taking $r_0 = 0$, which represents two tilted sets of domain walls meeting at the very bottom. Its free energy can always be lowered if all the domain walls go down vertically.
		
		Then we compare the expression for ${v_{\gamma} \in [v_d , (2-r) v_d/r]}$ and $2 s_{\rm eq} t (1-r)\mathcal{E}(0)$. Setting
		\begin{equation}
			r[\mathcal{E}_H( v_d)\frac{v_\gamma}{v_d}-\mathcal{E}_H(0)\frac{v_d+v_\gamma}{2v_d}] \le (1-r) \mathcal{E}_H(0), 
		\end{equation}
		gives $ v_{\gamma} \le {(2-r)} v_d'/r$ where 
		\begin{equation}
			v_d' := \frac{1}{2} \frac{\log \frac{d^2+1}{2d}}{\text{arctanh}(v_d)}. 
		\end{equation}
		We have $v_d' \le v_d$ and $\lim_{d\rightarrow \infty} (v_d' - v_d) = 0$. For $v_d < ({2-r}) v_d'/r$ to hold, we require $r \le {2 v_d'}/{(v_d + v_d')}$. 
		
		Finally the case of $v_{\gamma} \in [0,v_d)$. For the $Y$-shaped configuration to dominate we should have  
		\begin{equation}
			r [\mathcal{E}_H( v_\gamma) - \mathcal{E}_H(0)] \le (1-r) \mathcal{E}_H(0).
		\end{equation}
		This requires $r \le {\mathcal{E}_H(0)}/{\mathcal{E}_H(v_\gamma)}$. 
		
		In summary, we conclude that Haar random circuits have
		\begin{equation}
			\label{eq:v_TH_Haar}
			\bar S^{(2)}_{A}(\gamma) \simeq  s_{\rm eq} v_{\rm TE, H}^{(2)} t,
		\end{equation}
		where
		\begin{align}
			&v_{\rm TE, H}^{(2)} =  \label{eq:vTEH}\\
			&\begin{cases}
				2r [\mathcal{E}_H( v_\gamma) - \mathcal{E}_H(0)]  
				& r \le \frac{\mathcal{E}_H(0)}{\mathcal{E}_H(v_\gamma)} \,\, v_{\gamma} < v_d \\
				2(1-r) \mathcal{E}_H( 0)  
				& r > \frac{\mathcal{E}_H(0)}{\mathcal{E}_H(v_\gamma)} \,\, v_{\gamma} < v_d \\
				2r[\mathcal{E}_H( v_d)\frac{v_\gamma}{v_d}-\mathcal{E}_H(0)\frac{v_d+v_\gamma}{2v_d}] 
				& r \le  \frac{2 v_d'}{v_d + v_d'} \,\,\,\, v_{\gamma} \ge v_d \\
				2 (1-r) \mathcal{E}_H( 0)  
				& r >  \frac{2 v_d'}{v_d + v_d'}\,\,\,\, v_{\gamma} \ge v_d \\
			\end{cases}.\notag
		\end{align}
	}
	
	\section{R\'enyi entropies in dual unitary circuits}\label{app:secIVAcalc}
	
	{ 
		
		In this appendix we present the detailed calculations leading to the bounds on temporal higher R\'enyi entropies discussed in Sec.~\ref{sec:boundRenyis}. 
		
		\subsection{Upper bound on temporal R\'enyi entropies for generic quantum circuits}
		\label{app:upperboundRenyi}
		
		In this subsection we bound $S^{(\alpha)}_A(\gamma)$ in terms of the norm of the state $\bra{L_\gamma}$. We begin by 
		writing the Schmidt decomposition of the state $\bra{L_{\gamma}}$ between the region $A$ and the rest $\bar A$. Namely
		\begin{equation}
			\frac{\bra{L_{\gamma}}}{\sqrt{\braket{L_\gamma}}} = \sum_{r=1}^{\min(d^{|A|},d^{|{\bar A}|})} \!\!\!\!\Lambda_r  \bra{A_r}_A\otimes\bra{B_r}_{\bar A}\,,
		\end{equation}
		where 
		$\{\ket{A_r}_A\}$ and $\{\ket{B_r}_{\bar A}\}$ are orthogonal states, while the Schmidt values $\{\Lambda_r\}$ fulfil
		\begin{align}
			& 0 \leq \Lambda_{r}\leq \cdots \leq\Lambda_{r-1},\,\sum_{r=1}^{\min(d^{|A|},d^{|{\bar A}|})} \!\!\!\Lambda^2_r =  1. 
		\end{align}
		The integer 
		\begin{equation}
			n = {\rm min}\left\{r \quad \text{s.t.}\quad \Lambda_{r}=0\right\}, 
		\end{equation}
		is referred to as the Schmidt rank of the state. 
		
		Next, we invoke Eckart-Young Theorem~\cite{eckart1936approximation} to bound from below the largest Schmidt value. To this end we first recall the statement of the theorem
		\begin{theorem}[Eckart-Young]
			\label{thm:EY}
			The scalar product of an unnormalised state $\ket{\Phi_n}$ of Schmidt rank $n$ over the bipartition $B\bar B$ and a normalised state $\ket{\Phi_k}$ with rank ${k< n}$ fulfils the following lower bound
			\begin{equation}
				\abs{\braket{\Phi_n}{\Phi_k}} \leq \sqrt{\sum_{j=1}^k \Lambda_j^2},
				\label{eq:EYthm}
			\end{equation}
			where $\{\Lambda_r\}$ are the Schmidt values of $\ket{\Phi_n}$. The state saturating the bound is unique up to a global phase and reads as
			\begin{equation}
				\ket{\Phi^*_k} = \frac{1}{\sqrt{\sum_{j=1}^k \lambda_j^2}}\sum_{r=1}^{k} \lambda_r  \ket{a_r}_B\otimes\ket{b_r}_{\bar B}\,,
			\end{equation}
			where $\{\ket{a_r}_B\}$ and $\{\ket{b_r}_{\bar B}\}$ are sets of orthogonal states. 
		\end{theorem}
		\noindent This formulation of the Eckart-Young Theorem can be directly proven using the von Neumann trace inequality~\cite{mirsky1975trace}. 
		
		Using Theorem~\ref{thm:EY} we have that the largest Schmidt value $\Lambda_1$ of any state $\ket{\Phi_n}$ fulfils 
		\begin{equation}
			\Lambda_1 \geq \abs{\braket{\Phi_n}{\Phi_1}},
		\end{equation}
		for any normalised product state $\ket{\Phi_1}$. Specialising the theorem to our case, we consider a bipartition of the $2t$ sites in $\tA$ on the top and $\tAb=2t-\tA$ on the bottom halves.
		In particular, we  fix \begin{align}
			\frac{\tA}{2t}\equiv r,
		\end{align}
		and consider the following product state in this bipartition
		\begin{align}
			\bra*{\tilde{L}}=\bra*{\mcirc}^{\otimes \tA} \otimes \bra*{L_{\gamma_{\bar{A}}}},
		\end{align}
		where $\gamma_{\bar{A}}$ is the second part of the path $\gamma$, which comprises of $\tAb$ steps.
		Using only the unitarity of the gates, it is immediate to see that the scalar product of the state with $\tA$ $\ket{\mcirc}$ states leads to a cancellation of the first $\tA$ diagonal rows. Namely
		\begin{align}
			&	\bra{L_\gamma} \left(\ket{\mcirc}^{\otimes \tA}\right)=
			\scalebox{1}{\fineq[-0.8ex][0.6][0.6]
				{ \eigenDL[-1][0][l][3]
					\tsfmatD[3][3][l][2][parttr][no]
					\tsfmatD[2][2][l][2][parttr][no]
					\foreach \i in {0,1}
					{	
						\wcirc{2.5}{4.5-\i}
					}
					\draw[white,fill=white,shift={(1.5,1.5)}] (0,0) rectangle (1,1);
					\wcirc{1.5}{5.5} 
			}}
			=\notag\\
			&=	   \scalebox{1}{	\fineq[-0.8ex][0.6][0.6]
				{	\eigenDL[0][0][l][2]
					\tsfmatD[2][2][l][1][parttr][no]
			}}=\bra*{L_{\gamma_{\bar A}}}
		\end{align}
		
		So that we find
		\begin{align}
			\Lambda_1 \geq  	\frac{\braket*{{L_\gamma}}{\tilde{L}}}{\sqrt{\braket*{L_\gamma} \braket*{\tilde L}}}=\sqrt{\frac{\braket*{L_{\gamma_{\bar A}}}}{\braket*{L_\gamma}}}.
		\end{align}
		This gives 
		\begin{equation}
			\max_A S_A^{(\infty)}(\gamma) = - \log \Lambda_1^2 \le \log \frac{\braket{L_{\gamma_{\bar A}}}}{\braket{L_\gamma}}.
			\label{eq:Sinfinitybound}
		\end{equation}
		Next, we use the known inequality~\cite{wilming2019entanglement}
		\begin{equation}
			S^{(\alpha)}(\rho) \leq \frac{\alpha}{\alpha-1} S^{(\infty)}(\rho), \qquad \alpha>1,
			\label{eq:higherRenyibound}
		\end{equation}
		fulfilled by the function in Eq.~\eqref{eq:Sfun}, to obtain Eq.~\eqref{eq:boundRenyigeneral}. 
	}

	{ 
		\subsection{Norm of $\braket{L_\gamma}$ for dual unitary circuits}\label{app:normIMstatesDU}
		
		Here we compute $\braket{L_\gamma}$ in the special case of dual-unitary circuits. Using the dual unitarity relations for double gates 
		\begin{align}
			\fineq[-0.8ex][.85][1]{
				\tsfmatV[0][0][r][1][][][bertiniorange][topright]
				\tsfmatV[1.25][0][r][1][][][bertinigreen][bottomleft]
				\draw[ thick] (.5,1.5) -- (.75,1.5);
				\draw[ thick] (.5,0.5) -- (.75,0.5);
			}
			= 
			\fineq[-0.8ex][.85][1]{
				\tsfmatV[0][0][r][1][][][bertinigreen][topright]
				\tsfmatV[1.25][0][r][1][][][bertiniorange][bottomleft]
				\draw[ thick] (.5,1.5) -- (.75,1.5);
				\draw[ thick] (.5,0.5) -- (.75,0.5);
			}
			=
			\fineq[-0.8ex][.85][1]{
				\draw[ thick] (0,1.5) -- (1,1.5);
				\draw[ thick] (0,0.5) -- (1,0.5);
			}\,,
			\label{eq:DUfolded}
		\end{align}
		one can easily show that 
		\begin{align}
			&\braket{L_{\gamma}} = \notag\\
			&=\!\!\fineq[-0.8ex][0.5][0.5]
			{    \eigenDR[9][0][r][3][bertinigreen][bottomleft]        
				\eigenDL[0][0][l][3]
				\tsfmatD[3][2][l][2][parttr][no]	
				\tsfmatD[4][3][l][2][parttr][no]	
				\tsfmatD[6][2][r][2][parttr][no][bertinigreen][bottomleft]
				\tsfmatD[5][3][r][2][parttr][no][bertinigreen][bottomleft]
				\foreach \i in {0,3,4}
				{\draw[thick,shift={(1,0)}] (2.48,.48+\i)--(4.52,.48+\i);}
				\foreach \i in {1,2,5}
				{\draw[thick,shift={(1,0)}] (1.48,.48+\i)--(5.52,.48+\i);}
			}\notag\\
			&=\fineq[-0.8ex][0.5][0.5]
			{  \eigenDR[8][0][r][3][bertinigreen][bottomleft]        
				\eigenDL[0][0][l][3]
				\foreach \i in {0,...,3}
				{
					\draw[thick] (3.48-\i,0.5+\i)--(4.5+\i,0.5+\i);
				}
			}\notag\\
			& = \braket{L_{\gamma_{\rm lc}}}:= \mathcal{N}_{\ttot},
			\label{eq:norm_diagstate}
		\end{align} 
		where we introduced the diagonal path (cf.~\eqref{eq:lightconepath})
		\begin{equation}
			\gamma_{\rm lc}=\{+,+,\ldots,+\},
			\label{eq:gammastar}
		\end{equation}
		with length $\tau =({1+v_\gamma})t$ and denoted by $\mathcal{N}_{\ttot}$ the norm of $\bra*{L_{\gamma_{\rm lc}}}$. 
		
		We now observe that the latter quantity is directly related to spatial entanglement. Indeed, computing the purity of the regular density matrix $\rho_A(t)$ (cf.~\eqref{eq:DMfolded}) and choosing $A=[t-x,\infty]$ we find 
		\begin{align}
			\label{eq:state_purity_N}
			\mathcal P(t)=\frac{1}{d^{2t}}\mathcal{N}_{2t}.
		\end{align}
		Combining this equation with \eqref{eq:norm_diagstate}, \eqref{eq:Sinfinitybound}, and \eqref{eq:higherRenyibound} we recover the bound in Eq.~\eqref{eq:DUleftminent}.  
		
	}

	\section{Linear growth of temporal entanglement entropy in dual unitary circuits}
	\label{app:secIVBcalc}

	{ 
		In this appendix we present the detailed calculations leading to the bound on temporal entanglement entropy discussed in Sec.~\ref{sec:VNentropy}.

		\subsection{Reduction}
		\label{eq:proofboundLstar}
		
		Consider a generic bipartition of a state $L_\gamma$, $\gamma=\gamma_A\circ\gamma_{\bar A}$ 
		
		\begin{equation}
			\bra*{L_{\gamma}}=\!\!\!\fineq[-0.8ex][0.5][0.5]{
				\eigenDL[0][0][l][4]
				\tsfmatD[5][1][l][4][parttr][no]		
				\tsfmatD[5][3][l][3][parttr][no]		
				\tsfmatD[6][4][l][3][parttr][no]		
				\tsfmatD[5][7][l][1][parttr][no]
				\node[scale=2,red] at (6,2) {${\gamma_{\bar{A}}}$}; 
				\node[scale=2, blue] at (6,7) {$\gamma_A$};		
				\draw[line width=0.75mm, blue, rounded corners, stealth-,shift={(1,0)}] (4,4)--(5,5)--(3,7)--(4,8)--(3.5,8.5);				     \draw[line width=0.75mm, red, rounded corners, stealth-,shift={(1,0)}] (4,0)--(3,1)--(4,2)--(3,3)--(4,4);
			}.
			\label{eq:Lgamma2}
		\end{equation}
		Now we observe that, since the entanglement is invariant under local unitary transformations, the entanglement between $A$ and and $\bar A$ is not changed by the transformation 
		\begin{equation}
			\bra{L_\gamma}\quad\mapsto\quad  \bra{L_\gamma} (U^\dag_A\otimes U^\dag_{\bar{A}}), 
			\label{eq:LtoLprime}
		\end{equation}
		for any unitary matrices $U_A$ and $U_{\bar A}$ acting respectively only in $A$ and $\bar A$. We consider the transformations $U_A$ and $U_{\bar{A}}$ removing the largest number of gates; in the example shown in Eq.~\eqref{eq:Lgamma2}, this corresponds to:
		\begin{equation}
			\left[ U_{\bar A}\right]_{i,j}=\fineq[-0.8ex][0.5][0.5]{\tsfmatD[0][0][l][1][no][no]
				\node[scale=1.5] at (0.,1.5) {$j_2$};
				\node[scale=1.5] at (0.,.3) {$j_3$};		\node[scale=1.5] at (0.,2.5) {$j_1$};
				\node[scale=1.5] at (0.,-.7) {$j_4$};
				\node[scale=1.5] at (-1.8,1.5) {$i_2$};
				\node[scale=1.5] at (-1.8,.3) {$i_3$};
				\node[scale=1.5] at (-1.8,2.5) {$i_1$};
				\node[scale=1.5] at (-1.8,-.7) {$i_4$};
				\draw[thick] (-1.5,2.5)--(-.5,2.5);			\draw[thick] (-1.5,-.5)--(-.5,-.5);	
			}
			\qquad \left[U_{{A}}\right]_{i,j}=\fineq[-0.8ex][0.5][0.5]{\tsfmatD[0][0][l][3][no][no]\tsfmatD[-1][3][l][1][no][no]
				\node[scale=1.5] at (-1.8,.3) {$i_5$};
				\node[scale=1.5] at (-2.8,1.3) {$i_4$};
				\node[scale=1.5] at (-3.8,2.3) {$i_3$};
				\node[scale=1.5] at (-3.8,3.5) {$i_2$};
				\node[scale=1.5] at (-2.8,4.5) {$i_1$};
				\node[scale=1.5] at (-.2,.3) {$j_5$};
				\node[scale=1.5] at (-.2,1.3) {$j_4$};
				\node[scale=1.5] at (-1.2,2.3) {$j_3$};
				\node[scale=1.5] at (-1.2,3.5) {$j_2$};
				\node[scale=1.5] at (-1.2,4.5) {$j_1$};
			},
			\label{eq:Us} 
		\end{equation}
		where $i_a/j_a$ correspond to the $a$-th digit of $i/j$ in base $d^2$.
		The corresponding $\bra{L_{\gamma'}}$ state has  the following form \begin{equation}
			\bra*{L_{\gamma'}}=\fineq[-0.8ex][0.5][0.5]{
				\eigenDL[0][0][l][4]
				\tsfmatD[4][2][l][3][parttr][no]		
				\tsfmatD[5][3][l][3][parttr][no]		
				\node[scale=2,red] at (6,2) {$\gamma'_{\bar{A}}$}; 
				\node[scale=2, blue] at (6,7) {$\gamma'_A$};		
				\draw[line width=0.75mm, blue, rounded corners, stealth-,shift={(1,0)}] (4,4)--(1.5,6.5)--(3.2,8.2);				     \draw[line width=0.75mm, red, rounded corners, stealth-,shift={(1,0)}] (4,0)--(2,2)--(4,4);
				\foreach \i in {1,2}
				{
					\draw[thick] (1.5+\i,6.5+\i)--(1.8+\i,6.2+\i);
					\wcirc{1.5+\i}{6.5+\i}
				}
			},
			\label{eq:unitaryequivalentstate}
		\end{equation}
		where we highlighted the new paths $\gamma'_A$ and $\gamma'_{\bar A}$ forming the edge of $\bra*{L_{\gamma'}}$. 
		This new state has now effectively $\tA=|A| {(1+v_\gamma)}/{2}$ sites in the bipartition $|A|$, since the remaining product bullet states are disentangled with the rest.

		\subsection{Evaluation of $p_k$}
		\label{app:pkevaluation}
		Let us evaluate $\mel{\gamma'}{P_k}{L_{\gamma'}}$ in order to compute Eq.\eqref{eq:pkdef}. We are considering states $\bra{L_{\gamma'}}$ as the one shown in \eqref{eq:unitaryequivalentstate}, corresponding to a path (we ignore the bullet states disentangled from the rest) 
		\begin{align}
			\!\!\!\!\!\gamma'=\underbrace{\{+,+,\ldots,+\}}_{\tA}\circ\underbrace{\{-,-,\ldots,-\}}_{|\bar A|(1-v_{\gamma_{\bar A}})/2}\circ\underbrace{\{+,+,\ldots,+\}}_{|\bar A| (1+v_{\gamma_{\bar A}})/{2}},\label{eq:gammaprimedef}
		\end{align} of total length $\tA+|\bar A|$ .
		
		Graphically, it's easy to see that, using the unitarity  of the gates, any scalar product of the type $\bra{L_\gamma} \left(\ket{\mcirc}^{\otimes x}\right)$, where the $x$ bullet states are applied from the top, deletes the first $x$ main diagonal of the state $\bra{L_\gamma}$:\begin{align}
			\bra*{L_\gamma}\left( \ket{\mcirc}^{\otimes x}\right)=\fineq[-0.8ex][0.5][0.5]{
				\eigenDL[0][0][l][4]
				\tsfmatD[4][2][l][3][parttr][no]		
				\tsfmatD[5][3][l][3][parttr][no]
				\foreach \i in {1,2}
				{
					\draw[thick] (2+\i,8-\i)--(1.7+\i,7.7-\i);
					\wcirc{2+\i}{8-\i}
				}
				\draw [decorate, decoration = {brace}, thick,shift={(0.3,0.3)}]   (2+1,8-1)--(2+2,8-2);	
			}		 			=\\
			\fineq[-0.8ex][0.5][0.5]{
				\eigenDL[0][0][l][3]
				\tsfmatD[2][3][l][1][parttr][no]		
				\tsfmatD[3][4][l][1][parttr][no]
			}	=\bra{L_{\gamma\setminus x}},
			\label{eq:example1}
		\end{align}
		where with ${\gamma\setminus x}$ we indicate the path $\gamma$, where the first $x$ jumps have been deleted. 
		Using the definition of the projectors $P_k$ in Eq.\eqref{eq:projdefinition},
		we then find 
		\begin{align}
			&\mel{L_\gamma}{P_k}{L_\gamma}=\\
			&=\braket{L_{\gamma'\setminus \tA-k}}-\braket{L_{\gamma'\setminus (\tA-k+1)}},\notag
		\end{align}
		for $k>0$, and
		\begin{align}
			\mel{L_\gamma}{P_0}{L_\gamma}=
			\braket{L_{\gamma'\setminus \tA}}.
		\end{align}
		
		Finally, using equation \eqref{eq:norm_diagstate} and the shape of $\gamma'$ in \eqref{eq:gammaprimedef}, we find
		\begin{align}
			&	\mel{L_{\gamma'}}{P_k}{L_{\gamma'}}=\mathcal{N}_{\abs*{\bar A}{(1+v_{\gamma_{\bar{A}}})}/{2}+k}-\mathcal{N}_{\abs*{\bar A}{(1+v_{\gamma_{\bar{A}}})}/{2}+k-1}\notag\\
			&	\mel{L_\gamma}{P_0}{L_\gamma}=\mathcal{N}_{\abs*{\bar A}{(1+v_{\gamma_{\bar{A}}})}/{2}}.
		\end{align}
		Using again equation \eqref{eq:norm_diagstate}, we also have
		\begin{align}
			\braket{L_{\gamma'}}=\mathcal{N}_{(1+v_\gamma)t},
		\end{align}
		where we used the fact that \begin{align}
			\abs{A}\frac{1+v_{\gamma_A}}{2}+\abs{\bar A}\frac{1+v_{\gamma_{\bar A}}}{2}=\frac{1+v_\gamma}{2} 2t.
		\end{align}
		Finally, using Assumption \ref{asp:1}, we find the asymptotic scaling
		\begin{align}
			& \mel{L_{\gamma'}}{P_k}{L_{\gamma'}}\sim C >0, \\	
			&  \mel{L_{\gamma'}}{P_0}{L_{\gamma'}}\sim C \abs{\bar A}\frac{1+v_{\bar{A}}}{2} >0, \\
			& \braket{L_{\gamma'}}\sim C{(1+v)t}, 
		\end{align}
		which, put back in the definition of $p_k$ in \eqref{eq:pkdef}, gives
		\begin{equation}
			p_k = 
			\left\lbrace
			\begin{aligned}
				& \frac{1}{(1+v_\gamma)t} & \quad k \ne 0 \\
				& \frac{|\bar{A}|(1+v_{\gamma_{\bar A}})}{2(1+v_\gamma)t} & \quad k = 0 \\
			\end{aligned} \right. \label{eq:pkexpr2},
		\end{equation}
		in the main text we considered a constant local slope for the path $\gamma$, after coarse graining, which means substituting $v_{\gamma_A} =v_{\gamma_{\bar{A}}}=v_{\gamma}$ in Eq.\eqref{eq:pkexpr2}, obtaining Eq.\eqref{eq:pkexpr}.
		
		\subsection{Entropy of the state $|L'_k \rangle$ via membrane approach}
		\label{app:renyigrowth}
		
		In this appendix we use the entanglement membrane approach to compute the second R\'enyi entropy of the state in Eq.~\eqref{eq:k-state} which we call $\bra{L_k}$ and repeat the expression here: 
		\begin{equation}
			\bra{L_k}= \fineq[-0.8ex][1][1]{
				\draw (0,0)--(2.5,0)--++(-1.25/2,1.25/2)--++(1-2.5/4,1-2.5/4)--++(-1.25/2,1.25/2) --cycle;
				\foreach \x in {1,1.5}{
					\node () at (\x,-0.15) {$\cdots$};	
				}
				\foreach \x in {0,0.5,2,2.5}{
					\draw (\x-0.15,0) arc (180:360:0.15);
					\draw[fill] (\x,-0.15) circle (0.03);
				}
				\foreach \x in {0.05,0.25,1.3,1.5}{
					\draw (\x,\x)--++(0,0.2);
					\draw (\x,\x+0.25) circle (0.05);
				}
				\draw (-0.15,0)--++(0,0.05);
				\draw (-0.15,-0.15+0.25) circle (0.05);
				\foreach \x in {0.5,0.8,...,1.3}{
					\node () at (\x,1.03*\x+0.15) {$\udots$};	
				}
				\node[right] () at (2.65,0.5) {$\bar{A}$};
				\draw [decorate, decoration = {brace},shift={(0.15,0.2)}]   (2.5,0.5+0.2)  --++ (0,-0.8);
				\draw [decorate,decoration={brace,amplitude=5pt,raise=0pt,mirror},shift={(0.4,0.4)}]
				(2.5-1.25/2+1-2.5/4,1.25/2+1-2.5/4) -- ++(-1.25/2,1.25/2) node [black,midway,xshift=10pt,yshift=10pt] {$k$};
				\foreach \x in {1.3,1.1}{
					\draw (\x+0.05/1.414,3.25-\x+0.05/1.414)--++(0.2/1.414,0.2/1.414);
					\draw (\x,3.25-\x) circle (0.05);
				}
				\foreach \x in {1.9,2.1}{
					\draw (\x+0.05/1.414-0.08,3.25-\x+0.05/1.414)--++(0.2/1.414,0.2/1.414);
				}
				
				\draw (1.7+0.05/1.414-0.08,3.25-1.7+0.05/1.414)--++(0.2*1.414,0.2*1.414);
				\draw[fill=black,rotate around={45:(1.7+0.05/1.414-0.08+0.1*1.414,3.25-1.7+0.05/1.414+0.1*1.414)}]
				(1.7+0.05/1.414-0.08+0.05*1.414,3.25-1.7+0.05/1.414+0.05*1) rectangle (1.7+0.05/1.414-0.08+0.15*1.414,3.25-1.7+0.05/1.414+0.15*1.414);
				
				\foreach \x in {2.9, 3.05,3.2}{
					\draw (\x+0.05/1.414-0.8,3.25-\x+0.05/1.414)--++(0.2/1.414,0.2/1.414);
				}
				
				\foreach \x in {0,0.15,0.3}{
					\draw (2.6+\x+0.05/1.414-0.7,3.25-2.6+\x+0.05/1.414)--++(0.2/1.414,-0.2/1.414);
				}
			}.
		\end{equation}
		The projector (black box) is $\I_{d^2} - \ketbra{\mcirc}$. By using the dual unitary property, the action of the $\ketbra{\mcirc}$ is equivalent as replacing the left boundary state via a solvable EPR state: 
		\begin{equation}
			\label{eq:project_out}
			\fineq[-0.8ex][1][1]{
				\draw (0,0)--(2.5,0)--++(-1.25/2,1.25/2)--++(1-2.5/4,1-2.5/4)--++(-1.25/2,1.25/2) --cycle;
				\foreach \x in {1,1.5}{
					\node () at (\x,-0.15) {$\cdots$};	
				}
				\foreach \x in {0,0.5,2,2.5}{
					\draw (\x-0.15,0) arc (180:360:0.15);
					\draw[fill] (\x,-0.15) circle (0.03);
				}
				\foreach \x in {0.05,0.25,1.3,1.5}{
					\draw (\x,\x)--++(0,0.2);
					\draw (\x,\x+0.25) circle (0.05);
				}
				\draw (-0.15,0)--++(0,0.05);
				\draw (-0.15,-0.15+0.25) circle (0.05);
				\foreach \x in {0.5,0.8,...,1.3}{
					\node () at (\x,1.03*\x+0.15) {$\udots$};	
				}
				\node[right] () at (2.65,0.5) {$\bar{A}$};
				\draw [decorate, decoration = {brace},shift={(0.15,0.2)}]   (2.5,0.5+0.2)  --++ (0,-0.8);
				\draw [decorate,decoration={brace,amplitude=5pt,raise=0pt,mirror},shift={(0.5,0.5)}]
				(2.5-1.25/2+1-2.5/4,1.25/2+1-2.5/4) -- ++(-1.25/2,1.25/2) node [black,midway,xshift=10pt,yshift=10pt] {$k$};
				\draw (2,1.95)--++(0.2/1.414,0.2/1.414);
				\draw (1.95,1.9) circle (0.05);
				\foreach \x in {1.3,1.1}{
					\draw (\x+0.05/1.414,3.25-\x+0.05/1.414)--++(0.2/1.414,0.2/1.414);
					\draw (\x,3.25-\x) circle (0.05);
				}
				\foreach \x in {1.7,1.9,2.1}{
					\draw (\x+0.05/1.414-0.08,3.25-\x+0.05/1.414)--++(0.2/1.414,0.2/1.414);
				}
				\draw (1.82,1.75) circle (0.05);
				
				\foreach \x in {2.9, 3.05,3.2}{
					\draw (\x+0.05/1.414-0.8,3.25-\x+0.05/1.414)--++(0.2/1.414,0.2/1.414);
				}
				
				\foreach \x in {0,0.15,0.3}{
					\draw (2.6+\x+0.05/1.414-0.7,3.25-2.6+\x+0.05/1.414)--++(0.2/1.414,-0.2/1.414);
				}
			}
			=
			\fineq[-0.8ex][1][1]{
				\draw (0,0)--(2.5,0)--++(-1.25/2,1.25/2)--++(1-2.5/4,1-2.5/4)--++(-1.25/2,1.25/2) --cycle;
				\foreach \x in {1,1.5}{
					\node () at (\x,-0.15) {$\cdots$};	
				}
				\draw (-0.15,0) arc (180:360:0.15);
				\foreach \x in {0.5,2,2.5}{
					\draw (\x-0.15,0) arc (180:360:0.15);
					\draw[fill] (\x,-0.15) circle (0.03);
				}
				\foreach \x in {0.05,0.25,1.3,1.5}{
					\draw (\x,\x)--++(0,0.2);
					\draw (\x,\x+0.25) circle (0.05);
				}
				\draw (-0.15,0)--++(0,0.05);
				\draw (-0.15,-0.15+0.25) circle (0.05);
				\foreach \x in {0.5,0.8,...,1.3}{
					\node () at (\x,1.03*\x+0.15) {$\udots$};	
				}
				\node[right] () at (2.65,0.5) {$\bar{A}$};
				\draw [decorate, decoration = {brace},shift={(0.15,0.2)}]   (2.5,0.5+0.2)  --++ (0,-0.8);
				\draw [decorate,decoration={brace,amplitude=5pt,raise=0pt,mirror},shift={(0.2,0.2)}]
				(2.5-1.25/2+1-2.5/4,1.25/2+1-2.5/4) -- ++(-1.25/2,1.25/2) node [black,midway,xshift=10pt,yshift=10pt] {$k$};
				\foreach \x in {1.3,1.1}{
					\draw (\x+0.05/1.414,3.25-\x+0.05/1.414)--++(0.2/1.414,0.2/1.414);
					\draw (\x,3.25-\x) circle (0.05);
				}
				\foreach \x in {1.7,1.9,2.1}{
					\draw (\x+0.05/1.414-0.08,3.25-\x+0.05/1.414)--++(0.2/1.414,0.2/1.414);
				}
				
				\foreach \x in {2.9, 3.05,3.2}{
					\draw (\x+0.05/1.414-0.8,3.25-\x+0.05/1.414)--++(0.2/1.414,0.2/1.414);
				}
				
				\foreach \x in {0,0.15,0.3}{
					\draw (2.6+\x+0.05/1.414-0.7,3.25-2.6+\x+0.05/1.414)--++(0.2/1.414,-0.2/1.414);
				}
			}
			.
		\end{equation}
		Therefore this is the part to be projected out. We decompose the initial state on the bottom left of the diagram into a component of the solvable state (EPR state) and a remainder term (box state): 
		\begin{equation}
			\fineq[-0.8ex][2][0.5]{
				\draw (-0.15,0) arc (180:360:0.15);
				\draw[fill] (0,-0.15) circle (0.03);
			}
			= 
			\frac{1}{d}
			\fineq[-0.8ex][2][0.5]{
				\draw (-0.15,0) arc (180:360:0.15);
			}
			+ 
			\fineq[-0.8ex][2][0.5]{
				\draw (-0.15,0) arc (180:360:0.15);
				\draw[fill] (0-0.05,-0.15-0.05) rectangle (0+0.05,-0.15+0.05);
			}. 
		\end{equation}
		From Eq.~\eqref{eq:project_out}, $\bra{L_k}$ can be simplied to
		\begin{equation}
			\bra{L_k} 	 = 
			\fineq[-0.8ex][1][1]{
				\draw (0,0)--(2.5,0)--++(-1.25/2,1.25/2)--++(1-2.5/4,1-2.5/4)--++(-1.25/2,1.25/2) --cycle;
				\foreach \x in {1,1.5}{
					\node () at (\x,-0.15) {$\cdots$};	
				}
				\draw (-0.15,0) arc (180:360:0.15);
				\draw[fill] (0-0.05,-0.15-0.05) rectangle (0+0.05,-0.15+0.05);
				\foreach \x in {0.5,2,2.5}{
					\draw (\x-0.15,0) arc (180:360:0.15);
					\draw[fill] (\x,-0.15) circle (0.03);
				}
				\foreach \x in {0.05,0.25,1.3,1.5}{
					\draw (\x,\x)--++(0,0.2);
					\draw (\x,\x+0.25) circle (0.05);
				}
				\draw (-0.15,0)--++(0,0.05);
				\draw (-0.15,-0.15+0.25) circle (0.05);
				\foreach \x in {0.5,0.8,...,1.3}{
					\node () at (\x,1.03*\x+0.15) {$\udots$};	
				}
				\node[right] () at (2,0.5+0.2) {$\bar{A}$};
				\draw [decorate,decoration={brace,amplitude=5pt,raise=0pt,mirror},yshift=0pt]
				(2.5-1.25/2+1-2.5/4,1.25/2+1-2.5/4) -- ++(-1.25/2,1.25/2) node [black,midway,xshift=10pt,yshift=10pt] {$k$};
				\foreach \x in {1.3,1.1}{
					\draw (\x+0.05/1.414,3.25-\x+0.05/1.414)--++(0.2/1.414,0.2/1.414);
					\draw (\x,3.25-\x) circle (0.05);
				} 
			}.
		\end{equation}
		We then closely follow the discussion of Sec.~\ref{subsec:ev_analysis}. The evaluation follow exactly as in \eqref{eq:S_op_rho_A} when the state $\bra{L_{\gamma,x}}$ is replaced by $\bra{L_k }$, namely
		\begin{align}
			S^{(2)}(|L_k'\rangle) = F_1(x)-F_2(x), 
			\label{eq:Renyi2sigma}
		\end{align}
		with 
		\begin{equation}
			F_1(x)= - \log(\tr_{A}  ( \tr_{\bar A}( \ketbra*{L'_k} )^2)),
			\label{eq:appDUF1}
		\end{equation}
		and 
		\begin{equation}
			F_2(x)= - 2 \log \tr( \ketbra*{L'_k} )\,.
		\end{equation}
		Here $A$ is the subsystem formed by the top $\tA$ sites (the top $\tA - k$ sites are decoupled product states though) and, as in \eqref{eq:S_op_rho_A}. 
		
		The calculation follows the same lines as the one outlined in Sec.~\ref{subsec:ev_analysis} with one main difference: since the square state
		\begin{equation}
			\fineq[-0.8ex][2][0.5]{
				\draw (-0.15,0) arc (180:360:0.15);
				\draw[fill] (0-0.05,-0.15-0.05) rectangle (0+0.05,-0.15+0.05);
			}
		\end{equation}
		in the bottom left corner of $|L_k'\rangle$ is orthogonal to the loop state $\ket{\mcirc}$, its four-fold copy is orthogonal to the identity permutation in $S_4$. This means that the optimal domain wall configurations are not those reported in Fig.~\ref{fig:dw_config} but, instead, look like those reported in Fig.~\ref{fig:dw_L_tilde}. This gives 
		
		\begin{equation}
			F_1(x) = F_2(x) + 2 \min(k,|A|) \log d
		\end{equation}
		where we used that for dual-unitary circuits the line tension is equal to one. Plugging in \eqref{eq:Renyi2sigma} and using the monotonicity in $\alpha$ of the R\'enyi entropies we arrive at Eq.~\eqref{eq:membranepicturelowerb}. 
		
		In Fig.~\ref{plot:sigmaentropy}, we checked the validity of equation \eqref{eq:membranepicturelowerb}, by comparing the maximum value of $S^{(1)}(\rho_k)$ as a function of $\ell\equiv k+\abs{\bar{A}}$ (i.e. the number of sites definining the corresponding Hilbert space) for $v_\gamma=1$. From the membrane theory (cf. Eq \eqref{eq:membranepicturelowerb}) we expect an asymptotic growth of the peak to be equal to 
		\begin{align}
			\sup_{k, k+\abs{\bar A}={\rm const}}\!\!\!\!\!\!\!\!\!\!\! \min(k,\abs{\bar A})\log(d^2)&=\frac{k+\abs{\bar A}}{2}\log(d^2)\notag\\
			&=\ell\log(d).\label{eq:maximumentropygrowth}
		\end{align}		
		This prediction agrees with the data for the higher values of the entanglement power $p$. For lower values of the entangling power we expect the asymptotic form to arise at larger system sizes.
		The growth rate of the sup in Eq. \eqref{eq:maximumentropygrowth} is actually a necessary and sufficient condition the validity of Eq. \eqref{eq:membranepicturelowerb}.
		This is because, using well known properties of the entanglement entropy (for example, the positivity of the mutual information), one has:\begin{align}
			\abs{S(\rho^{(\ell)}_k)-S(\rho^{(\ell)}_{k-1})}\le \log(d^2)
			\label{eq:mutualinfobound},
		\end{align}
		where we stressed for clarity that here we consider a density matrix with $\ell$ sites and a bipartition (on which entanglement is defined) with $k, \ell-k$ sites.
		Combining Eq. \eqref{eq:mutualinfobound} with  Eq. \eqref{eq:maximumentropygrowth}, one must have 
		\begin{align}
			S(\rho^\ell_{k})=\min(k,\ell-k) \log(d^2) + o(k) + o(\ell),
		\end{align}
		which is indeed Eq. \eqref{eq:membranepicturelowerb}.
		\begin{figure}
			
	\ifdef{0}{}		{	\begin{tikzpicture}[scale=1,remember picture]
					\begin{axis}[
						grid=major,
						legend columns=2,
						legend style={at={(0.7,0.7)},anchor=south east,font=\scriptsize,draw= none, fill=none},
						mark size=1.3pt,	
						xlabel=$\tau$,
						ylabel=$\max_k S^{(1)}(\rho_k)$,
						y label style={at={(axis description cs:.1,.5)},anchor=south,font=\normalsize	},		
						tick label style={font=\normalsize	},	
						x label style={font=\normalsize	},	
						]
						\addplot[
						smooth,
						thick,
						mark=*,
						blue,
						dashed
						] table[x index=0,y index=1] {plots_data/sigmastate_S.dat};
						\addlegendentry{p=0.27}
						\addplot[
						smooth,
						thick,
						mark=*,
						orange,
						dashed
						] table[x index=0,y index=2] {plots_data/sigmastate_S.dat};
						\addlegendentry{p=0.33}	
						\addplot[
						smooth,
						thick,
						mark=*,
						green,
						dashed
						] table[x index=0,y index=3] {plots_data/sigmastate_S.dat};
						\addlegendentry{p=0.40}	
						\addplot[
						smooth,
						thick,
						mark=*,
						red,
						dashed
						] table[x index=0,y index=4] {plots_data/sigmastate_S.dat};
						\addlegendentry{p=0.47}	
						\addplot[
						smooth,
						thick,
						mark=*,
						purple,
						dashed
						] table[x index=0,y index=5] {plots_data/sigmastate_S.dat};
						\addlegendentry{p=0.53}	\addplot[
						smooth,
						thick,
						mark=*,
						brown,
						dashed
						] table[x index=0,y index=6] {plots_data/sigmastate_S.dat};
						\addlegendentry{p=0.60}
						\addplot[
						smooth,
						thick,
						mark=*,
						pink,
						dashed
						] table[x index=0,y index=7] {plots_data/sigmastate_S.dat};
						\addlegendentry{p=0.67}
						\addplot[dashed,domain=6:12,very thick]{x * ln(2)-2};
						\addlegendentry{$x \log(2)+\text{const}$}
					\end{axis}	
			\end{tikzpicture}}\includegraphics{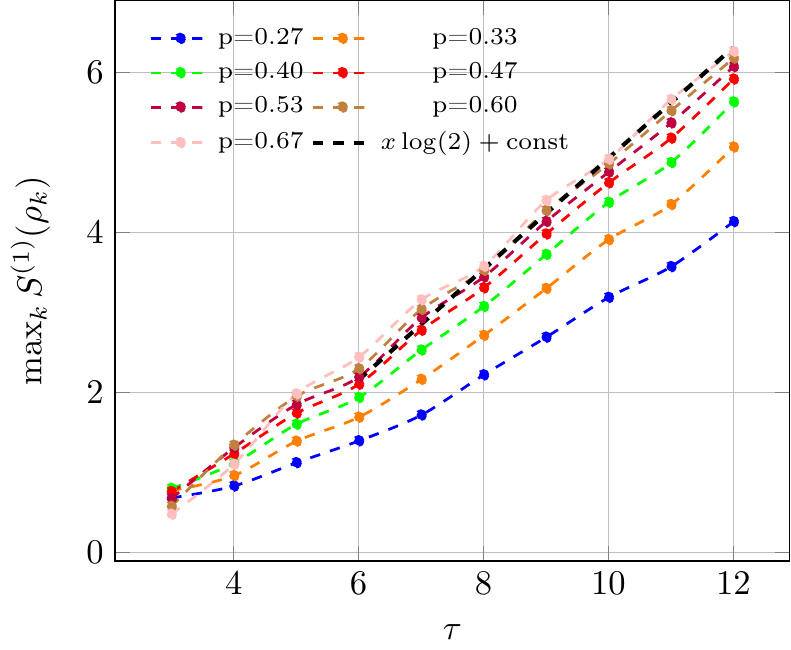}
			\caption{Entanglement entropy for the matrix $\rho_{k}$, obtained by maximizing the value over all possible choices of $k$, keeping $\tau=\abs{\bar A}+k$ fixed, for generic dual unitary gates of different entangling power, local Hilbert space dimension $d=2$, and $v_\gamma=1$.}
			\label{plot:sigmaentropy}
		\end{figure}
		
	}
	
	\begin{figure}
		\includegraphics[width=0.9\columnwidth]{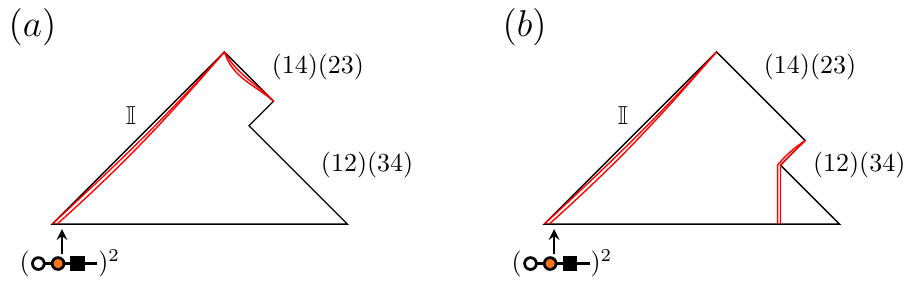}
		\caption{Domain wall configuration giving the leading contribution to Eq.~\eqref{eq:appDUF1}. (a) When $k < |A|$, the two domain walls $(14)(23)$ split into four domain walls $(12)(34) \times (13)(24)$. Two of them go to the left to contract with the orthogonal states. The other two $(12)(34)$ go to the right and cancel with the two domain walls at the interactions of $A$ and $\bar{A}$. (b) The domain walls at the tip do not split. They go to the left to contract with the orthogonal states.} 
		\label{fig:dw_L_tilde}
	\end{figure}

	\section{Asymptotic behaviour of $\mathcal{P}(t)$ under random dual-unitary gates}
	\label{app:Masymptotics}
	
	{ 
		In this appendix we characterise the asymptotic behaviour of the purity in dual unitary circuits with random local gates. In particular, following Ref.~\cite{foligno2022growth} we consider local gates of the form
		\begin{equation}
			u_{+}(\tau,x)\otimes u_{-}(\tau,x) \cdot U \cdot v_{+}(\tau,x)\otimes v_{-}(\tau,x),
			\label{eq:gates}
		\end{equation}
		where $U$ is a fixed two-site dual-unitary gate and $u_{\pm},v_\pm$ are random single-site matrices ${\in {\rm U}(d)}$ distributed independently in the spacetime. In this setting, Ref.~\cite{foligno2022growth} proved that if 
		\be
		p(U) \geq \frac{d^2-1}{d^2} \left(1-\frac{1}{\sqrt{2d+2}}\right).
		\label{eq:pUbound}
		\ee
		(cf.~\eqref{eq:entanglingpowerdef}), then   
		\begin{equation}
			d^{x}\dualave{\mathcal P(x/2)} \leq  A + B x\,,\qquad A, B \geq 0, 
		\end{equation}
		where $\dualave{\cdot}$ is the average over ${\boldsymbol u}$ drawn from the full ${\rm U}^{\otimes 4Lt}(d)$ group (Haar average). 
		
		Here we want to show that for non-solvable states $d^{x}\dualave{\mathcal P(x/2)}$ is bounded by a linearly growing function also from below. Namely
		\be
		d^{x}\dualave{\mathcal P(x/2)} \geq C x. 
		\ee
		Defining the convenient auxiliary quantity 
		\be
		\mathcal{M}_x = d^{x} \mathcal P(x/2)  - d^{x-1} \mathcal P(x/2-1/2),
		\ee
		our goal is to show 
		\begin{equation}
			\lim_{x\to\infty}\dualave{\mathcal{M}_x} = \dualave{\mathcal{M}_\infty}  >0. 
		\end{equation} 
		This proves Assumption.~\eqref{asp:1} in the random dual-unitary setting.} 
	
	First, we note that $\dualave{\mathcal{M}_x}$ can be related to the function $\mathcal{Q}_x$ --- introduced in Eq.~(85) of Ref.~\cite{foligno2022growth} --- as follows
	\begin{align}
		\dualave{\mathcal{M}_x}=d^x \frac{c-1}{\sqrt{d^2-1}}\mathcal{Q}_x.\label{eq:Mxcorrespondence}
	\end{align}
	Here the parameter $c$ is defined in terms of the initial state matrix $m$ (cf. Eq.~\eqref{eq:initialstate})
	\begin{align}
		c=\frac{1}{d}{\tr[\left(m m^\dagger\right)^2]}\in[1,d]. 
	\end{align}
	In particular, the value ${c=1}$ corresponds to a unitary initial state matrix $m$, i.e., a solvable initial state. In this case, noting that $\mathcal{Q}_x$ is finite for ${c=1}$, Eq.~\eqref{eq:Mxcorrespondence} gives $\mathcal{M}_x=0$. This is the expected result for solvable states: the norm of $\bra{L_\gamma}$ is equal to one and its increment ${\mathcal{M}_{t^*}}$ is zero. From now on we consider $c>1$ and argue that in this case $\dualave{\mathcal{M}_{x}}$ is always strictly larger than zero. 
	
	We begin by noting that a direct application of Eq.~(86) of Ref.~\cite{foligno2022growth} gives 
	\begin{align}
		\dualave{\mathcal{M}_x}=\dualave{\mathcal{M}}_{x_0}+\sum_{i=x_0+1}^x \mathcal{S}_i.
		\label{eq:recrel}
	\end{align}
	Here we introduced 
	\begin{equation}
		\mathcal S_x = d^x \frac{c-1}{\sqrt{d^2-1}} \mathcal R_x,
	\end{equation}
	where $\mathcal R_x$ is the function defined in Eq.~(87) of Ref.~\cite{foligno2022growth}. 
	
	Next, using Eq.~(103) of Ref.~\cite{foligno2022growth} we conclude 
	\begin{align}
		|\mathcal{S}_x|\le A a^x,
		\label{eq:expdecay}
	\end{align}
	where $p$ is the entangling power of $U$ (cf. Eq.~\eqref{eq:entanglingpowerdef}) and we introduced
	\begin{align}
		&\lambda =(1-p)^2+\frac{p^2}{d^2-1},\\
		&a =\frac{d+c}{d+1} d \lambda \label{eq:adef},\\
		&A =\frac{(c-1)^2}{(d+c)^2}\frac{d+1}{d-1} \sqrt{\frac{d^2-1}{d^2 \lambda^3}}.
		\label{eq:Adef} 
	\end{align}
	For high enough values of the entangling power
	\begin{align}
		p>\bar{p}(d)=\frac{d^2-1}{d^2}\left(1-\frac{1}{\sqrt{2d+2}}\right),\label{eq:entanglingpowerboound}
	\end{align}
	it is immediate to verify that $a<1$ for any value of $c$ in the range $[1,d]$, which allows to find an upper bound for $	\dualave{\mathcal{M}_x}$. Namely
	\begin{align}
		\dualave{\mathcal{M}_x}&\le \dualave{\mathcal{M}_{x_0}}+\sum_i \abs{\mathcal{S}_i}\notag\\
		&\le \dualave{\mathcal{M}_{x_0}}+ \frac{A}{1-a}  a^{x_0+1}
		\label{eq:condition0}
	\end{align}
	This bound has been first presented in Ref.~\cite{foligno2022growth}. Our goal here is to bound $\dualave{\mathcal{M}_x}$ also from below, showing that it is always strictly larger than $0$. To this end we combine \eqref{eq:recrel}, \eqref{eq:expdecay}, and the triangle inequality to write 
	\begin{align}
		\dualave{\mathcal{M}_x}&\ge \dualave{\mathcal{M}_{x_0}}-\sum_i \abs{\mathcal{S}_i}\notag\\
		&\ge 	\dualave{\mathcal{M}_{x_0}}-A \frac{a^{x_0+1}-a^{x+1}}{1-a}.
		\label{eq:condition1}
	\end{align}
	For $p$ fulfilling the bound \eqref{eq:entanglingpowerboound} one has $a<1$. This implies that, if we find an $x_0$ such that 
	\begin{align}
		\dualave{\mathcal{M}_{x_0}}> A \frac{a^{x_0+1}}{1-a},
		\label{eq:boundMx}
	\end{align}
	then 
	\begin{align}
		\lim_{x\rightarrow \infty}\dualave{\mathcal{M}_{x}}=\dualave{\mathcal{M}_{\infty}}> 0.
	\end{align}
	In order to get some intuition it is useful to consider two limiting cases. First we fix the values of $x$, $p$, $d$, and restrict ourselves to a neighbourhood of the solvable case, which corresponds to $c=1$. We choose the neighbourhood to be small compared to the other parameters so that we can treat everything perturbativelly around the lowest nontrivial order of the solvable case
	\begin{align}
		c\in[1,1+\epsilon], \quad \epsilon \ll \frac{1}{x},{1-a}.
	\end{align} 
	In this situation, it is easy to see that 
	\begin{align}
		\dualave{\mathcal{M}_x}= O(c-1), \qquad A =O((c-1)^2), \label{eq:cexpansion}
	\end{align}
	which immediately imply the validity of Condition \eqref{eq:boundMx} if $c\ne 1$. This shows that some properties of the solvable case are not stable under perturbations.
	
	The other useful limit is $d\gg 1$, which makes the expressions \eqref{eq:Adef}, \eqref{eq:adef} much easier to handle. We consider $\dualave{\mathcal{M}_4}$ at leading order in $d$, expanded at the first relevant order for $1-p$. We consider gates with entangling power close to the one of the Hadamard gate, or larger, meaning that\begin{align}
		1-p\lessapprox O(d^{-1})
	\end{align} 
	The asymptotic expression at the lowest relevant order is:
	\begin{align}
		&\dualave{\mathcal{M}_4}=(c-1)(1+c(1-p)) > (c-1)=O(d^0)
	\end{align}
	Expanding Eqs \eqref{eq:Adef},\eqref{eq:adef}, we find
	\begin{align}
		&A=\frac{(c-1)^2}{(d+c)^2}\frac{d^3}{((1-p)^2d^2+1)^\frac{3}{2}} \lessapprox O(d^3)
		\\
		&a=\frac{d+c}{d}\frac{(1-p)^2 d^2+1}{d}=O(d^{-1})
	\end{align}
	
	Putting together everything in condition \eqref{eq:boundMx}, we see the left side is $O(d^0)$, and the right side is $O(d^-2)$, so the condition is respected.
	Finally, to address the general case, we compute $\dualave{\mathcal{M}_{x_0}}$ numerically for a high value of $x_0$, showing that the bound \eqref{eq:boundMx} holds for a non-trivial interval of entangling powers. 
	
	In particular, we consider the case $d=2$. In this case, the values of the entangling power $p$ fulfilig the bound \eqref{eq:entanglingpowerboound} are given by 
	\begin{align}
		\frac{2}{3}>p>\bar{p}(2)\approx 0.4438.
		\label{eq:fullinterval}
	\end{align}
	Where we used that for $d=2$ the maximal attainable value of $p$ is $2/3$~\cite{foligno2022growth}.

	Computing $\dualave{\mathcal{M}_x}$ for $x=20$, we verify the inequality \eqref{eq:boundMx} holds for any $c$ if we pick $p$ in the interval  
	\begin{align}
		\frac{2}{3}>p\ge \bar n \approx 0.47660548\,.
	\end{align}
	
\begin{figure}[h!]
		\includegraphics{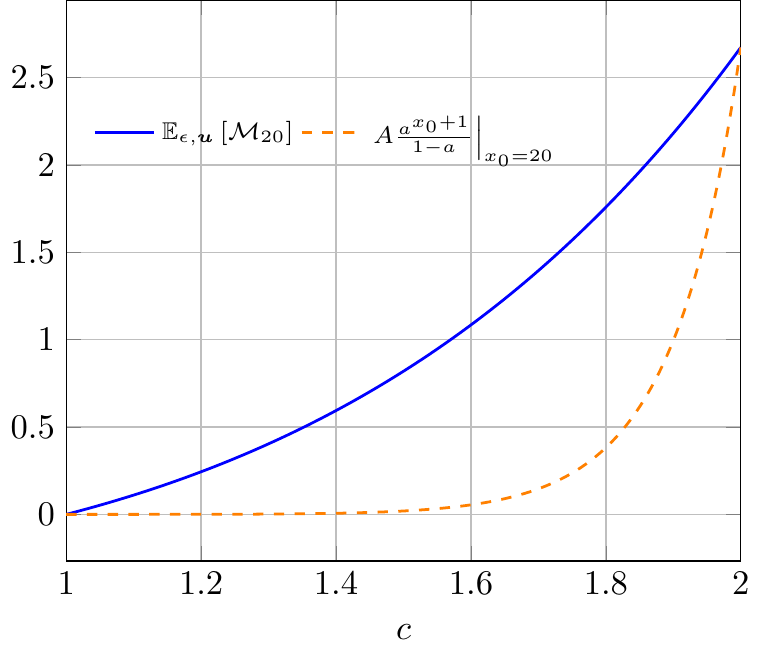}
		\ifdef{0}{}{	\begin{tikzpicture}[scale=1,remember picture]
				\begin{axis}[grid=major,
					legend columns=2,
					legend style={at={(0.75,0.71)},anchor=south east,font=\scriptsize,draw= none, fill=none},
					no markers,
					ytick distance=0.5,
					xmin=1,
					xmax=2,
					xlabel=$c$,
					y label style={at={(axis description cs:.1,.5)},anchor=south,font=\normalsize	},		
					tick label style={font=\normalsize	},	
					x label style={font=\normalsize	},	
					]
					\addplot[
					smooth,
					thick,
					mark=*,
					blue,
					] table[x index=0,y index=1] {plots_data/databound.dat};
					\addlegendentry{$\dualave{\mathcal{M}_{20}}$}
					\addplot[
					smooth,
					thick,
					mark=*,
					orange,
					dashed
					] table[x index=0,y index=2] {plots_data/databound.dat};
					\addlegendentry{$\left.A \frac{a^{x_0+1}}{1-a}\right|_{x_0=20}$}
				\end{axis}	
		\end{tikzpicture}}
		\caption{$\dualave{\mathcal{M}_{x_0}}$ and $A \frac{a^{x_0+1}}{1-a}$ versus $c$ for $x_0=20$, $p=\bar{n}$ and $d=2$. Note the linear and quadratic growth around $c=1$ according to \eqref{eq:cexpansion}.}
		\label{plot:bound}
	\end{figure}

	Note that, to verify the equality for $p$ close to $\bar{p}(d)$ one would need to consider arbitrarily large values of $x_0$. Indeed the denominator $({1-a})^{-1}$ diverges at $p=\bar{p}(d)$. As an example, we show in Fig \ref{plot:bound} the value of $A {a^{x_0+1}}/({1-a})$ versus $\dualave{\mathcal{M}_{x_0}}$, for $x_0=20$ as a function of $c$. 
	
	\section{Parameterisation of dual-unitary gates for numerical experiments}
	\label{sec:parameterisation}
	
	To produce the data presented in plots involving dual unitary gates.(i.e. Figs \ref{plot:correction_maximal_growth}, \ref{plot:higherrenyibound},
	\ref{plot:sigmaentropy2},
	\ref{fig:pk},
	\ref{plot:Von_neumann_growth_r}, \ref{plot:fitteddata} and \ref{plot:Von_neumann_growth}) we parameterised the gates as in Eq.~\eqref{eq:gates}, with fixed one-site unitaries $u_\pm,v_{\pm}$:\begin{align}
		u_+=	\begin{pmatrix}
			0.204-0.971 i &-0.108-0.068i\\
			0.125+0.0254i & -0.524+0.842i
		\end{pmatrix}\\
		u_-=	\begin{pmatrix}
			-0.279-0.921i&0.238+0.132i\\-0.272+0.017i & -0.649+0.710i
		\end{pmatrix} \\
		v_+=	\begin{pmatrix}
			-0.025-0.367i
			& -0.921-0.127i
			\\ 0.908-0.202i
			& 0.005+0.368i
		\end{pmatrix}\\
		v_-=	\begin{pmatrix}
			0.380-0.321i
			& 0.436+0.750i
			\\0.807+0.318i
			&0.260-0.424i
		\end{pmatrix},
	\end{align}
	and two-site dual unitary given by  
	\begin{equation}
		U(p)=\begin{pmatrix}
			e^{-i J(p)} & 0& 0& 0\\
			0&0&- i e^{i J(p)}&0\\
			0&-i e^{i J(p)}&0&0\\
			0& 0& 0&e^{-i J(p)}
		\end{pmatrix},
	\end{equation}
	where 
	\be
	J(p)=\arcsin(\sqrt{1-\frac{3p}{2}}),
	\ee
	and $p\in[0,1]$. Using the definition of entangling power in \eqref{eq:entanglingpowerdef} one can immediately verify that 
	\begin{equation}
		p(U(p))=p. 
	\end{equation}

	\section{Recurrence relation for $\overline{\braket{L_\gamma}}$}\label{app:averagedunitaries}
	In this Appendix, we consider the Haar averages in Eq. \eqref{eq:ratio}.
	A doubly folded averaged unitary is projected on $2-$dimensional local vector space, spanned by the normalized vectors\begin{align}
		\ket{\mcirc_2}=\frac{\ket{\I}}{d}\qquad 
		\ket{\msquare_2}=\frac{\ket{(12)}}{d},
	\end{align}
	where $\ket{(12)}, \ket{\I}$ refer to the permutation vectors represented in Fig \eqref{fig:bd_cond} (a).

	Moreover, an Haar averaged unitary gate fulfils the following relations
	\begin{align}
		&\overline{\fineq[-0.8ex][1][1]{
				\roundgate[0][0][1][tr][bertiniblue]
				\roundgate[0][0.20][1][tr][bertinired]
				\roundgate[0][0.4][1][tr][bertiniblue]
				\roundgate[0][0.60][1][tr][bertinired]
		}}=\fineq[-0.8ex][1][1]{
			\roundgate[0][0][1][tr][violet]
		},\notag\\
		&\fineq[-0.8ex][1][1]{
			\roundgate[0][0][1][tr][violet]
			\wcircc{.5}{.5}
			\wsqrr{-.5}{.5}
		}=\frac{d}{d^2+1}\left(\fineq[-0.8ex][1][1]{
			\draw[thick](.2,.5)--(.2,0);
			\draw[thick](-.5,.5)--(-.5,0);
			\wcircc{.2}{.5}
			\wcircc{-.5}{.5}
		}+\fineq[-0.8ex][1][1]{
			\draw[thick](.2,.5)--(.2,0);
			\draw[thick](-.5,.5)--(-.5,0);
			\wsqrr{.2}{.5}
			\wsqrr{-.5}{.5}
		}\right),\notag\\
		&\fineq[-0.8ex][1][1]{
			\roundgate[0][0][1][tr][violet]
			\wcircc{.5}{.5}
			\wcircc{-.5}{.5}
		}=\fineq[-0.8ex][1][1]{
			\draw[thick](.2,.5)--(.2,0);
			\draw[thick](-.5,.5)--(-.5,0);
			\wcircc{.2}{.5}
			\wcircc{-.5}{.5}
		}\qquad
		\fineq[-0.8ex][1][1]{
			\roundgate[0][0][1][tr][violet]
			\wsqrr{.5}{.5}
			\wsqrr{-.5}{.5}
		}=\fineq[-0.8ex][1][1]{
			\draw[thick](.2,.5)--(.2,0);
			\draw[thick](-.5,.5)--(-.5,0);
			\wsqrr{.2}{.5}
			\wsqrr{-.5}{.5}
		}.
		\label{eq:rec2}
	\end{align}
	Here, we choose a  specific normalization for the initial state such that 
	\begin{align}
		\fineq[-0.8ex][1][1]{
			\draw[thick](-.5,.5)--(-.5,0);
			\wcircc{-.5}{.5}
			\fill (-0.5,0) circle (0.08);
		}=\fineq[-0.8ex][1][1]{
			\draw[thick](-.5,.5)--(-.5,0);
			\wsqrr{-.5}{.5}
			\fill (-0.5,0) circle (0.08);
		}:= 1 \label{eq:rec1normal}
	\end{align}
	
	We can define the  quantity\begin{align}
		\mathcal{A}_{x,y}= \fineq[-0.8ex][0.4][0.4]{
			\foreach \x in {0,1,...,5}{
				\foreach \y in {0,...,\x}{	
					\roundgate[2*\x-\y][\y+0.00][1][tr][violet]
					\ifthenelse{\x>2}
					{ 				\roundgate[10-\y][\y+10-2*\x][1][tr][violet]}
					{}
				}
			}
			\foreach \x in {0,...,7}{
				\wcirc{\x-.5}{.48+\x}
			}
			\foreach \x in {0,...,3}{
				\wsqr{\x+7.5}{7.5-\x}
			} 	\foreach \x in {0,...,4}{
				\wsqr{10.5}{3.5-\x}
			}
			\foreach \x in {0,1,...,10}{
				\foreach \y in {0}{
					\fill (\x-0.5,-0.5+\y) circle (0.08);
				}
			}
			\draw [decorate, decoration = {brace}, thick,shift={(-1.3,0)}]   (0,0) -- (5,5);	 		
			\draw [decorate, decoration = {brace}, thick,shift={(4.7,6)}]   (-1+0.2,-1+0.2) -- (2,2);
			\node[scale=2] at (01,4) {$x$} ;
			\node[scale=2] at (4.5,7.5) {$y$} ;
		}\: d^{x+y},
	\end{align}
	with $y\le x+1$. It is immediate to see that
	\begin{align}
		\overline{\braket{L_{\gamma_{t+1}}}}=\mathcal{A}_{t,t+1}.
	\end{align}
	Using \eqref{eq:rec2}-\eqref{eq:rec1normal}, one finds the following recurrence relations
	\begin{align}
		&\mathcal{A}_{x,y}=\begin{cases}
			\frac{d^2}{d^2+1}\left(\mathcal{A}_{x-1,y-1}+\mathcal{A}_{x,y-1}\right)\qquad y=x+1,\\
			\frac{d^2}{d^2+1}\left(\mathcal{A}_{x-1,y}+\mathcal{A}_{x,y-1}\right)\qquad \quad 0 < y\le x,\\
			\left(\frac{2d^2}{d^2+1}\right)^x \qquad \qquad \qquad \qquad\qquad y=0.
		\end{cases}
		\label{eq:recA}
	\end{align}
	The treatment for the average $\overline{\bra{L_\gamma}\hspace{1pt}\ket{L_{\gamma/2}}\otimes\ket{L_{\gamma/2}}}$ is similar: the average corresponds to the diagram
	\begin{align}
		d^{2t-1} \phantom{a}\fineq[-0.8ex][0.4][0.4]{
			\foreach \x in {0,1,...,2}{
				\foreach \y in {0,...,\x}{	
					\roundgate[2*\x-\y][\y+0.00][1][tr][violet]
					\roundgate[4-\y][\y+4-2*\x+0.00][1][tr][violet]
				}
			}
			\foreach \x in {0,...,4}{
				\wcirc{\x-.5}{.5+\x}
			}
			\foreach \x in {0,1,...,4}{
				\foreach \y in {0}{
					\wcirc{\x-0.5}{-0.5+\y}
				}
			}
			\foreach \x in {0,1,...,2}{
				\foreach \y in {0,...,\x}{	
					\roundgate[2*\x-\y][-6+\y+0.00][1][tr][violet]
					\roundgate[4-\y][\y-2-2*\x+0.00][1][tr][violet]
				}
			}
			\foreach \x in {0,...,4}{
				\wcirc{\x-.5}{.5+\x-6}
			}
			\foreach \x in {0,1,...,4}{
				\foreach \y in {0}{
					\fill (\x-0.5,-0.5+\y-6) circle (0.08);
				}
			}
			\foreach \x in {0,...,11}{
				\wsqr{4.5}{4.5-\x}}
		} \:=
		d^{t}d^{t-1}
		\fineq[-0.8ex][0.4][0.4]{
			\foreach \x in {0,1,...,2}{
				\foreach \y in {0,...,\x}{	
					\roundgate[4-\y][\y+4-2*\x+0.00][1][tr][violet]
				}
			}
			\foreach \x in {2,...,4}{
				\wcirc{\x-.5}{.5+\x}
			}
			\foreach \x in {2,...,4}{
				\wcirc{\x-0.5}{-0.5-\x+4}
			}
			\foreach \x in {0,1,...,2}{
				\foreach \y in {0,...,\x}{	
					\roundgate[2*\x-\y][-6+\y+0.00][1][tr][violet]
					\roundgate[4-\y][\y-2-2*\x+0.00][1][tr][violet]
				}
			}
			\foreach \x in {0,...,4}{
				\wcirc{\x-.5}{.5+\x-6}
			}
			\foreach \x in {0,1,...,4}{
				\foreach \y in {0}{
					\fill (\x-0.5,-0.5+\y-6) circle (0.08);
				}
			}
			\foreach \x in {0,...,11}{
				\wsqr{4.5}{4.5-\x}}
		}\,.
	\end{align}
	The top diagram can be expressed again with recursive relation, defining the quantity $\mathcal{B}_{x,y}$
	\begin{align}
		\mathcal{B}_{x,y}=d^{x+y} \fineq[-0.8ex][0.4][0.4]{
			\foreach \x in {2,...,4}{
				\foreach \y in {0,...,\x}{	
					\roundgate[4-\y][\y+4-2*\x+0.00][1][tr][violet]
				}
			}
			\foreach \x in {0,...,2}{
				\wcirc{\x-.5}{.5+\x}
			}
			\foreach \x in {0,...,4}{
				\wcirc{\x-0.5}{-0.5-\x}
			}
			\foreach \x in {4,...,9}{
				\wsqr{4.5}{4.5-\x}}
			\foreach \x in {0,1,2}{
				\wsqr{4.5-\x}{0.5+\x}}
			\draw [decorate, decoration = {brace}, thick,shift={(-1.3,0)}]   (0,0) -- (3,3);	 	
			\draw [decorate, decoration = {brace}, thick,shift={(-1.5,-0.2)}]   (5,-5)--(0,0);	 		
			\node[scale=2] at (-0.5,2) {$y$} ;
			\node[scale=2] at (0.5,-3.5) {$x$} ;
		}\,,
	\end{align}
	which fulfils
	\begin{align}
		&\mathcal{B}_{x,y}=\begin{cases}
			\frac{d^2}{d^2+1}\left(\mathcal{B}_{x-1,y-1}+\mathcal{B}_{x,y-1}\right)\qquad y=x,\\
			\frac{d^2}{d^2+1}\left(\mathcal{B}_{x-1,y}+\mathcal{B}_{x,y-1}\right)\qquad \quad y<x,\\
			1 \qquad  \qquad  \qquad  \qquad  \quad  \qquad\quad\:\:\, y=0.
		\end{cases}\label{eq:recB}
	\end{align}	
	Eq.~\eqref{eq:ratio} can then be expressed in terms of these quantities as (assuming a generic bipartition of the temporal state at time $t$ in $t=t_1+t_2$)
	\begin{align}
		\bar r_t	=\frac{\mathcal{B}_{t_1,t_1}\sqrt{\mathcal{A}_{t_2,t_2+1}}}{\sqrt{\mathcal{A}_{t,t+1}\mathcal{A}_{t_1,t_1+1}}}\,.
		\label{eq:equation1}
	\end{align}

	Interestingly, we can map the recurrence relation 
	\eqref{eq:recA} in a different problem. First we slightly change normalization by defining \begin{align}
		\mathcal{A}_{x,y}=	\widetilde{\mathcal{A}}_{x,y}\left(\frac{d^2}{d^2+1}\right)^{x+y},
	\end{align}
	then, the quantity $\widetilde{A}_{x,y}$ can be thought of as the number of paths connecting the two black dots in the following grid, without crossing the dashed line  $y=x+2$ and in the minimum number of steps
	\begin{align}
		\fineq[-0.8ex][0.5][0.5]{
			\def\k{8};
			\foreach \i in {0,...,\k}
			{
				\draw[thick] (0,\i)--(\k-1,\i);
				\ifthenelse{\i<\k}
				{
					\draw[thick] (\i,0)--(\i,\k);
				}{}
			}
			\foreach \i in {2,...,\k}
			{
				\draw[fill=blue] (\i-2,\i)circle (.08);
			}
			\draw[dashed] (-.2,1.8)--(\k-1.8,\k+0.2);
			\draw[very thick, red] (0,0)--(\k-1,0);
			\draw[fill=black] (\k-1,\k)circle (0.14);
			\draw[fill=black] (0,0)circle (0.14);
			\draw [decorate, decoration = {brace}, thick,shift={(0,-0.4)}]    (\k-1,0)--(0,0);
			\draw [decorate, decoration = {brace}, thick,shift={(0+.4,0)}]    (\k-1,\k)--(\k-1,0);
			\node[scale=2] at (0.5*\k-0.5,-1) {$x$};
			\node[scale=1.5] at (-1,-.5) {$(0,0)$};
			\node[scale=2] at (\k,0.5*\k) {$y$};
			\node[scale=1.5] at (-1+\k+0.5,\k+0.5) {$(x,y)$};
		}.
	\end{align}
	Each paths gains a weight $1+1/{d^2}$ every time it touches the top boundary and a factor $2$ for every crossed red link of the bottom boundary.
	To compute the asymptotic scaling of this quantity we can ignore the $1+1/{d^2}$ weight, which does not change the scaling for  $d$ large enough, since the number of paths touching the top boundary exactly $p$ times is exponentially suppressed in $p$ with respect  to the total number of paths, which balances the 
	\begin{equation}
		\left(\frac{d^2+1}{d^2}\right)^p,
	\end{equation}
	weight (this holds for $d>1$). 
	This statement can be made more precise using theorem 2 of \cite{latticepaths}. It is possible to show then that the number of  paths touching the boundary $p$ times is the following (the convention is to set a binomial coefficient to  0 if the top argument is lower than the bottom one, or the latter is $<0$): \begin{align}
		\binom{x+y-p}{x}-\binom{x+y-p}{x+2},
	\end{align}
	so that, considering the appropriate weight for these paths, and setting $x=y-1=t$ (we ignore the weights for the red links on the bottom for the sake of this argument), the total is\begin{align}
		\sum_{p=0}^t \left(\binom{2t+1-p}{t}-\binom{2t+1-p}{t+2}\right) \left(1+\frac{1}{d^2}\right)^p.
	\end{align}
	Studying the asymptotic scaling of this sum(using the Stirling formula and approximating the sum with an integral, expanded around the maximum), we can see the scaling at leading order in $t$  is unaffected as long as $d>1$.\\
	We can call $a_n$ the number of paths connecting $(x,y)$ to $(n,0)$, then we can write
	\begin{align}
		\centering\widetilde{\mathcal{A}}_{x,y}\sim \sum_{n=0}^{x} (a_n-a_{n+1}) 2^n,	
	\end{align}
	with 
	\begin{align}
		\!\!\!\!\!\!a_n\!=\!\begin{cases}
			\binom{x+y-n}{x-n}\!-\!\binom{x+y-n}{x+3} \,\, 0\le n \le x,\, n\le y-3,	\\
			\binom{x+y-n}{y} \hspace{48pt} 0\le n\,, n\ge y-2,	\\
			0 \hspace{79pt} n<0\,,  n>x.
		\end{cases}
	\end{align}
	
	In particular,  we can rewrite  \eqref{eq:equation1} as
	\begin{align}
		\sum_{n=0}^{x} (a_n-a_{n+1}) 2^n=	
		a_0  + \sum_{n=1}^{x} a_n 2^{n-1},\label{eq:tosum}
	\end{align}
	and find an asymptotic expression for $a_n$ using Stirling's formula (we take $(x,y)=(t,t+1)$)
	\begin{align}
		&a_n 2^{n-1} \sim \frac{1}{2} f\left(\frac{n}{t}\right)\\
		&f(z)=\frac{2^{zt}}{\sqrt{2\pi t }}\frac{(2-z)^{(2-z)t}}{(1-z)^{(1-z) t}}\left[\frac{6}{t}+2z-z^2+O\left(\frac{z}{t}\right)\right]\!.\notag
	\end{align}
	Finally, we can estimate the sum in \eqref{eq:tosum} with an integral, computed with the saddle point approximation
	\begin{align}
		\sum_{n=1}^{t} a_n 2^{n-1} =O \left(t \int_0^{\infty} f(z) \text{d} z \right).
	\end{align}
	We expand $\log(f(z))$ around its minimum $z_0$, at the leading orders in $t$: \begin{align}
		&z_0= \sqrt{\frac{2}{t}}-\frac{11}{4t}+o\left(\frac{1}{t}\right),\notag\\
		&\left.\dv{\log(f)}{z}\right|_{z=z_0}=O\left(\frac{1}{\sqrt{t}}\right),\\
		&\left.\dv[2]{\log(f)}{z}\right|_{z=z_0}=-t +o(t),\notag\\
		&\log(f(z_0))=	t\log(4)-\log(t)+\frac{1}{2}\left(\log(\frac{4}{\pi})-1\right).\notag
	\end{align}
	Using the saddle point approximation, for large $t$ we find\begin{align}
		\int_0^{\infty} f(z) \text{d} z \sim  \frac{4^{t}}{t^{-\frac{3}{2}}} \frac{2}{\sqrt{e\pi}} \int_{-\sqrt{2}}^\infty e^{-\frac{x^2}{2}}\text{d} x,
	\end{align}
	which gives 
	\begin{align}
		\mathcal{A}_{t,t+1}=O\left( \left(\frac{2 d^2}{d^2+1}\right)^{2t}{t^{-\frac{1}{2}}}\right).	
		\label{eq:equation2}
	\end{align}
	We can compute $\mathcal{B}_{t,t}$ with a similar approach:
	in this case we need to consider the paths connecting a point $(x,y)$ to the origin without crossing the line $x=y+1$. As before we approximate this quantity by  ignoring the weights obtained touching the top boundary $1+1/{d^2}$. We can then write
	\begin{align}
		\mathcal{B}_{x,y}=O\left(\left(\frac{d^2}{d^2+1}\right)^{x+y}\sum_{n=0}^x b_n \left(\frac{d^2+1}{d^2}\right)^n\right),
		\label{eq:equation4}
	\end{align}
	where $b_n$ is the number of paths connecting $(x,y)$ to $(n,1)$
	\begin{align}
		\!\!\!\!\!b_n\!\!=\!\!\begin{cases}
			\binom{x+y-n-1}{y-1}\!-\!\binom{x+y-n-1}{x+1}\,\, 0\!\le \!n\! \le\! x,\, n\!\le\! y\!-\!2	\\
			\binom{x+y-n-1}{y-1} \hspace{58pt} 0\!\le \!n\!\le\! x,\, n\!\ge\! y\!-\!1	\\
			0 \hspace{97pt} n\!<\!0,\,  n\!>\!x.
		\end{cases}
	\end{align}
	As before we estimate the sum \eqref{eq:equation4} with an integral, in the case $(x,y)=(t,t)$
	\begin{align}
		&\mathcal{B}_{t,t}	= O\left(t\int_{0}^\infty g(z) \text{d} z\right),\notag\\
		& g\left(\frac{n}{t}\right)\sim b_n \left(\frac{d^2+1}{d^2}\right)^n\\
		& g(z)=\left(\frac{d^2}{d^2+1}\right)^{zt}\frac{1}{t\sqrt{2\pi t}}\frac{(2-z)^{(2-z)t}}{(1-z)^{(1-z)t}}\left(4+2z+O(z^2)\right)\notag
	\end{align}
	The difference with the previous case is that $g'(z)\ne 0$ in the domain we are interested in, as long as $d>1$. The maximum value attained is $g(z=0)$, thus we can estimate the integral expanding $\log(g)$ around $z=0$ at first order 
	\begin{align}
		\int_{0}^\infty g(z)	\text{d} z\sim \frac{g^2(0)}{g'(0)}=O({4^t}{t^{-\frac{5}{2}}}),
	\end{align}
	finding the scaling
	\begin{align}
		\mathcal{B}_{t,t}=O\left(\left(\frac{2d^2}{d^2+1}\right)^{2t}{t^{-\frac{3}{2}}}\right).\label{eq:equation3}
	\end{align}
	Plugging \eqref{eq:equation2}-\eqref{eq:equation3}
	into \eqref{eq:equation1}, we find 
	\begin{align}
		\bar r_t=O(t^{-\frac{5}{4}}).
		\label{eq:eqfinal}
	\end{align}

	\bibliography{bibliography}

\end{document}